\documentclass[entropy,article,moreauthors,10pt,a4paper]{mdpi}
\usepackage{amsmath,amssymb}

\firstpage{1} 
\articlenumber{x}
\doinum{10.3390/------}
\pubvolume{xx}
\pubyear{2016}
\copyrightyear{2016}

\newcommand{\ie}{\emph{i.e., }}
\newcommand{\reff}[1]{(\ref{#1})}
\newcommand{\eref}[1]{Eq.\reff{#1}}
\newcommand{\erefs}[1]{Eqs.\reff{#1}}
\newcommand{\figref}[1]{FIG.\ref{#1}}

\newcommand{\np}{n_p}
\newcommand{\nb}{n_B}

\newcommand{\Nu}{N_1}

\newcommand{\Nn}{N_n}

\newcommand{\kj}{k_j}

\newcommand{\lj}{\ell_j}
\newcommand{\lu}{\ell_1}

\newcommand{\omp}{\omega_p}

\newcommand{\phij}{\varphi_j}

\newcommand{\phija}{\phi_j}

\newcommand{\etab}{\bar{\eta}}

\newcommand{\al}{\alpha}
\newcommand{\Na}{N_\al}
\newcommand{\sa}{\sigma_{\al}}
\newcommand{\ch}{\mathcal{S}}
\newcommand{\lam}{\langle\delta\bar{\xi}^{'2}\rangle}
\newcommand{\lamal}{\langle\delta\bar{\xi}^{'2}_\al\rangle}
\newcommand{\vres}{v_{res}}
\newcommand{\dvnl}{\Delta v^{NL}}
\newcommand{\dvsep}{\Delta v^{SEP}}
\newcommand{\kk}{K}

\Title{Mixed Diffusive-Convective Relaxation of a Warm Beam of Energetic Particles in Cold Plasma}
\Author{Nakia~Carlevaro$^{1,2}$, Alexander~V.~Milovanov$^{2,3}$, Matteo~V. ~Falessi$^{4,2}$, Giovanni~Montani$^{2,5}$, Davide~Terzani$^{6}$ and Fulvio~Zonca$^{2,7}$*}
\address{%
$^{1}$CREATE Consortium, Via Claudio, 21 (80125) Napoli, Italy.\\
$^{2}$ENEA, C. R. Frascati, Via E. Fermi, 45 (00044) Frascati (RM), Italy.\\
$^{3}$Dep. of Space Plasma Physics, Space Research Inst., Russian Academy of Sciences, 117997 Moscow, Russia.\\
$^{4}$Dep. of Physics, ``RomaTre'' University, Via della Vasca Navale, 84 (00146) Roma, Italy.\\
$^{5}$Dep. of Physics, ``Sapienza'' University of Rome, P.le Aldo Moro, 5 (00185) Roma, Italy.\\
$^{6}$Dep. of Physics, University of Naples ``Federico II'', Via Cinthia, I-80126 Napoli, Italy. \\
$^{7}$Inst. for Fusion Theory and Simulation and Dep. of Physics, Zhejiang Univ., Hangzhou 310027, P.R.China.}
\corres{fulvio.zonca@enea.it}

\abstract{This work addresses the features of fast particle transport in the bump-on-tail problem for varying width of the fluctuation spectrum, in the view of possible applications to studies of energetic particle transport in fusion plasmas. Our analysis is built around the idea that strongly shaped beams do not relax through diffusion only and that there exists an intermediate time scale where the relaxations are convective (ballistic-like). We cast this idea in the form of a self-consistent nonlinear dynamical model, which extends the classic equations of the quasi-linear theory to ``broad'' beams with internal structure. We also present numerical simulation results of the relaxation of a broad beam of energetic particles in cold plasma. These generally demonstrate the mixed diffusive-convective features of supra-thermal particle transport essentially depending on nonlinear wave-particle interactions and phase-space structures. Taking into account modes of the stable linear spectrum is crucial for the self-consistent evolution of distribution function and the fluctuation intensity spectrum.}



\begin{document}

\section{Introduction}\label{intro}

The relaxation of supra-thermal particle beams in plasmas is a problem of fundamental significance. Dating back to the 1960s, it provides a paradigm for the quasi-linear theory of weak plasma turbulence \cite{Vedenov,Pines}, which finds applications in a wide field ranging from astrophysics and cosmical geo-physics \cite{Zaslavsky,Sagdeev} to fusion plasma \cite{Sagdeev,Kadomtzev}. Its original important promotion is due to Bernstein, Greene and Kruskal (BGK), whose seminal work in Ref. \cite{Kruskal} has come to be known as the classic ``bump-on-tail'' (BoT) problem. Specific implications concern Landau damping \cite{oneil65,mazitov65} and nonlinear behavior of the beam-plasma instability driven by wave-particle interactions \citep{Kadomtzev,OWM71,OM68} (see also \citep{Sh63,OL70,SS71,Th71,L72,Ma72}). The ``perturbation theory of strong plasma turbulence'', originally proposed by Dupree \cite{dupree66} and illuminating resonance broadening by fluctuation induced stochastic particle motion also ought to be mentioned for its importance and crucial implications in plasma research.

The relevance of the BoT problem for fusion plasmas research was revived in the 1990's by Berk and Breizman \citep{BB90a,BB90b,BB90c}, who proposed it as a paradigmatic model to investigate and understand nonlinear interaction of supra-thermal particles with Alfv\'en fluctuations \citep{CZ07,CZ13,BB90c,BS11}. In fact, nonlinear interplay of energetic particles (EPs) with Alfv\'enic fluctuations (such as Alfv\'en eigenmodes (AEs), EP modes and drift Alfv\'en waves \citep{ZCrmp,CZ07,CZ13,Fa07}) and their consequences for fluctuation induced cross-field transport constitute basic phenomena for thermonuclear plasmas \citep{Fa07,H08,BS11,La13,VB13,WB11,ML13,ML12,GPT14,ZC15ppcf,ZC15njp,Pi15}.

The fluctuation spectrum of AEs, EP driven modes and drift Alfv\'en waves in fusion plasmas covers disparate spatiotemporal scales and can have both ``broad'' features, such as those of typical plasma turbulence, as well as an almost ``coherent'' (``narrow'') nearly periodic component \cite{ZC15ppcf,ZCrmp}. A
line-broadened quasi-linear approach has been proposed in Refs. \citep{BB95b,BB96b} for computing EP transport 
by means of a diffusion equation, which could address not only overlapping resonant Alfv\'enic fluctuations but also the 
broadening of the resonant spectrum for isolated instabilities in the case of multiple AEs. This model has been extended and compared with experimental observations \citep{GG12}
and with numerical solutions of the BoT paradigm \citep{GBG14}. In general, however, understanding the complex features of EP transport in fusion plasmas requires going beyond the local description of fluctuation induced fluxes and extending the diffusive transport paradigm \cite{ZC15ppcf,ZCrmp}. Accounting for modes of the linear stable spectrum is also crucial \cite{lauber15,schneller16}. Thus, posing these issues for the BoT problem becomes an interesting and relevant research topic, in the light of its possible implications as paradigm for Alfv\'enic fluctuation-induced supra-thermal particle transport in fusion plasmas near marginal stability. In particular, a relevant question to pose is assessing the changing behavior of EP transport when properties of the fluctuation spectrum are modified from ``narrow'' (nearly periodic or coherent) to ``broad'' (turbulent). This is the main focus of the present work.
Non-diffusive processes could also be of interest in various other areas of research. For example in the energy extraction from plasmas (see, \emph{e.g.}, \cite{Fi14,HSF15} and references therein for a recent discussion of the energy extraction problem); and in the light power absorption by plasmas \cite{LW14,LW13}, where meso-scale convective (coherent) phenomena may generate relevant effects.

The issues addressed in the present work clearly have common aspects with important and well-known results of the vast BoT literature. More precisely, we deal with 
resonant wave-coupling mediated by resonant particles, which is summarized and explained in the review work by Laval and Pesme \cite{Laval}. For sufficiently strong nonlinearity, such that resonance broadening dominates over the linear growth rate, renormalization of the propagator yields to wave-coupling terms that are of the same order of quasi-linear diffusion and must be taken into account, as explained by the turbulent trapping model (TTM) \cite{laval84}. In general, phase-space structures and, thereby, nonlinear wave-particle interactions are known to be important, \emph{e.g.}, in enhancing velocity space diffusion due to longitudinal plasma turbulence \cite{CD92,EEbook,EE08}. Most of existing literature deals with ``broad'' turbulent fluctuation spectra.
In this work, meanwhile, we address the role of nonlinearity in supra-thermal particle transport by a fluctuation spectrum that is not necessarily ``broad''. We revisit the implications of the classic BoT problem and demonstrate the occurrence of relaxation dynamics of the convective type on intermediate (meso-) time scales (other than the familiar asymptotic velocity-space diffusion and the formation of the plateau \cite{Vedenov,Pines,Kadomtzev}). Our results elucidate the crucial role played by the wave-particle nonlinear interaction in determining transport in phase space both on meso- and asymptotic time scales.

The paper is organized as follows. First, we touch on the classic BoT problem in brief (Sec. 2), with a particular attention to the causes and origins of stochastic behavior of the resonant particles. In Sec. 3, we formulate a mixed convection-diffusion model of velocity space transport, which extends the familiar quasi-linear diffusion to relaxation processes in the presence of many overlapping resonances. We discuss similarities and differences of our approach with the well-established theoretical framework
of the classic BoT problem, including its aforementioned subtleties connected with a sufficiently strong nonlinearity. Mathematically, the model in Sec. 3 relies on the formalism of the Klein-Kramers equation with force-field (if external or effective self-organized) and incorporates the various exclusions from pure diffusive style models. Numerical simulation results are collected in Sec. 4, where one also finds a Hamiltonian formulation of the BoT problem based on Ref. \cite{CFMZJPP}. There, four different cases are presented and discussed, changing the strength of the nonlinearity parameter and the width of the fluctuation spectrum. It is shown numerically that convective relaxation takes place on meso-scales for strong nonlinearity and a sufficiently broad spectrum, 
in contrast to the narrow spectrum case, dominated by coherent structures. Convective relaxation is also found to occur for small nonlinearity parameter as a result of a self-consistent evolution of the fluctuation spectrum on the same time scales of particle transport. We support these findings with a simplified theory model.
Finally, Sec. \ref{conclusions} is devoted to summary and conclusions.

\section{The classic ``bump-on-tail'' problem and the Brownian random-walk paradigm}\label{botparadigm}

We will assume that the reader is familiar with the classic BoT problem \cite{Kruskal} of a distribution function that excites electrostatic waves. The basic insight is that the velocity space gradient drives (damps) the instabilities via the Cherenkov resonance $v=\vres=\omega_{k_j}/k_j$ with the plasma species (customarily associated with the energetic electrons). Here $v$ is the particle velocity and $\omega_{k_j}$ and $k_j$ are respectively the frequency and the wave-vector of the resonant mode. Instability occurs when the gradient is positive ($\partial f / \partial v > 0$) and is damped when it is negative ($\partial f / \partial v < 0$). The latter phenomenon is known as the Landau damping and in many ways is the inversion of the BoT problem (by ``velocity distribution" we mean the probability density, $f=f(v,t)$, of finding a particle at time $t$ with a velocity value between $v$ and $v+dv$ regardless of its position in real space).

In the quasi-linear theory \cite{Vedenov,Pines,Kadomtzev} of weak plasma turbulence, one assumes that the electron distribution function contains a small ``bump'' in the parameter range of large energies (much larger than the characteristic thermal energy) and that the dispersion of the electron velocities is also large compared with the thermal velocity. The bump being small implies that the level of the excited electrostatic noises is so low that the different unstable modes do not interact, hence one neglects any nonlinearities in the wave-field. Then the only nonlinearity one indeed takes into account is the feedback effect of the waves onto the averaged velocity distribution function (hence the name of this theory: quasi-linear). The linear instability growth rate, which is frequency and wave-vector dependent, is given by 
\begin{equation}
\gamma_{k_j} = \frac{\pi \omega_{p}^3}{2k_j^2n_p}\frac{\partial f}{\partial v}|_{\vres}\;,
\label{1} 
\end{equation} 
where $\omega_{p}$ is the unperturbed Langmuir frequency (\ie $\omp\equiv\sqrt{4\pi\np e^2/m_e}$, with $n_p$ the thermal plasma density; and $m_e$ and $e$ denoting the mass and charge
of electrons, respectively) and the gradient in the velocity space is taken at the Cherenkov resonance exactly. The large velocity dispersion within the bump is at the basis of another important assumption of the quasi-linear theory, namely, that the number of the excited modes is so large that their phases are represented by the random functions. Then the kicks received by the plasma particles via the resonant interactions with the waves will be also random in the limit $t\rightarrow\infty$ (here $t$ is the time). It is generally believed that this quasi-linear effect of the many excited waves onto the plasma is contained in a Brownian random walk of the energetic particles toward the thermal core and the formation of a characteristic ``plateau'' ($\partial f / \partial v = 0$) in the equilibrium velocity distribution function \cite{Vedenov,Pines}.

The conclusion that the transport in the velocity space is diffusive or Brownian random walk-like is however not at all trivial and should be addressed. It occurs as a consequence of the idea that the resonant particles are not caught by one single resonance on time scales longer than a certain characteristic time (thought as the typical bouncing time in the potential well of the wave). Instead these particles will hop in a random fashion between the many overlapping resonances, hence their motion is statistical, rather than deterministic. This statistical approach to the particle random motion in the velocity space is well documented in books and reviews (\emph{e.g.} Ref. \cite{Zaslavsky,Sagdeev} and \cite{Chirikov_USP,Report} due to the Novosibirsk school of nonlinear science).

Consider the exact equations of motion of a charged particle in the potential electric field of a wave packet  
\begin{equation}
\ddot{x} = \frac{e}{m_e} E(x,t) = \frac{e}{m_e}\sum_{j} \left[E_{k_j} e^{i(k_jx-\omega_{k_j} t)} + c.c.\right]\;,
\label{2} 
\end{equation} 
where $\dot{x} = v$ is the particle velocity, and $E_{k_j}$ is the Fourier amplitude of a mode with wave-vector $k_j$. The electric field is assumed to be ``small'' in the sense of quasi-linear theory; and here it is represented as the sum of a large number of Fourier harmonics with frequencies $\omega_{k_j}$ and wave-vectors $k_j$. Hence, it is shown, following Zaslavsky \cite{Zaslavsky}, that barely trapped particles $-$ those staying close to separatrices in phase space $(x, \dot{x})$ $-$ may occasionally become detrapped by jumping onto an open trajectory. These phenomena of occasional trapping-detrapping occur because the integrals of the motion are essentially destroyed by the perturbation $E(x,t)$ within a small layer surrounding the separatrix (known as the ergodic layer) \cite{Zaslavsky,Sagdeev}. For not too small perturbations, the width of the ergodic layer with respect to frequency is of the order of 
\begin{equation}
\Delta\omega_{k_j} = \sqrt{\frac{ek|E_{k_j}|}{m_e}}\;.
\label{OT} 
\end{equation} 
The diffusion behavior occurs, when the width $\Delta\omega_{k_j}$ exceeds by a large margin the distance between the resonances $\delta\omega_{k_j}$ (for a statistically significant number of $k$'s), implying that the resonances strongly overlap within a certain interval of wave-vectors $k_j$. This is usually quantified by saying that the Chirikov parameter is large \cite{Ch79}, \ie
\begin{equation}
\ch = \Delta\omega_{k_j}/\delta\omega_{k_j} \gg 1\;.
\label{SQ} 
\end{equation} 
One sees that the Chirikov parameter $\ch\propto\sqrt{|E_{k_j}|}$ being much larger than 1 implies that the level of the excited electrostatic waves in turn cannot be as small as one likes. It might not be taken for granted that this level just lies within the assumptions of the quasi-linear theory discussed above, even though we know {\it by experience} that the conflict does not normally occur in the classic BoT setting.

\section{Revising the classic ``bump-on-tail'' problem: Diffusion-convection model}\label{dcmodel}

We are now in position to revisit the implication of velocity space diffusion as a paradigmatic model of the beam relaxation in cold plasma. The main idea here is that large-amplitude or strongly shaped beams do not relax through diffusion only and that there exists an intermediate time scale where the relaxations are ``convective'' (ballistic-like). We cast this idea in the form of a self-consistent nonlinear dynamical model, \ie mixed diffusion-convection model, which aims at generalizing the classic equations of the quasi-linear theory to ``broad'' (warm) beams with internal structure.

We shall assume for simplicity, without loss of generality, that the density of the beam particles is small compared with the density of the background plasma and we neglect, accordingly, the possible perturbations to quasi-neutrality (and the associated ambipolar electric field) due to the beam injection. 

Next, we consider a ``broad'' beam as composed of a large number of overlapping single cold beams, each having a characteristic nonlinear width $\dvnl$, and the spacing between the beams represented by $\dvsep$. The value of $\dvnl$ is determined by the kinematics of nonlinear resonance broadening in the coupled system of electrostatic waves and beam-plasma particles (\emph{e.g.}, Refs.\cite{shapiro63,Sagdeev}). The behavior crucially depends upon the value of the nonlinearity parameter 
\begin{equation}
\kk=\dvnl/\dvsep \; . \label{KNL}
\end{equation} 
Here, $\dvnl$ can be estimated as $\dvnl \sim (\Delta \omega_k^2/k) \tau_s$, with $\tau_s = 1/(k\Delta v)$ the ``autocorrelation time'', that is the time for particles to be scattered by randomization with respect to wave phases. Note that the nonlinearity parameters in Eqs.~(\ref{SQ}) and~(\ref{KNL}) are connected as $\kk \simeq \ch/Q \equiv \ch^2/\Delta \ell$, with $\Delta \ell = (L/2\pi) \Delta k$
being a measure of the fluctuation spectrum width ($\Delta k$), $L$ the macroscopic system size,  and 
\begin{align}
Q = (\tau_s \Delta \omega_k)^{-1} = \Delta \ell/\ch \; . 
\end{align}  
We also note that another nonlinearity parameter $R=(\gamma_k \tau_{nl} )^{-3}$ (with $\tau_{nl} = (k^2 D_{QL})^{-1/3}$ \cite{dupree66,Ch79} and $D_{QL}$ denoting the quasi-linear diffusion coefficient) can be introduced for measuring the strength of  wave-coupling effects \cite{Laval,laval84}, and could also be expressed, here, as $R \sim 2\pi \ch (\Delta \omega_k/\gamma_k)^3$
(cf. also Sec. 4.2). For large $R$ (and $\ch$), Laval and Pesme derived the TTM model for a ``broad'' turbulent spectrum ($\Delta \ell \gg 1$) \cite{laval84}, and demonstrated that growth rate and diffusion coefficient are expected to be increased with respect to the quasi-linear estimate. These results are typically meant to describe the condition
$\kk \ll 1$, where the propagator renormalization yields the corresponding correction to $D_{QL}$ \cite{Laval}.

The behavior changes if $\kk\gtrsim 1$. In this parameter range, the beams strongly overlap as a consequence of the strong nonlinearity. 
The main effect the overlap between the beams has on the instability growth rate is amplification of the number density of the resonant particles. Summing across all overlapping beams (and dropping the $j$-index in this whole Section for the sake of simplicity), we have, instead of Eq.(\ref{1}),   
\begin{equation}
\Gamma_k \simeq \frac{\pi \omega_{p}^3}{2k^2n_p}\left(\frac{\partial f}{\partial v}|_{\vres} + \frac{\Delta f}{\dvnl}\right)\;, 
\label{Gamma} 
\end{equation} 
where $\Delta f$ is calculated in vicinity of the resonant velocity $\vres=\omega_k/k$ within the velocity spread $\dvnl$. Eq.(\ref{Gamma}) suggests that the instability growth rate in the broad-beam problem is incremented by 
\begin{equation}
\Delta\gamma_k = \Gamma_k -\gamma_k \simeq \frac{\pi \omega_{p}^3}{2k^2n_p} \frac{\Delta f}{\dvnl}\;, 
\label{Gamma+} 
\end{equation} 
as compared to the classic BoT problem, with respect to the growth rate $\gamma_k$. {\it A priori}, one may expect that the amplification $\Delta f$ is proportional to the local number density, hence to the velocity distribution itself; that is, $\Delta f \simeq \chi f$, where the coefficient $\chi$ quantifies the coupling properties among the beams. Following Refs.\cite{Mio12,Mio14}, we write this coefficient as the Boltzmann factor  
\begin{equation}
\chi = \exp[-\dvsep/\dvnl]\;. 
\label{Boltzmann} 
\end{equation} 
The implication is that $\Delta f$ is limited to the transition probabilities of resonance particles on a grid, with the spacing $\dvsep$, and the effective ``temperature'' of nonlinear interaction $\dvnl$. Using the $\kk$ parameter, we write $\chi=\exp[-1/\kk]$, which interpolates between the classic BoT problem ($\kk\ll1$; $\chi\simeq 0$) and the regime of strong nonlinearity of interest here ($\kk\gtrsim 1$; $\chi\simeq 1$). Here, as in the classic BoT problem, we include the twists and controversies summarized by Laval and Pesme in their review \cite{Laval}. We also note in passing that the functional dependence in the $\chi$ value is non-perturbative in that it goes as exponential of $-1/\kk$ and not as $1/\kk$, this being a small parameter of the model. Given that the Chirikov parameter is large, \ie that the condition in Eq.(\ref{SQ}) holds, the relaxation current in velocity space $J_k$ (\ie the flux of the energetic particles in the direction of the thermal core) will be proportional to the instability growth rate $\Gamma_k$. We have $J_k = -\zeta\Gamma_k$, where $\zeta$ is a numerical normalization parameter and will be obtained below. The minus sign indicates that the flux goes against the $v$ axis. With the aid of Eq.(\ref{Gamma}) one obtains
\begin{equation}
J_k = - \zeta \frac{\pi \omega_{p}^3}{2k^2n_p} \left(\frac{\partial f}{\partial v}|_{\vres} + \chi \frac{f}{\Delta v^{NL}}\right)\;.
\label{Flux} 
\end{equation} 
Note that the resonance part of $J_k$, \ie the first term on the right-hand-side of Eq.(\ref{Flux}), has the sense of Fick's second law in velocity space (the Fick paradigm infers that internal fluxes are driven by point-wise gradients with local coefficients: diffusivities and conductivities. Models which are based on such assumptions are referred to as local transport models). The continuity of the flux-function implies 
\begin{equation}
\frac{\partial f}{\partial t} + \nabla_v\cdot J_k = 0,\ \ \ \nabla_v \equiv \partial/ \partial v\;.
\label{Cont} 
\end{equation} 
Combining with Eq.(\ref{Flux}), one is led to a Fokker-Planck equation  
\begin{equation}
\frac{\partial f}{\partial t} = \frac{\partial}{\partial v} {{\Lambda_v}} \delta (v - \omega_k / k) \frac{\partial f}{\partial v} + \frac{\partial}{\partial v} \chi \Lambda_v \frac{f}{\dvnl}\;,
\label{FPE} 
\end{equation} 
where we have denoted for simplicity $\Lambda_v = \zeta (\pi \omega_{p}^3 / 2k^2n_p)$. Clearly, the first term on the right-hand-side of Eq.(\ref{FPE}) represents the well-known quasi-linear diffusion in the limit $t\rightarrow+\infty$. In a basic theory of weak plasma turbulence one writes the quasi-linear diffusion coefficient as \cite{Kadomtzev}
\begin{equation}
D_{QL} = \frac{\pi e^2}{m_e^2} \sum_k |E_k|^2 \delta (\omega_k - kv) = \frac{\pi e^2}{k m_e^2} \sum_k |E_k|^2 \delta (v - \omega_k/k) \;.
\label{Diff} 
\end{equation} 
Comparing with Eq.(\ref{FPE}), and remembering the expression of $\Lambda_v$, one arrives at
\begin{equation}
\zeta \frac{\pi \omega_{p}^3}{2k^2n_p} \delta (v - \omega_k / k) \simeq \frac{\pi e^2}{k m_e^2} \sum_k |E_k|^2 \delta (v - \omega_k/k) ,
\label{Match} 
\end{equation} 
from which the $\zeta$ value can be inferred via an asymptotic matching procedure. 
At this point, no free parameters have remained in the Fokker-Planck model in Eq.~(\ref{FPE}). 

In the discussion above, we have implicitly assumed that the distribution function $f=f(v,t)$ does not involve any coordinate dependence in real space. 
Nor have we included the polarization response of the background plasma to the beam injection (this is because the density of the beam particles has been assumed to be very small from the outset: much smaller than the background plasma density).
These and other exclusions, such as for instance the possible presence of external fields, can be addressed in a usual way for statistical physics of complex systems \cite{Metzler2004} by upgrading the Fokker-Planck model in Eq.~(\ref{FPE}) to a transport model of the Klein-Kramers type, \ie  
\begin{equation}
\frac{\partial f}{\partial t} + v \frac{\partial f}{\partial x} = \frac{\partial}{\partial v} {{\Lambda_v}} \delta (v - \omega_k / k) \frac{\partial f}{\partial v} + \frac{\partial}{\partial v} \left[\chi \frac{\Lambda_v}{\dvnl} + \frac{e}{m}\frac{dV(x)}{dx}\right] f (x,v,t)\;.
\label{KKT} 
\end{equation} 
The Klein-Kramers equation \cite{Metzler2004,Wax,Kampen,Risken} determines the dynamical evolution of the bivariate probability density $f=f (x,v,t)$ of finding a passive test particle with a velocity value between $v$ and $v+dv$ at time $t$ in position $x$. In the above $m$ is the particle mass, $e$ is the electric charge (negative for the electrons), and $V(x)$ is electrostatic potential, corresponding to the potential force field $F(x) = - edV(x)/dx$. One sees that the potential field naturally contributes to convection, however, the sign of this contribution relies on the sign of $edV(x)/dx$. The resulting change of the probability density due to convection is therefore sum of two terms, these being the influence convection term due to $F(x)$, on the one hand, and an internally induced convection term generated self-consistently through the intrinsic wave-particle nonlinearities, on the other hand. It is noted that the internally induced convection may both enhance (if $edV(x)/dx > 0$) or suppress (if $edV(x)/dx < 0$) the potential force term. This paves the way for a competition between the two terms, and to some kind of interference between them, which may be both constructive (for $edV(x)/dx > 0$) or destructive (for $edV(x)/dx < 0$). These interference phenomena will be discussed elsewhere.

It is worth noting that the Klein-Kramers equation in Eq.~(\ref{KKT}) is written for a fully ``collisionless'' plasma. This being relaxed, Eq.~(\ref{KKT}) (and its Fokker-Planck reduction in Eq.~(\ref{FPE})) should be supplemented by collisional drag term and moreover by the entropy-based Laplacian term caused by these collisions. 
The closure of the Klein-Kramers equation~(\ref{KKT}) is obtained through
\begin{equation}
\frac{\partial |E_k|^2}{\partial t} + v_g \frac{\partial |E_k|^2}{\partial x} = 2\Gamma_k |E_k|^2\;,
\label{Clos} 
\end{equation} 
which generalizes the well-known quasi-linear equation \cite{Kadomtzev} for spectral energy density in that it uses the amplified increment $\Gamma_k$ in place of $\gamma_k$. Here, $v_g = \partial\omega_k / \partial k$ is group velocity. Note that the instability growth rate on the right-hand-side of Eq.(\ref{Clos}) steps in with the well-known front factor 2. Eqs.(\ref{KKT}) and~(\ref{Clos}) represent the basic system of dynamical equations for our model. These equations describe the beam relaxation and the evolution of the wave spectrum as coupled processes and extend via the $\Gamma_k$ value the known equations of weak plasma turbulence to the case of a broad spectrum with $\kk \gtrsim 1$. This result is qualitatively similar to that of the TTM theory \cite{laval84}, which, as noted above and summarized in Ref. \cite{Laval}, predicts that growth rate and diffusion coefficient are increased with respect to the quasi-linear estimate for sufficiently strong nonlinearity parameter $R$. The novelty of the present approach is the prediction of a convective relaxation term for
$\kk \gtrsim 1$, recovering the classic BoT problem and theory for a sufficiently broad turbulent fluctuation spectrum ($\kk \ll 1$).

\subsection{Towards multi-scale dynamics: Comparing the relaxation times}\label{dcmodel1}

An essentially new element of our model is the case of ``convective relaxation'' $-$ contained in the last term on the right-hand-side of Eq.(\ref{KKT}). Indeed the convection term enters on an equal footing with the classic diffusion term via the generalization of the instability growth rate in Eq.(\ref{Gamma}). The characteristic relaxation time in the convection domain is easily seen to be given by 
\begin{equation}
\tau_{\rm conv} \simeq \Delta v^{NL}\Delta v_{b}/\chi\Lambda_v\; ,
\label{RTime} 
\end{equation} 
where $\Delta v_{b}$ is the broad (\ie warm) beam  width in velocity space. 
This should be compared with the characteristic relaxation time via the quasi-linear diffusion, which we assess as $\tau_{\rm diff} \simeq \Delta v_{b}^2/\Lambda_v$. In general, we can assume $\Delta v_{b} \gg \dvnl$, suggesting that the convective relaxation is a meso-scale process. We interpret this result as follows.

\subsection{Mixed diffusive-convective behavior and the asymptotic character of the diffusion}\label{dcmodel2}

For $\kk\gtrsim 1$ (\ie $\chi\simeq 1$, strong-overlap limit), the relaxation dynamics are completely dominated by the nonlinear amplification of the instabilities via the share in the resonance particle population. The coupling processes among the beams are such as to increase the density of the resonant particles where the second derivative of the distribution function is positive, \ie $\partial^2 f / \partial v^2 > 0$, and, via the conservation of the total number of the particles, act to decrease, at the same time, the resonant density where the second derivative is negative ($\partial^2 f / \partial v^2 < 0$). This generates an unstable propagating front in velocity space directed to the thermal core of the distribution function comprising the beams. Moreover the front is self-reinforcing: its strength, \ie the density pedestal, initially grows with time as more resonance particles are trapped in it. The process is analogous to a snow avalanche or self-amplifying chain reaction (The idea that the avalanching dynamics occur via an amplification of instabilities in the parameter range of large nonlinearity is indeed very general and have been discussed for ``strong'' electrostatic turbulence in Refs.\cite{PLA,JPP}). 

As more and more resonant particles are caught by the propagating front, the local number density on its left edge continues to increase, and so does the velocity gradient, which emits the waves. These emissions have feedback on the transport process in velocity space favoring entropy based diffusion through the $\partial f / \partial v$ term. The velocity diffusion, in its turn, will try to wash out the eventual singularities, and to pose a smoother relaxation dynamics, which is quasi-linear-like. That will be a mixed regime, \ie convective (transferring the resonant particles from inside the bump on tail onto its left edge) competing with the diffusion. The competition will stop as soon as the majority of the beam particles have been deposited on the front, from which they diffuse away via the random walks. It is in this sense that we say the convective relaxations occur on intermediate (\ie meso-) scales. These are dictated by the velocity span of the beam and in the time domain by the duration of the amplification processes. The asymptotic ($t\rightarrow+\infty$) relaxation process will be {\it always} diffusive quasi-linear (with the caveats discussed in the previous section \cite{Laval}).

One sees that the relaxation process acquires multi-scale features in the broad-beam problem. It begins initially as a convective process, with time scale in Eq.(\ref{RTime}), followed by an asymptotic diffusion process on very long time scales. Note that the convective relaxation occurs despite the condition that the Chirikov parameter is large, \ie $\ch\gg 1$, and is attributed to the fact that the nonlinearity of the interaction, contained in the ordering $\kk\gtrsim 1$, is also large in its turn, giving rise to a Fokker-Planck generalization of the velocity-space diffusion equation. We should stress that by ``generalizing'' the quasi-linear theory we mean, in fact, the inclusion of the meso-scale relaxation of the convective type (via the coupling term in the respective Klein-Kramers equation), without touching on the asymptotic, diffusive behavior and the formation of the plateau (cf. discussion at the end of the previous section). The meso-scale nature of convective relaxation is also reflected by the condition
on the width of the fluctuation spectrum $\kk\gtrsim 1$.

\subsection{Boltzmann's H-theorem and the entropy growth rates}\label{dcmodel3}

Although obvious, it should be emphasized that the Chirikov parameter being much larger than 1 implies that the dynamics are random (chaotic-like) on micro-, respectively, wave-particle interactions, scales. It is convenient to characterize this implication of the chaotic dynamics in terms of Boltzmann's $H$-theorem and assess the corresponding entropy growth rates through respectively the diffusion and the convective relaxation processes. 

For weakly interacting classical systems, the entropy $S = S(t)$ is related to the probability density function through 
\begin{equation}
S (t) = -\int_{-\infty}^{+\infty} f(t, v) \log f(t, v) dv\;.
\label{Ent} 
\end{equation} 
We shall assume for simplicity, without lacking generality, that the distribution function $f$ represents solely the beam particles, so it is identically zero outside the bump region. So we differentiate the functional dependence in $S(t)$ over the time, then substitute the time derivative $\partial f(t, v) / \partial t$ with the sum of the diffusion and the convection terms using the Fokker-Planck equation~(\ref{FPE}), and integrate by parts over the velocity variable under the assumption that the velocity gradients vanish on both sides of the integration domain. Making use of the resonance condition, after a simple algebra one obtains 
\begin{equation}
\frac{d}{d t} S (t) = \Lambda_v f^{-1} \left(\frac{\partial f}{\partial v}\right)^2|_{\vres} \geqslant 0\;.
\label{Ent+} 
\end{equation}
One sees that the time derivative of the entropy is always non-negative, \ie $dS/dt \geqslant 0$, and is moreover restricted to the relaxation process of the diffusive type, as it should. Indeed the convective relaxation does not as a matter of fact contribute to the entropy growth rate and in this case can be considered adiabatic.

\section{Numerical simulations of broad beam relaxation in cold plasma}\label{sims}

In this Section, in support of the theoretical framework introduced above, we discuss numerical simulation results of the mixed diffusive-convective relaxation of a tenuous broad beam of energetic particles, whose density $n_B$ is much smaller than that of the 1D background plasma ($n_p$), which is considered as a cold linear dielectric medium. In particular, we adopt the Hamiltonian formulation of the problem described in Ref.\cite{CFMZJPP}, where the broad supra-thermal particle beam is discretized as superposition of $n\gg1$ cold beams self-consistently evolving in the presence of $m\geqslant n$ modes nearly degenerate with Langmuir waves, \ie with frequency $\omega_{k_j} \simeq \omp$ for $j=1,...,m$. This ensures that the dielectric function of the cold background plasma is nearly vanishing and allows casting the Poisson equation for each plasma oscillation into the form of a simple evolution equation, while particles trajectories are solved from the equations of motion \reff{2} \cite{OWM71}.

In the following, we first briefly summarize the Hamiltonian formulation of the broad beam relaxation in cold plasma derived in Ref.\cite{CFMZJPP} (Sec.\ref{sims1}). Then, we discuss the dimensionless parameters that are used to characterize the different nonlinear dynamic regimes (Sec.\ref{sims2}), which will be studied numerically with four different set-ups of initial conditions (Sec.\ref{sims3}). Numerical simulation results are then presented and discussed, adopting also a test particle transport analysis to illustrate the mixed diffusive-convective nature of beam relaxation (Sec.\ref{sims4}). Finally, we discuss a simple toy-model of diffusion-convection relaxation to illustrate the important role played by the self-consistent evolution of fluctuation intensity on the same time scale of beam transport (Sec.\ref{sims5}).

\subsection{The multi beam approach: BoT Hamiltonian formulation}\label{sims1}

This model is described in Ref. \cite{CFMZJPP}, which we refer to for details and discussion of its relationship to the vast literature on Hamiltonian formulation of the BoT problem. The model consists in the description of $n$ cold beams evolving in a self-consistent way with $m\geqslant n$ modes with $\omega_{k_j}\simeq\omp$. Among the waves, we choose $n$ modes which are resonantly driven by the beams (linearly unstable), \ie $k_\al=\omp/v_\al$ (where $v_\al$ are the initial beam velocities and $\al=1,...,n$) and $m-n$ linearly stable ones. The nonlinear evolution of this system, as shown in Ref.\cite{CFMZJPP}, is equivalent to a broad beam and $m$ Langmuir waves for properly chosen initial conditions \cite{VK12}. More specifically, the $n$ cold beams can be considered as an initial discretization of the broad beam, when particles are conveniently distributed.

The dynamics of beam electrons is described by the Newton's law, while Poisson's equation gives the the self-consistent mode evolution. The 1D cold plasma equilibrium is taken as a periodic slab of length $L$, and the position along the $x$ direction for each beam is labeled by $x_\al$. Following Ref.\cite{OWM71}, we assume that single beams consist of $\Na$ particles located at $x_{\al i}$ (\ie $x_{1i}$, $x_{2i}$, ..., $x_{ni}$; $i=1,...,\Na$), so that $\sa\nb$ is the particle density of the ``$\alpha$''-beam, with $\sa=\Na/N$ and $N=\Nu+..+\Nn$ the total particle number. For the sake of convenience, beam particle positions are represented as $x_{\al i}= \bar{v} t+\xi_{\al i}(t)$, with $\bar{v}=(v_1+..+v_n)/n$ the initial average beam speed. Meanwhile, the Langmuir wave scalar potential $\varphi(x,t)$ is expressed in terms of the Fourier components $\phij(k_j,t)$ (with $j=1,...,m$), yielding the corresponding electric fields $E_{k_j} (k_j,t) = - i k_j \phij(k_j,t)$, consistent with Eq.(\ref{2}). Finally, consistent with Ref.\cite{OWM71}, the following scaled variables are introduced: $\lj=(2\pi/L)^{-1}\kj$, $\bar{\xi}_{\al i}=2\pi{\xi}_{\al i}/L$, $\etab=(\nb/2\np)^{1/3}$, $\tau=t\omp\etab$, $\phija=(\phij\,e\kj^2)/(m\etab^{2}\omp^2)$,  $\beta_j=\big(\kj\,\bar{v}/\omp-1\big)/\etab$. Noting that the plasma dielectric constant, $\epsilon_p(\omega_k)$, is nearly vanishing, and adopting time scale separation between $\tau$ evolution and $\omp^{-1}$ \cite{OWM71}, 
\begin{equation}
\epsilon_p(\omega_{k_j})=1-\frac{\omp^2}{\omega_{k_j}^2} \simeq 2 i \etab \frac{\partial}{\partial \tau}\;. \label{eq:tsep}
\end{equation}
Noting Eqs.(\ref{2}) and (\ref{eq:tsep}), the system of equation governing the interaction between $m$ Langmuir modes and $n$ beams finally reads as \cite{CFMZJPP}
\begin{subequations}\label{mainsys}
\begin{align}
\bar{\xi}_{\al i}''&=\sum_{j=1}^{m}\big(i\,\lj^{-1}\;\phija\;e^{i\lj\bar{\xi}_{\al i}+i\beta_j \tau}+c.c.\big)\;,\\
\phija'&=ie^{-i\beta_{j}\tau}\;\sum_{\al=1}^{n}\,\frac{\sa}{\Na}\sum_{i=1}^{\Na} e^{-i\lj\bar{\xi}_{\al i}}\;,
\end{align}
\end{subequations}
where the prime denotes the derivative with respect to $\tau$. Momentum and energy conservation properties of Eqs.\reff{mainsys} are discussed in Ref.\cite{CFMZJPP}.

In the simulations, each beam has an initial velocity expressed as $v_\al=\Theta_\al v_1$; and we get $\bar{v}=\Theta v_1/n$, with $\Theta=\Theta_1+...+\Theta_n$. As we assume that the first $n$ modes are resonantly driven by the beams ($k_\al=\omp/v_\al$), one obtains $\Theta_\al=\ell_1/\ell_\al$. The initial conditions for the beam nonlinear shifts $\bar{\xi}_{\al i}$ are given random for each beam between $0$ and $2\pi$, while the initial velocities are $\bar{\xi}_{\al i}'(0)=(\Theta_\al-\Theta/n)/(\ell_1\etab)$. Furthermore, the resonance mismatch terms read $\beta_{j=1,...,m}=(\Theta\lj/n\lu\;-1)/\etab$. Finally, we note that the broad beam distribution function can be described by a suitable choice of the $n$ beam partial densities $\sigma_{\alpha}$. In the following, we solve \erefs{mainsys} using a Runge-Kutta (fourth order) algorithm and $N=10^5$ total particles. For the considered time scales and for an integration step $h=0.01$, both the total energy and momentum (for the explicit expressions, see \cite{CFMZJPP}) are conserved with relative fluctuations of about $1.4\times10^{-5}$.

\subsection{Characterization of the nonlinear dynamic regime}\label{sims2}

Numerical simulations, discussed here, illustrate different nonlinear dynamic behaviors, which can be characterized introducing 
proper dimensionless parameters (as already discussed). In particular, we can introduce the Chirikov parameter \cite{Ch79} that characterizes the resonance overlap of the fluctuation spectrum as in Eq. (\ref{SQ}). For each of the $n$ cold beams, we can evaluate the corresponding 
Chirikov parameter $\ch_{\al}$ for the initial conditions discussed in the previous subsection and 
consistent with Eq. (\ref{SQ}). Assuming $\ell_{\al+1}=\ell_\al+1 \gg 1$, we get the following scaled expression of the Chirikov parameters
\begin{align}
\ch_\al \simeq \ell_\al \etab\sqrt{|\phi^{sat}_\al|} \;. \label{eq:salpha}
\end{align}
As already stated in the previous Section, for $\ch_{\al}<1$, the beams can be treated as independent, while resonance overlap occurs for $\ch_{\al}>1$. These two different situations correspond to different transport processes in the velocity space. The case with $\ch_\al\lesssim 1$ was studied in Ref.\cite{CFMZJPP} with the multi beam approach adopted here. In the present work, we focus on the $\ch_{\al}>1$ case,
when many resonances are overlapping. Our aim, in fact, is to assess the different role of diffusion and convection in the relaxation of a broad beam, interacting with multiple Langmuir fluctuations in a cold background plasma. In particular, one of the main goals of our numerical simulations, presented below, is to clarify how linear stable modes affect particle transport in the BoT problem and how it depends on the Chirikov parameter and width of the fluctuation spectrum.

As already discussed in Secs. \ref{botparadigm} and \ref{dcmodel}, we remind that large Chirikov parameter, intuitively, corresponds to diffusive beam relaxation in velocity space \cite{CD92,EEbook} and to the typical condition of applicability of quasi-linear theory \cite{Vedenov,Pines}, with the twists and subtleties discussed in the review paper 
by Laval and Pesme \cite{Laval}. However, such a case also requires the fluctuation spectrum be sufficiently broad; \ie characterized by a sufficiently large $\Delta k/\bar k$, with $\Delta k = |k_m - k_1|$ and $\bar{k}=\omp/\bar{v}$ the resonant wave-vector at the initial average beam velocity. This is because the characteristic autocorrelation time
\begin{equation}
 \tau_s \simeq \frac{1}{\bar v \Delta k} \simeq \frac{1}{\bar k \Delta v} \; , \label{eq:taus}
\end{equation}
must be sufficiently short compared to the characteristic time $\tau_{nl} = (k^2 D_{QL})^{-1/3}$ (cf. Section 3). Here, we have introduced the notation $\Delta v =|\omp(1/k_1 - 1/k_m)|$ for the spectral width in the velocity space. Using quasi-linear theory \cite{Vedenov,Pines}, we can estimate the quasi-linear diffusion time, $\tau_{\rm diff}^{NL}$, that, by means of Eq. (\ref{Diff}), can be cast as
\begin{align}
\frac{\tau_s}{\tau_{\rm diff}^{NL}} \simeq \frac{D_{QL}}{k(\Delta v)^3} 
\simeq \frac{\tau_s^3}{\tau_{nl}^3} \simeq 2 \pi \ch \Delta \omega_k^3 \tau_s^3
\simeq 2 \pi \sum\limits_\al|\phi^{sat}_\al|^{2}\etab^4\ell_\alpha^4 \big(\Delta \ell)^{-4}\;.
\label{eq:taunl}
\end{align}
Here, we have denoted by $\ch$ a characteristic value of $\ch_\alpha$ from Eq.(\ref{eq:salpha}) (typically referred to the central beam); and $\Delta \omega_k$ the characteristic wave-particle trapping frequency, estimated from Eq. (\ref{OT}) for an isolated resonant mode at saturation, which, in normalized form, can be expressed as $\omega_B = (\etab\omp)^{-1} \big(ek^{2}\varphi^{sat}/m\big)^{1/2}$.
Meanwhile, adopting the theoretical framework of Secs. \ref{botparadigm} and \ref{dcmodel}, $\tau_{\rm conv}^{NL}$  in the convection domain can be  estimated from Eqs. (\ref{Diff})
and (\ref{RTime}) as
\begin{align}
\frac{\tau_s}{\tau_{\rm conv}^{NL}}  \simeq \frac{\chi D_{QL}}{k (\Delta v)^2\Delta v^{NL}} 
\simeq 2\pi \chi \ch \Delta \omega_k \tau_s
\simeq 2 \pi \chi  \sum\limits_\al|\phi^{sat}_\al| \etab^2 \ell_\alpha^2 \big(\Delta \ell)^{-2} \;.
\label{eq:taunlconv}
\end{align}
The characteristic parameter $Q$ (cf. Section 3) can be written as
\begin{align}
Q\equiv \left(\Delta \omega_k \tau_s\right)^{-1} \simeq (\Delta \ell/{\cal S}) \;, \label{eq:Qdef}
\end{align}
whose reciprocal physically represents the fraction of a trapping time over which a particle
is accelerated by any of the fluctuating fields; and we see that larger $\ch$ corresponds to smaller $Q$ for fixed $\Delta \ell$.
Note that
\begin{align}
\big( \tau_s/\tau_{\rm diff}^{NL} \big) \simeq 2\pi \ch/Q^3 \ll 1 \; , \label{eq:QLcond2}
\end{align}
for quasi-linear theory to be valid \cite{Laval}. Furthermore, 
\begin{align}
\big( \tau_s/\tau_{\rm conv}^{NL} \big) \simeq 2\pi \chi \ch/Q \simeq 2 \pi \chi \kk  \; , \label{eq:QLcondconv}
\end{align}
and the ratio of characteristic time scales is $\tau_{\rm diff}^{NL}/\tau_{\rm conv}^{NL} \sim \chi Q^2$. Here, we note that if $Q$ is not sufficiently large, the nonlinear behavior may be significantly coherent as consequence of phase space structures. 
In fact, even when resonances are significantly overlapped ($\mathcal{S}\gg1$), a sufficiently large $\Delta\ell$, \ie a sufficiently broad spectrum, is needed to prevent the characteristic nonlinear shift in particle velocity from covering a significant fraction of the whole fluctuation spectrum. In other words, for $Q\simeq1$, the spectrum behaves as ``quasi-periodic'' even for $\mathcal{S}\gg1$; and wave particle trapping becomes a relevant process.
Meanwhile, characteristic time scales of convective and diffusive relaxation of the particle beam become comparable.
In the following, we will discuss four different cases with varying combinations of $Q$ and ${\cal S}$, illustrating the mixed diffusive-convective relaxation of a broad beam.

\subsection{Numerical simulation results}\label{sims3}

\begin{figure}[ht!]
\includegraphics[width=.285\textwidth,clip]{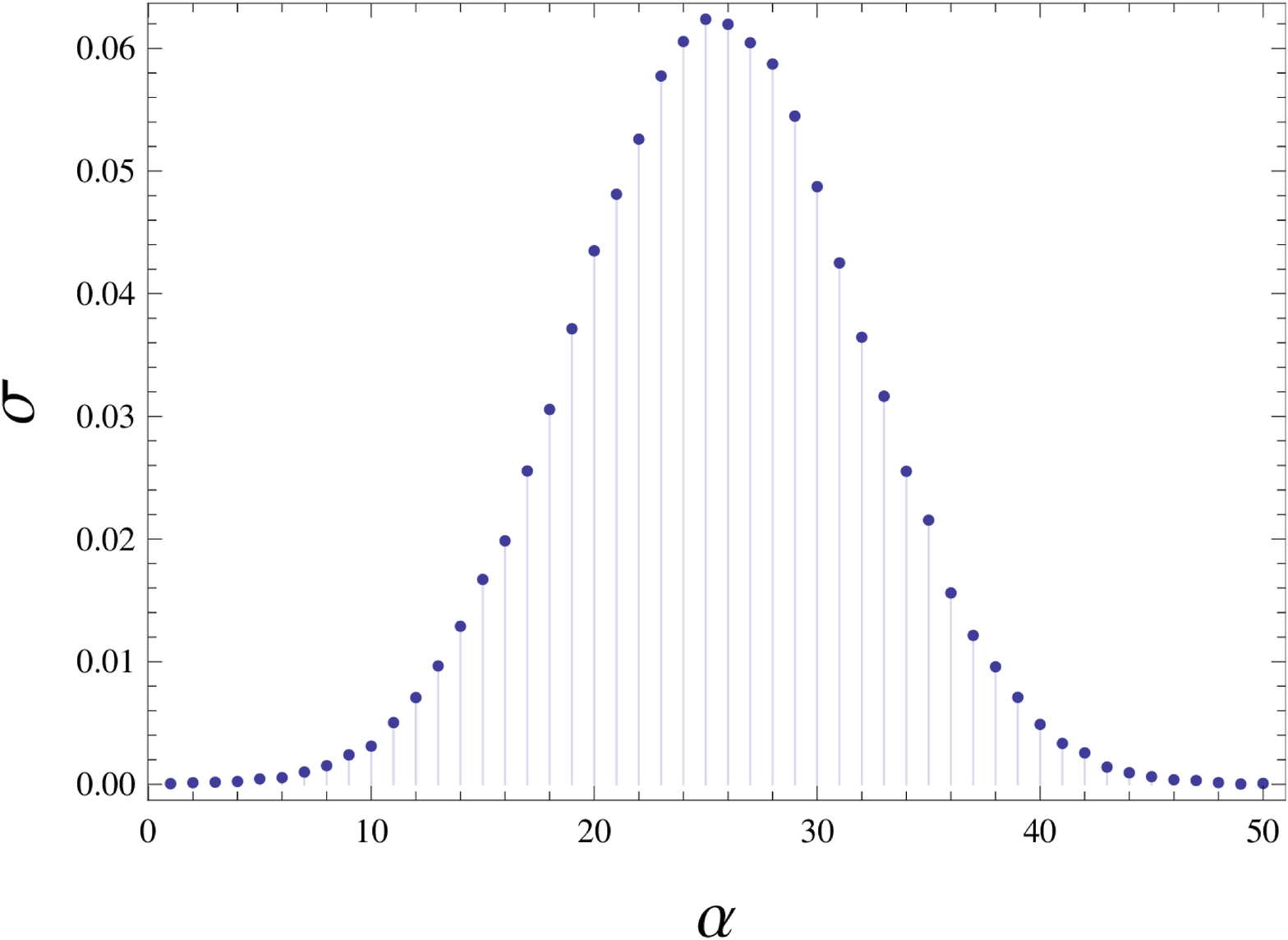}
\includegraphics[width=.3\textwidth,clip]{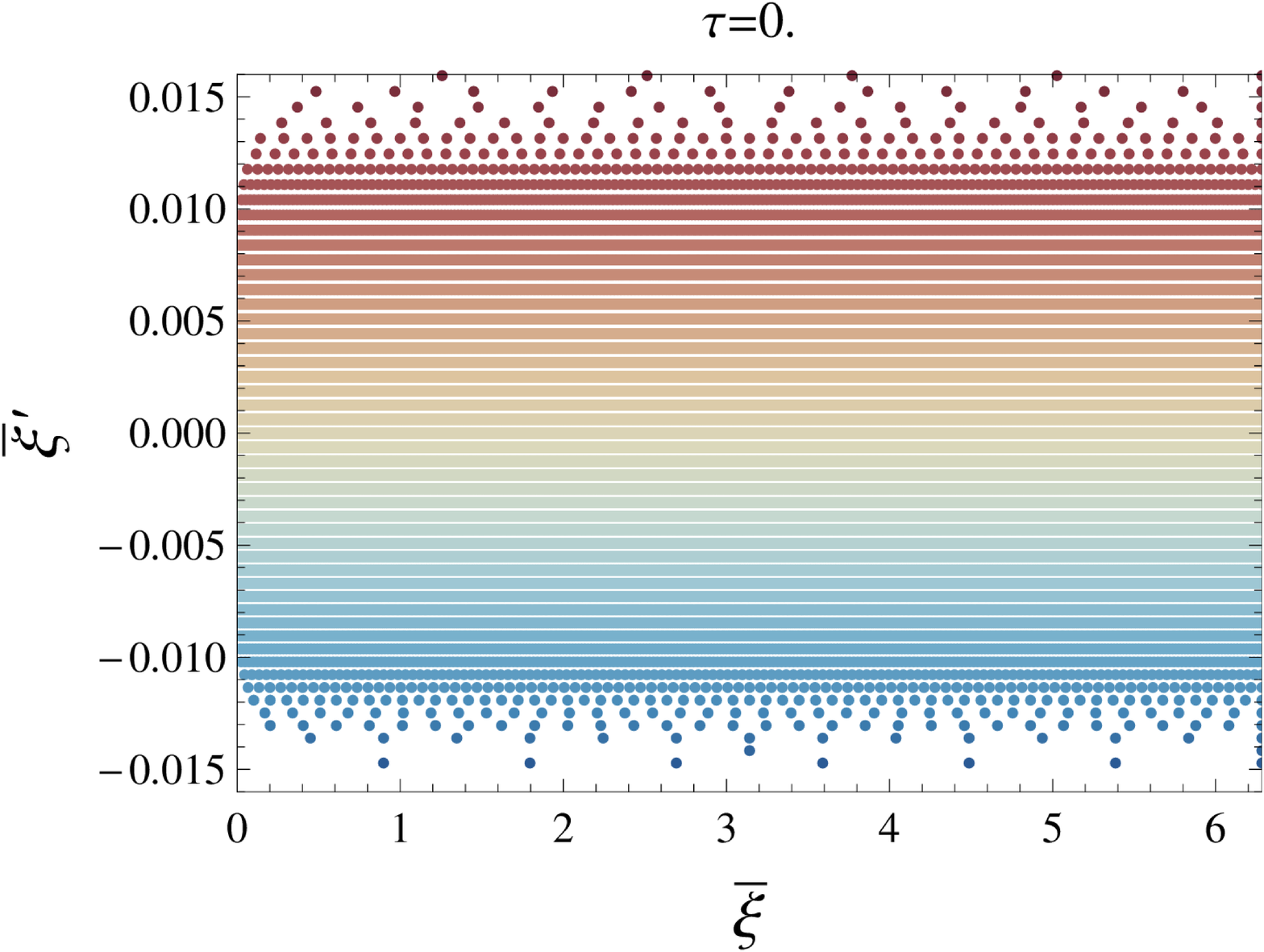}
\includegraphics[width=.3\textwidth,clip]{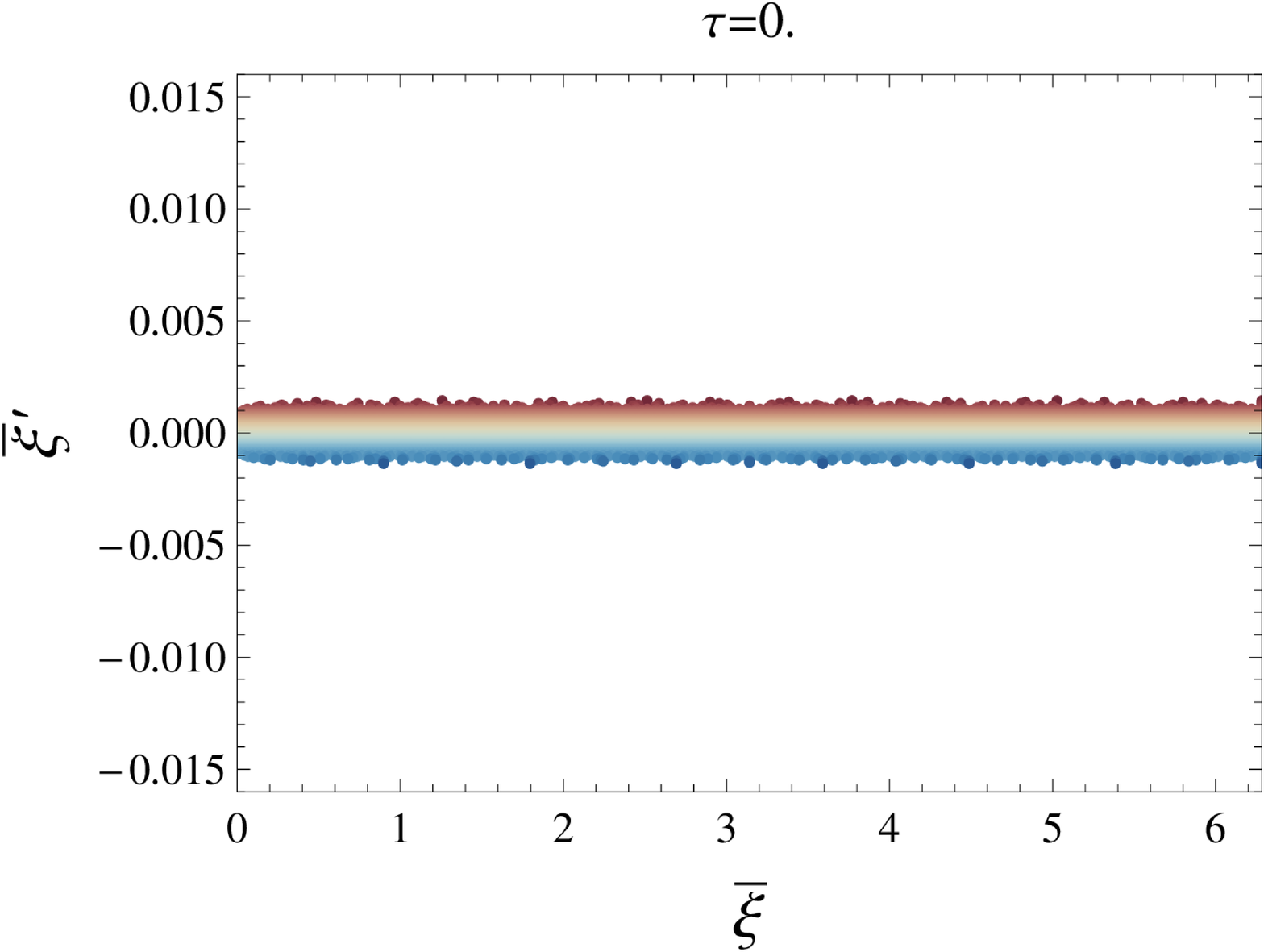}
\caption{\label{fig1} Left-hand panel: densities of the beams ($\sigma_\alpha$) as a function of the beam index $\alpha=1,...,n$. Initial ($\tau=0$) phase space $(\bar{\xi},\bar{\xi}')$ for the cases \emph{(i)}, \emph{(ii)} (central panel) and \emph{(iii)}, \emph{(vi)} (right-hand panel): each point corresponds to a single charge sheet and different colors denote different beams.}
\end{figure}

We report, here, numerical simulation results with $n=50$ beams for $4$ distinct cases, which differ for the number of modes $m$ and for the magnitude of the relevant Chirikov parameters: \emph{(i)} $\ch\simeq2$ and $m=n=50$ (only resonant modes); \emph{(ii)} $\ch\simeq2$ and $m=120$ ($50$ resonant modes and $70$ linearly stable modes); \emph{(iii)} $\ch\simeq15$ and $m=n=50$; \emph{(iv)} $\ch\simeq15$ and $m=120$. The initial values of the modes are set with random phases $\psi_j$ and with random amplitude of $\mathcal{O}(10^{-2})$ for the resonant (linearly unstable) $n$ modes, while the random amplitude of non-resonant (linearly stable) modes is of $\mathcal{O}(10^{-3})$. The initial densities of the beams, \ie the $\sigma_\alpha$ values ($\alpha=1,...,n$), are assumed Gaussian distributed in the velocity space as depicted in \figref{fig1} (left-hand panel), where we also plot the initial phase space for the cases \emph{(i)}, \emph{(ii)} (central panel) and \emph{(iii)}, \emph{(vi)} (right-hand panel).

Let us now discuss the underling motivation for the choice of the selected numerical studies.
Cases \emph{(i)}, \emph{(ii)} are chosen to illustrate, as a result, the diffusive as well as convective character of a broad beam under the effect of a broad fluctuation spectrum and weak nonlinearity on the temporal meso-scale. In fact, case  \emph{(i)} has $Q \simeq 30$; thus, we expect that the addition of non-resonant modes in case  \emph{(ii)} may change the beam relaxation quantitatively but not qualitatively. Meanwhile, cases \emph{(iii)}, \emph{(iv)} are selected to show the effect of strong nonlinearity and the possible crucial qualitative effect of including the stable part of the fluctuation spectrum. More precisely, $Q \simeq 3.6$ in case \emph{(iii)} and, thus, significant coherent behavior can be expected; while in case \emph{(iv)}, the inclusion of stable modes more than doubles the spectrum width, restoring the time asymptotic diffusive behavior. Finally, we note that only case \emph{(iv)} is characterized by $K\gtrsim1$ and a broad spectrum. In fact, for cases \emph{(i)}, \emph{(ii)}, \emph{(iii)}, we have $K\simeq0.07$, $0.03$, $4.5$, respectively, but in case \emph{(iii)} the spectrum is narrow. Thus, only case \emph{(iv)} is suitable for comparison to the analysis in Sec.\ref{dcmodel}. Indeed, for these parameters, we show that the convection contribution to the particle dynamics becomes comparable to diffusion.

Let us now analyze separately these 4 distinct cases, discussing their features and the different underlying physics processes.
In the following, times are expressed in $(\etab\omp)^{-1}$ units. 

\subsubsection{Case \emph{(i)}: $\ch\simeq2$ and $376\leqslant\ell_{res}\leqslant425$}
\begin{figure}[ht!]
\includegraphics[width=.34\textwidth,clip]{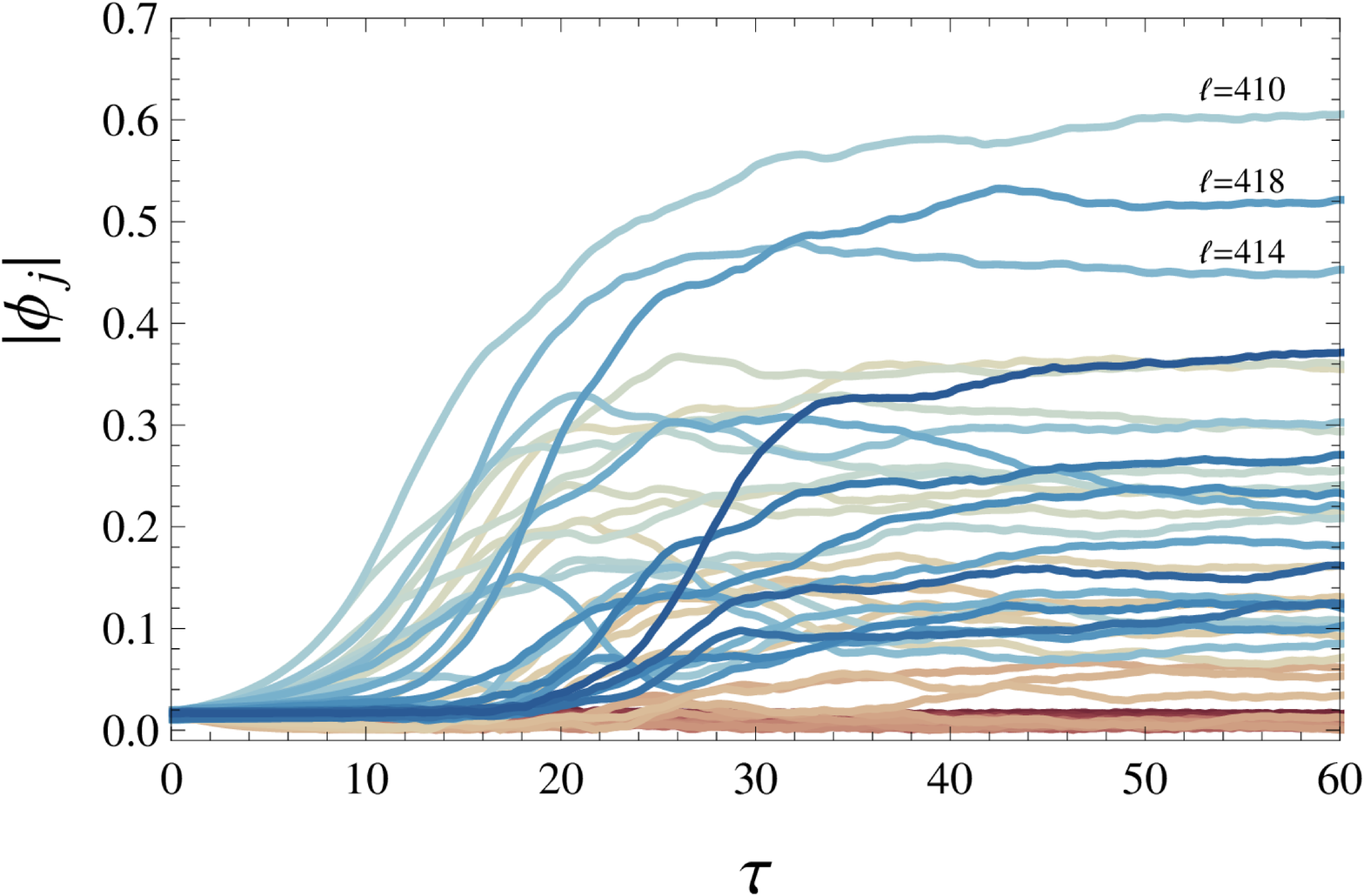}
\includegraphics[width=.3\textwidth,clip]{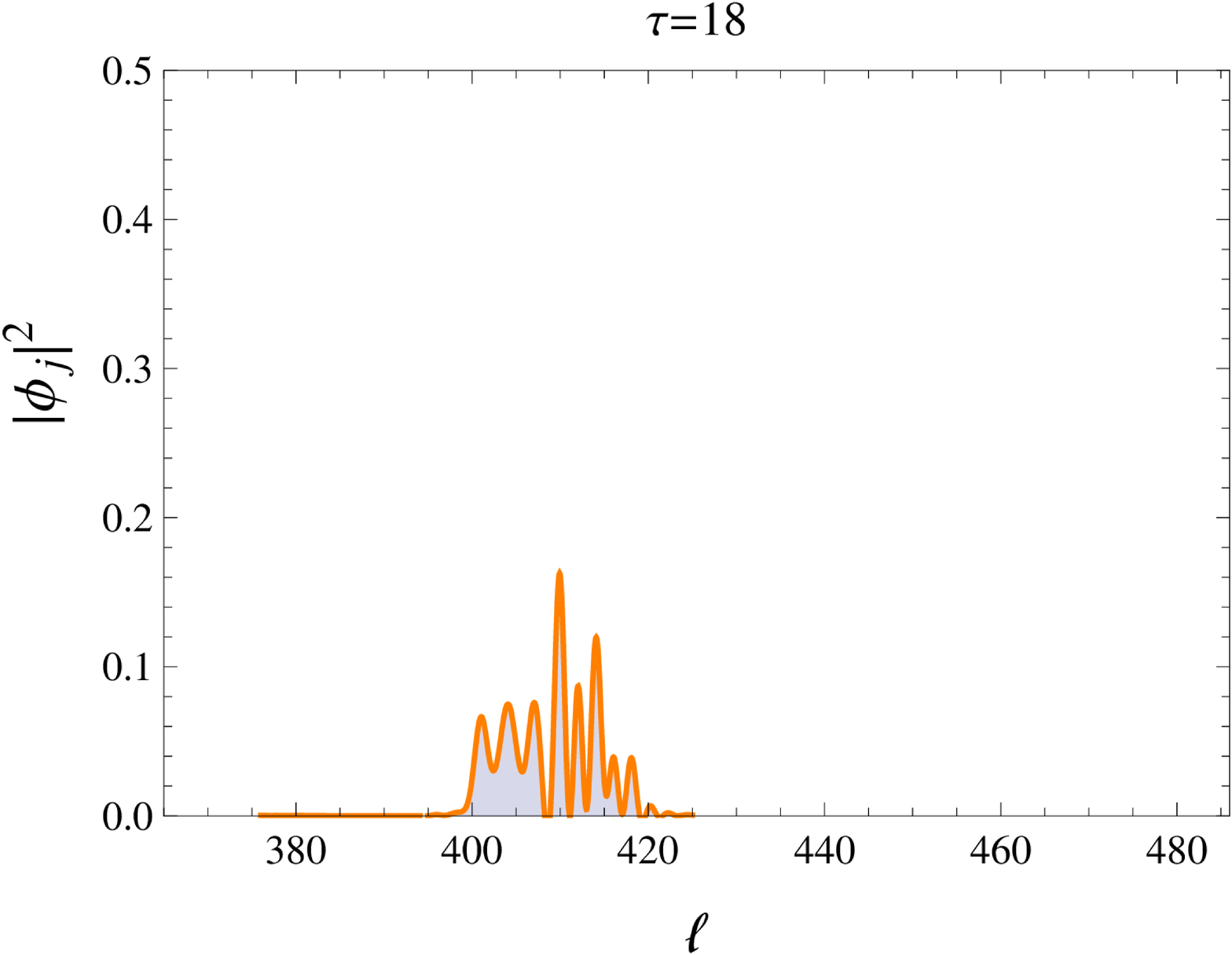}
\includegraphics[width=.3\textwidth,clip]{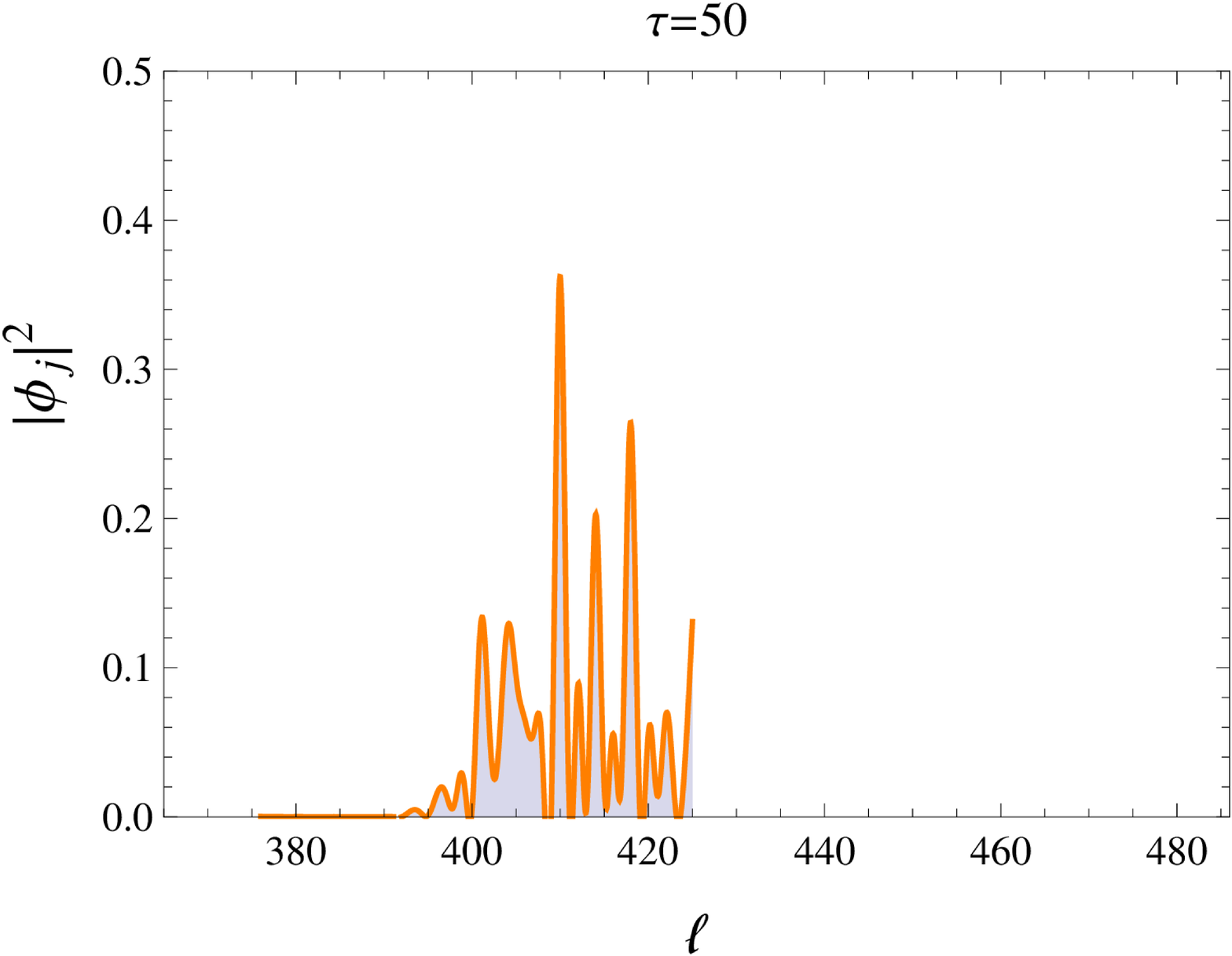}
\caption{\label{fig1.1}Case \emph{(i)}. Left-hand panel: temporal evolution of the mode amplitude $|\phi_j|$; colors are fixed from blue (largest $\ell$) to red (smallest $\ell$), and time-asymptotic dominant wave-numbers are indicated. Central and right-hand panels: intensity spectrum at different times.}
\end{figure}

\begin{figure}[ht!]
\includegraphics[width=.23\textwidth,clip]{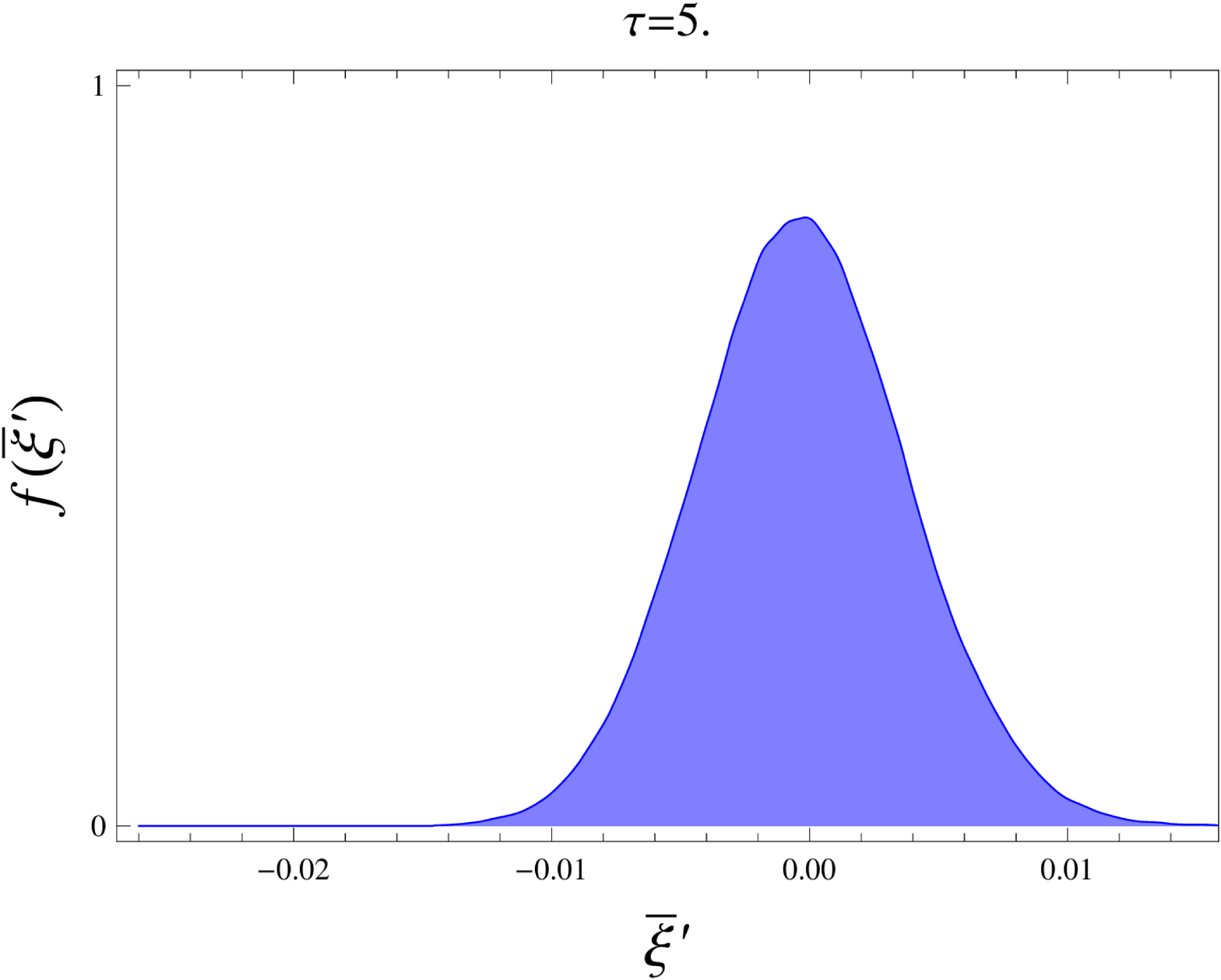}
\includegraphics[width=.23\textwidth,clip]{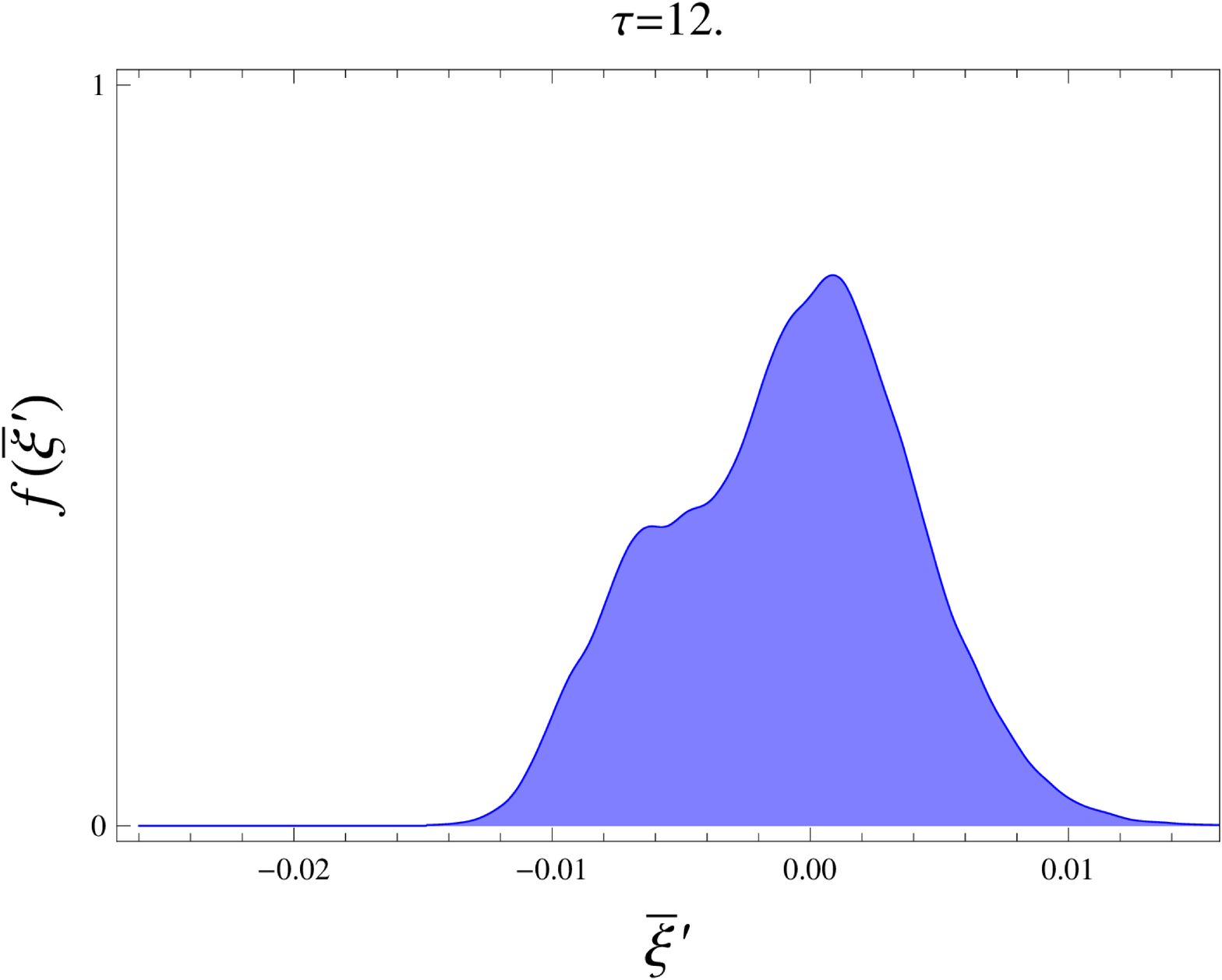}
\includegraphics[width=.23\textwidth,clip]{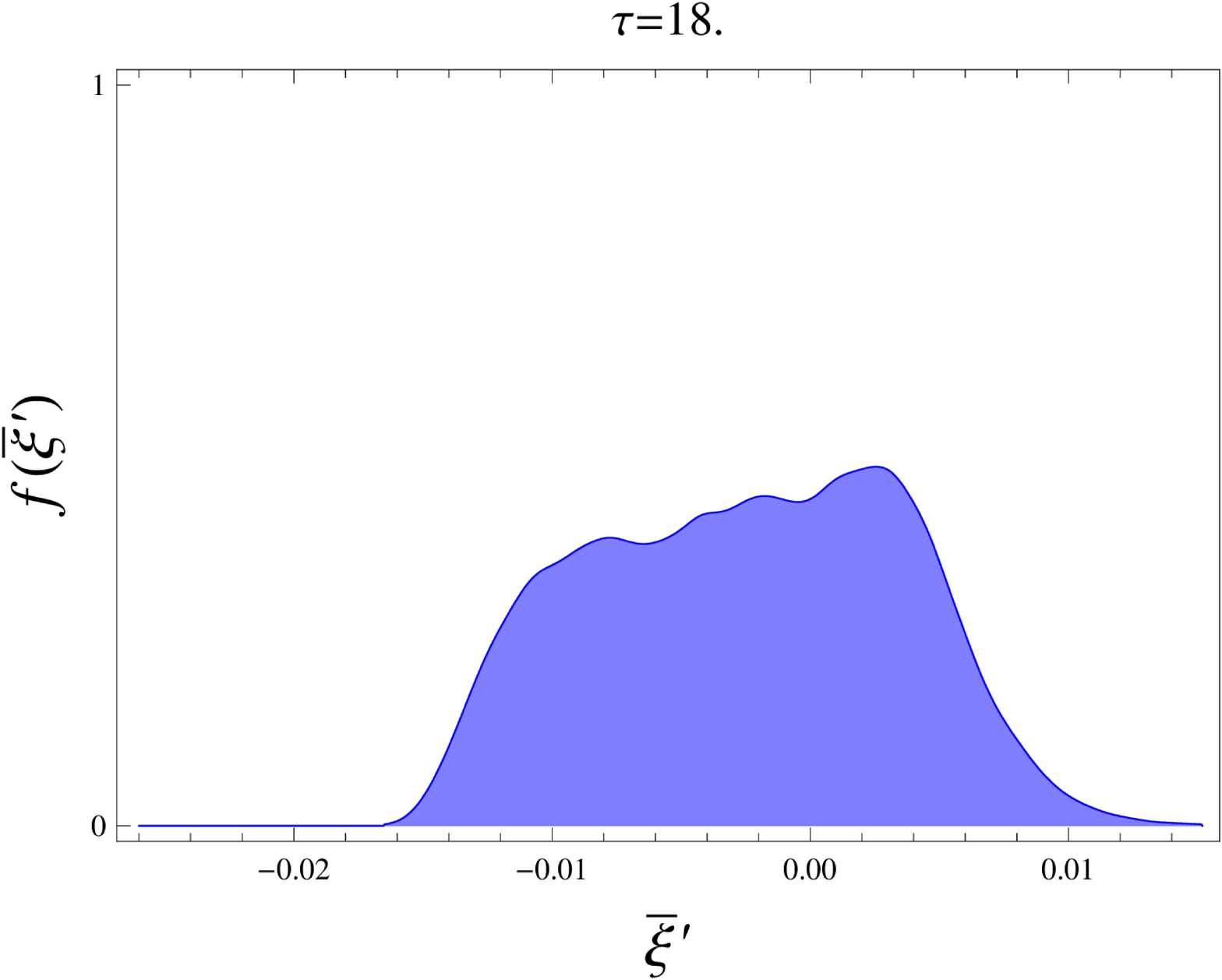}
\includegraphics[width=.23\textwidth,clip]{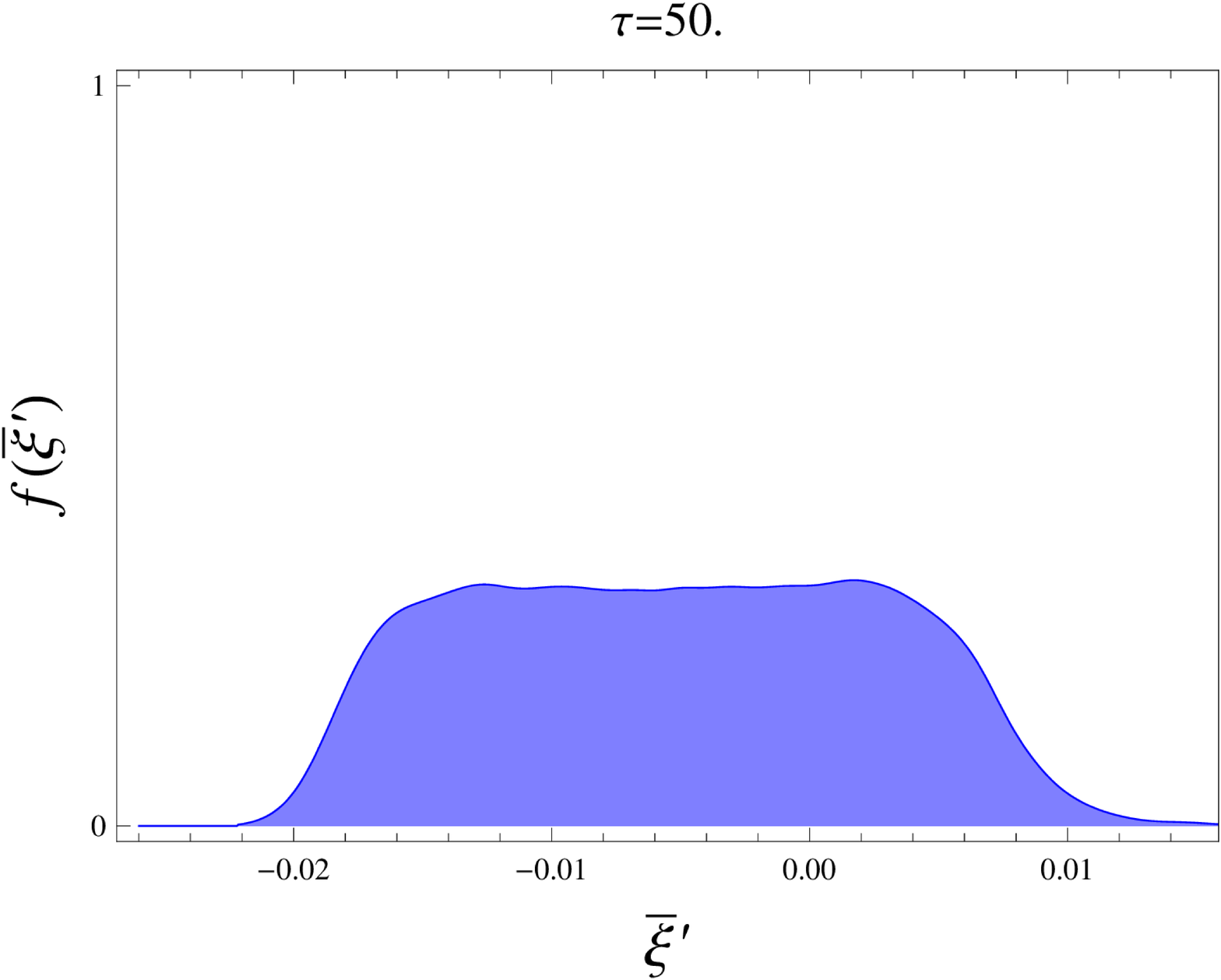}
\caption{\label{fig1.2} Case \emph{(i)}. Evolution of the velocity distribution function $f(\bar{\xi}')$ (in arbitrary units), shown at different times.}
\end{figure}

In this case, the mode evolution and the corresponding particle velocity distribution function, for sufficiently large time, are consistent with the quasi-linear model prediction. In the left panel of \figref{fig1.1}, we can observe the saturation of each spectral component, while the other two panels show the intensity spectrum at different times; with a peak structure shifting towards large $\ell$, consistent with total energy conservation.
This suggests, at least in the initial phase of the evolution, the presence of a convection (drag) in velocity space, reflecting the effect of weak coherent structures. Time asymptotically, \figref{fig1.2} shows how the velocity distribution function evolves towards a plateau, consistent with quasi-linear diffusion.
In other words, the dominant diffusion process is due to particles interacting with multiple modes with random phases. It is also worth observing how the coarse-grained features of this distribution function are consistent with the results obtained in \cite{VK12}. Moreover, in that work, the energy spectrum is shown to have irregularities associated with significant jumps in the particle velocities, characterized by Gaussian distribution. Similar irregularities can also be observed in our energy spectrum and, therefore, we can conclude that they do not depend on the morphology of the initial fields phase (in \cite{VK12}, the initial phases are not taken as random).

In this case, we get the following estimations: $\tau_s=0.5/(2\pi)\simeq 0.08$, $\tau_{\rm diff}^{NL}\simeq113$, $Q\simeq30$ and $K=0.07$ ($\tau_B=2\pi\omega_B^{-1}\simeq16$). This is consistent with the broad character of the considered spectrum and condition \reff{eq:QLcond2} is properly satisfied. Thus, we can conclude that case \emph{(i)}, time asymptotically, can be described by the well known quasi-linear approximation. 

\subsubsection{Case \emph{(ii)}: $\ch\simeq2$ and $366<\ell<376\leqslant\ell_{res}\leqslant425<\ell<485$}
\begin{figure}[ht!]
\includegraphics[width=.34\textwidth,clip]{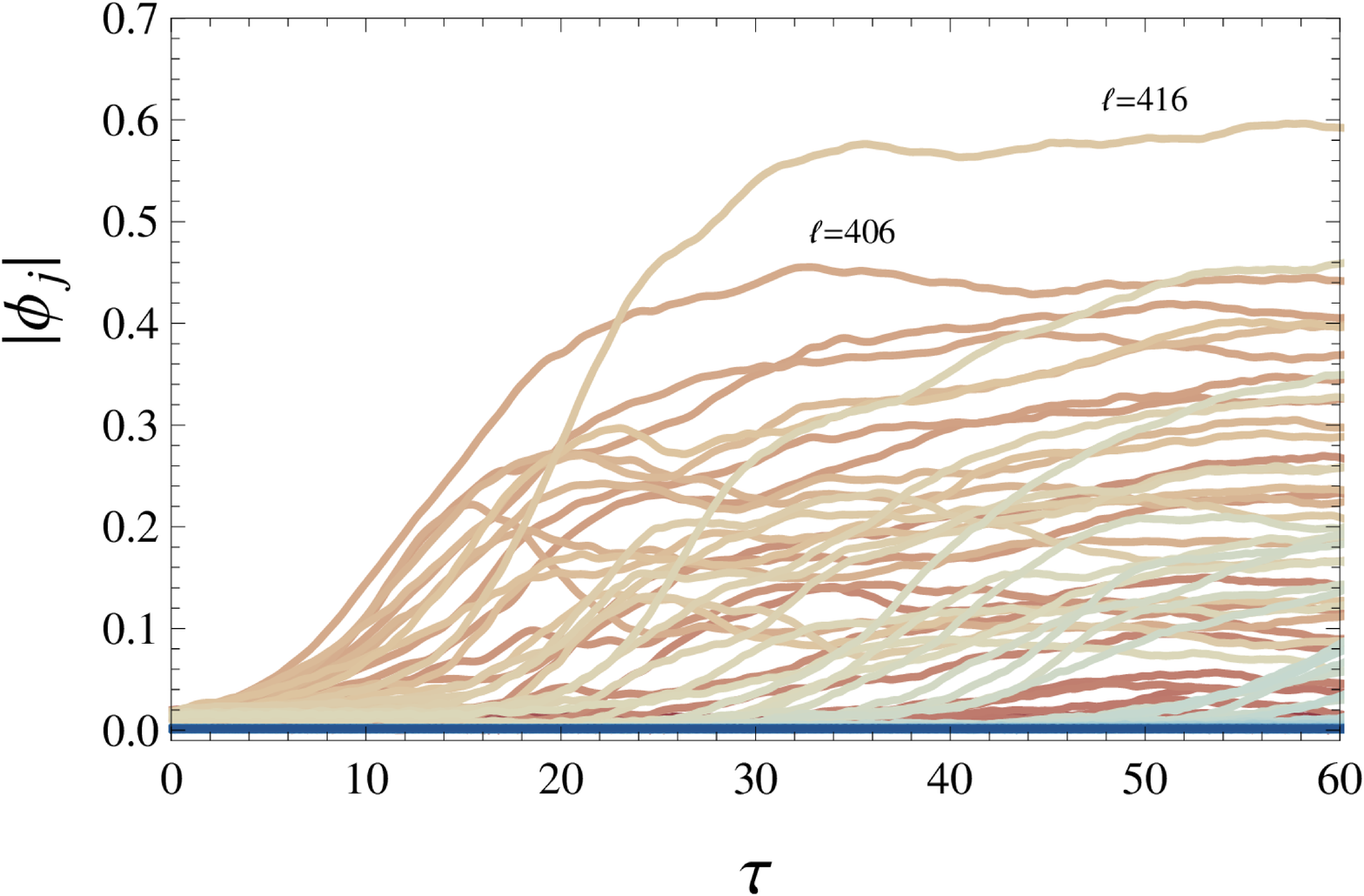}
\includegraphics[width=.3\textwidth,clip]{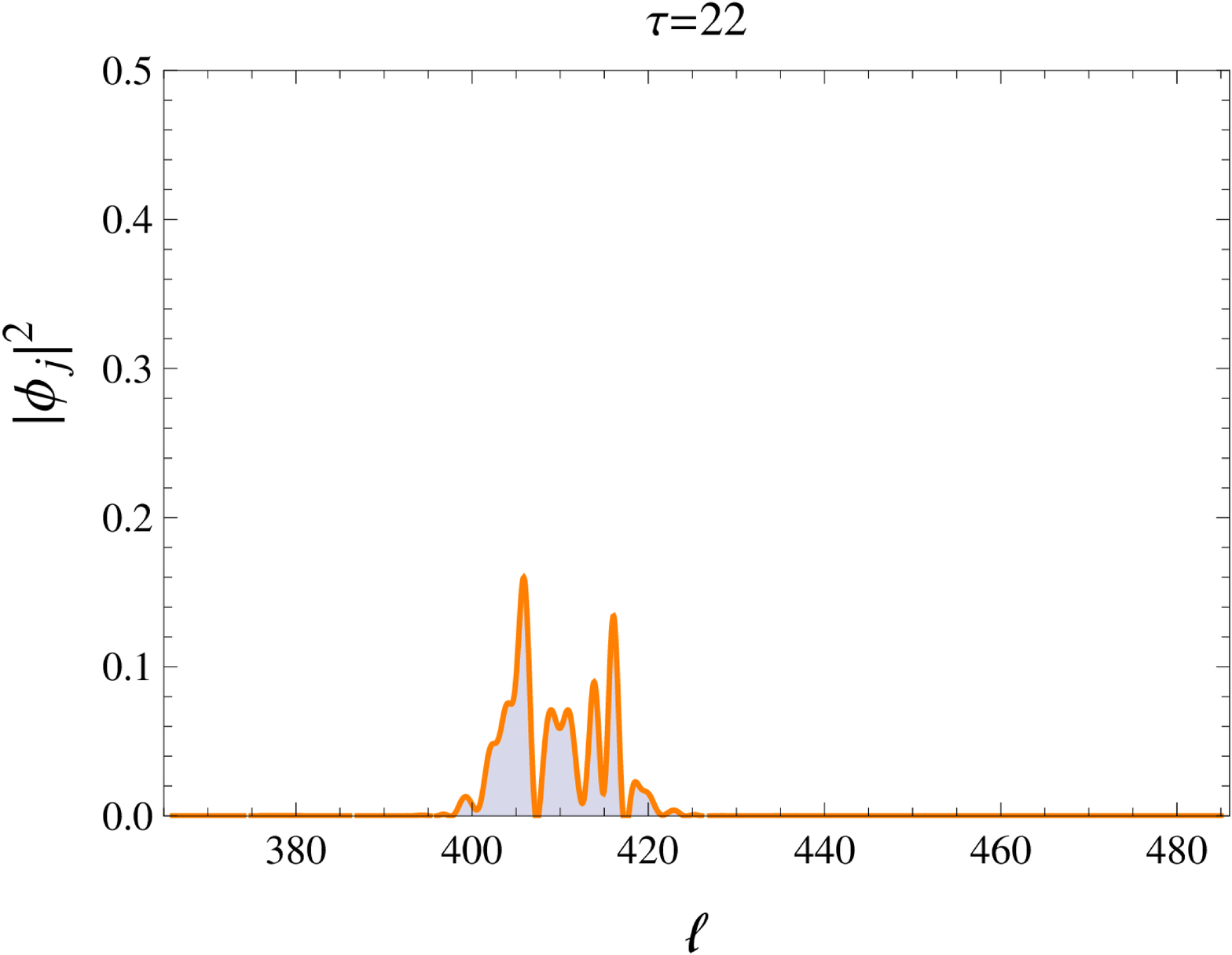}
\includegraphics[width=.3\textwidth,clip]{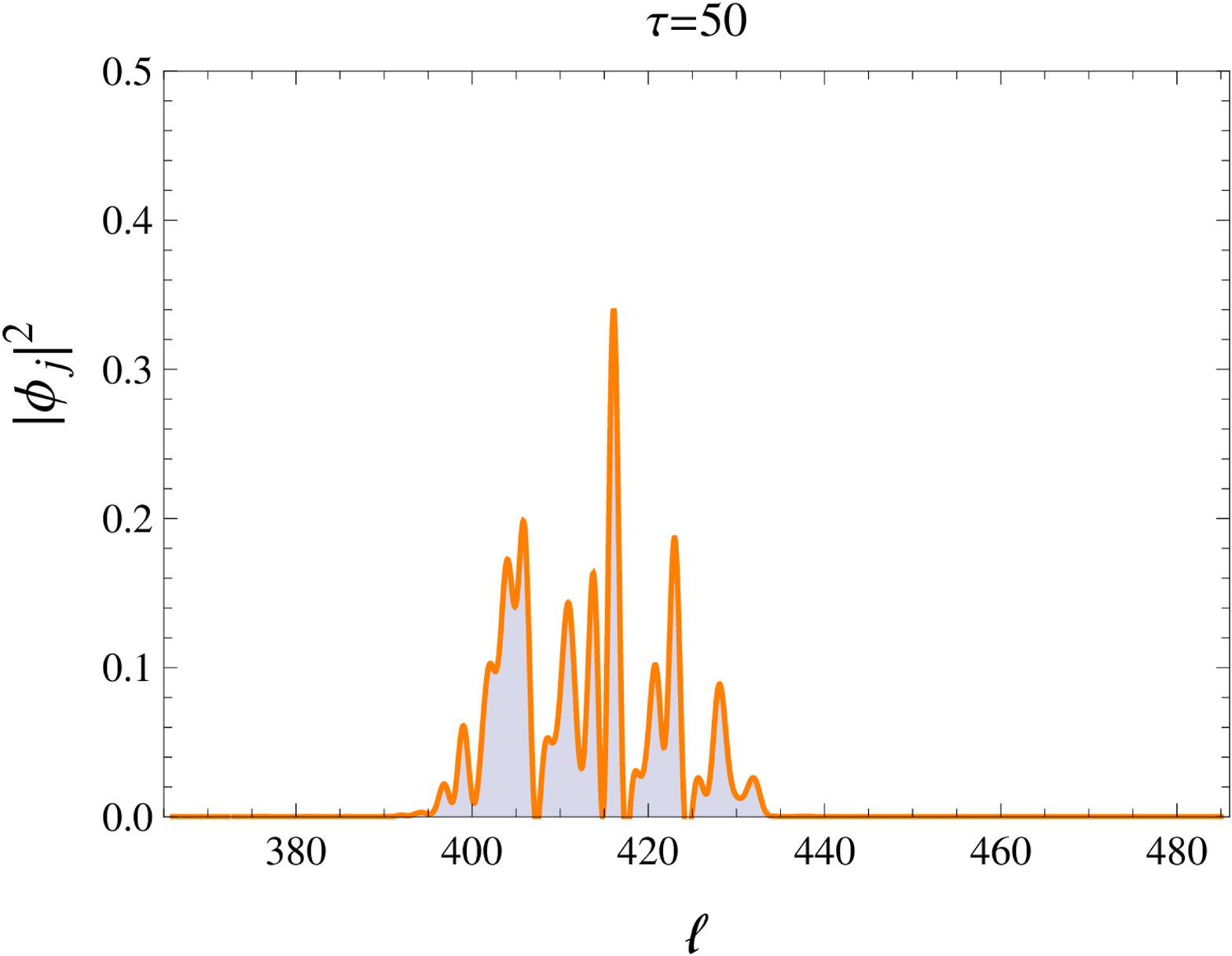}
\caption{\label{fig2.1}Case \emph{(ii)}. Left-hand panel: temporal evolution of $|\phi_j|$. Central and right-hand panels: intensity spectrum at different times.}
\end{figure}

In this case ($Q\simeq60$, $K=0.03$), which is the same as case \emph{(i)} but with linear stable modes included, we see in \figref{fig2.1} how the fluctuation spectrum is effectively broadened and the behavior of the saturated modes is significantly different from that of isolated resonant modes. A significant convection (drag) in velocity space is associated with the energy transfer from fast particles to the initially linear stable modes.
\begin{figure}[ht!]
\includegraphics[width=.23\textwidth,clip]{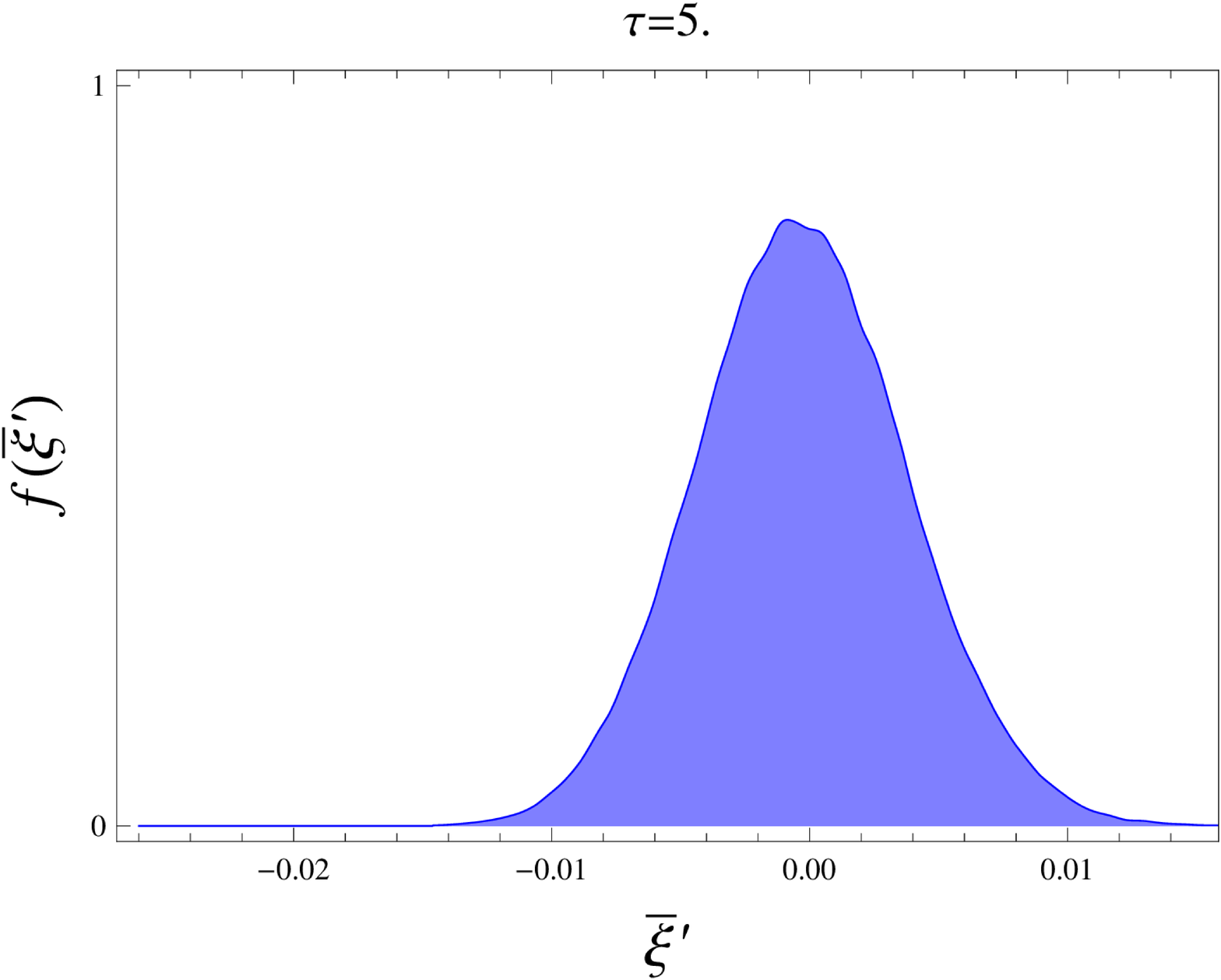}
\includegraphics[width=.23\textwidth,clip]{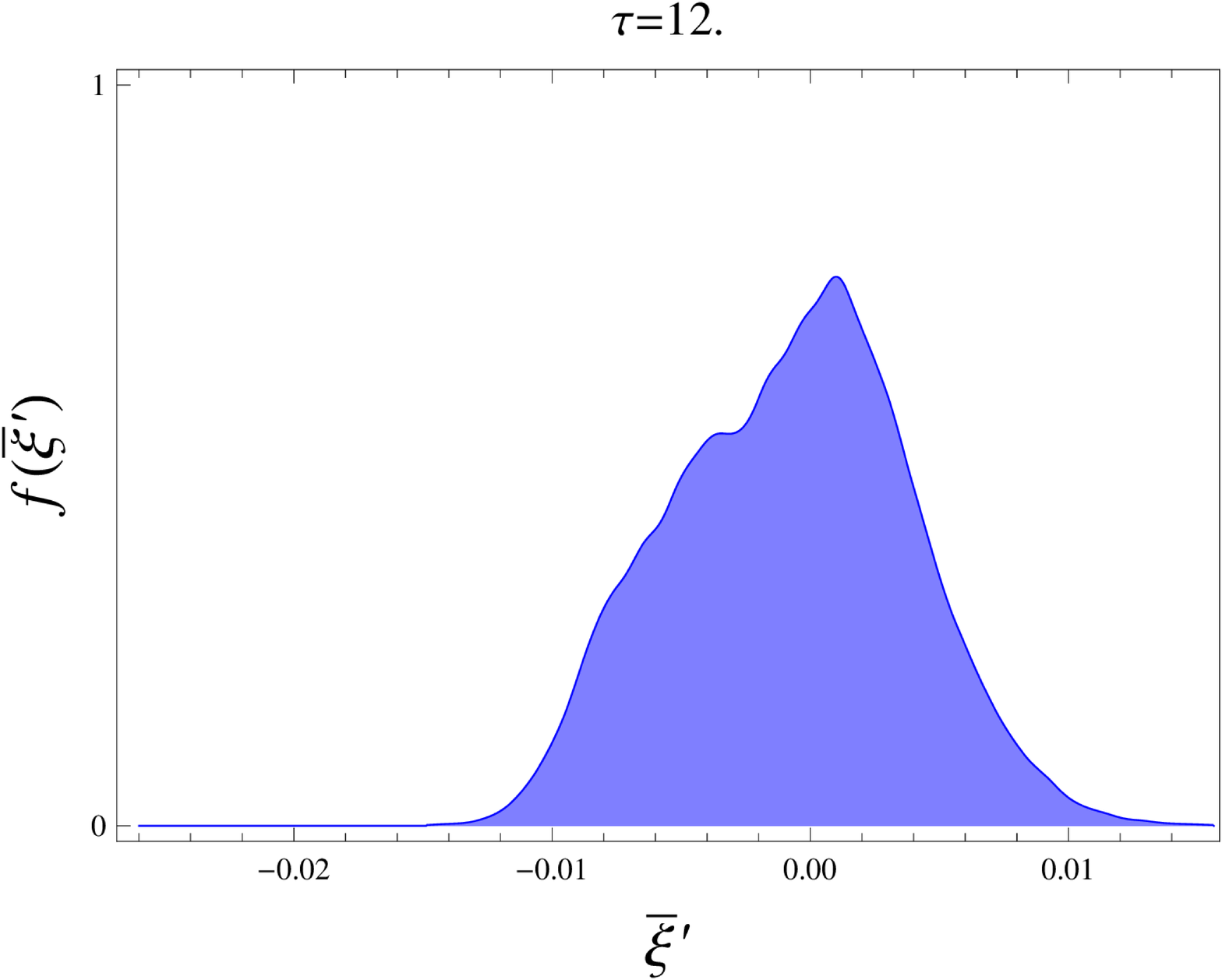}
\includegraphics[width=.23\textwidth,clip]{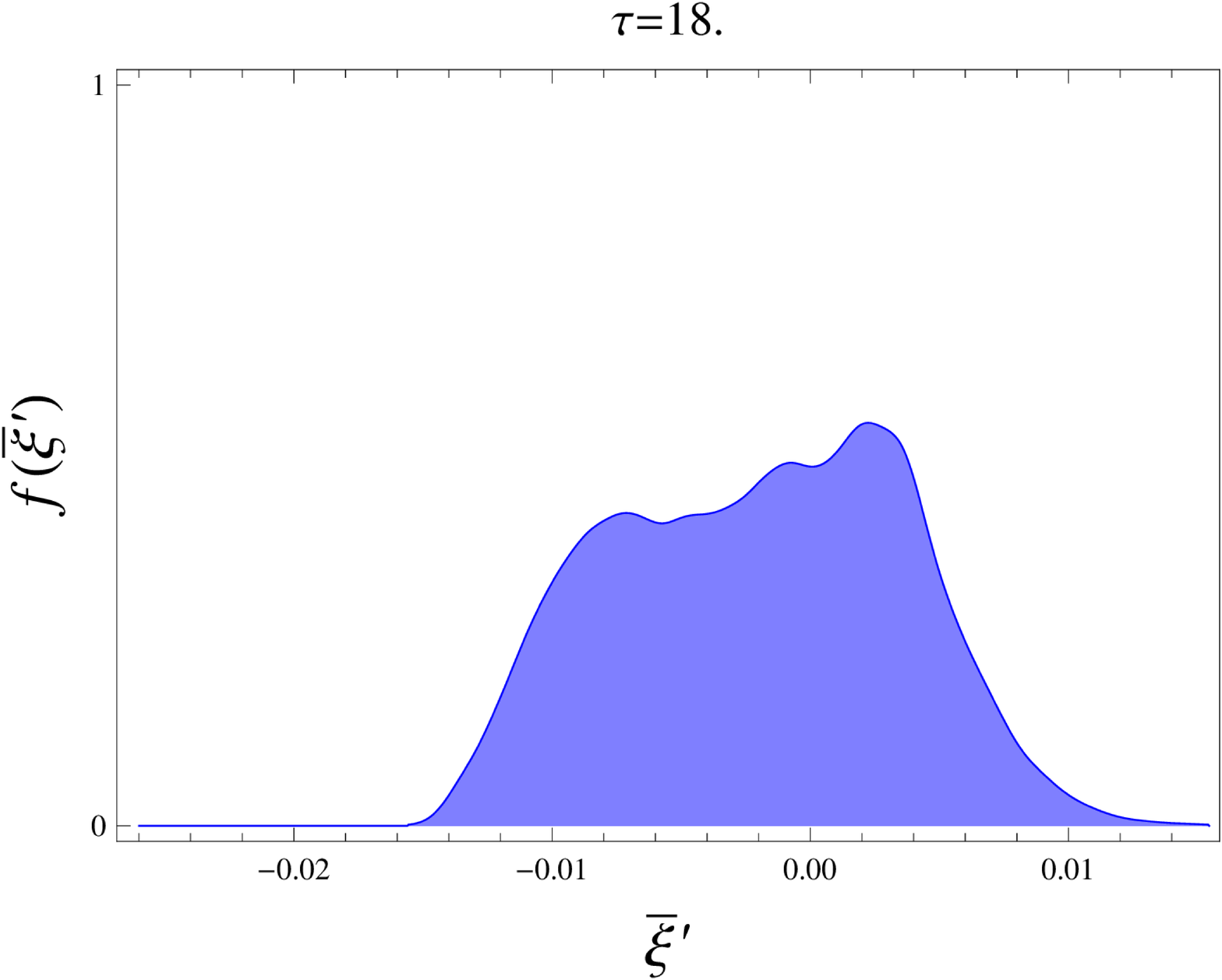}
\includegraphics[width=.23\textwidth,clip]{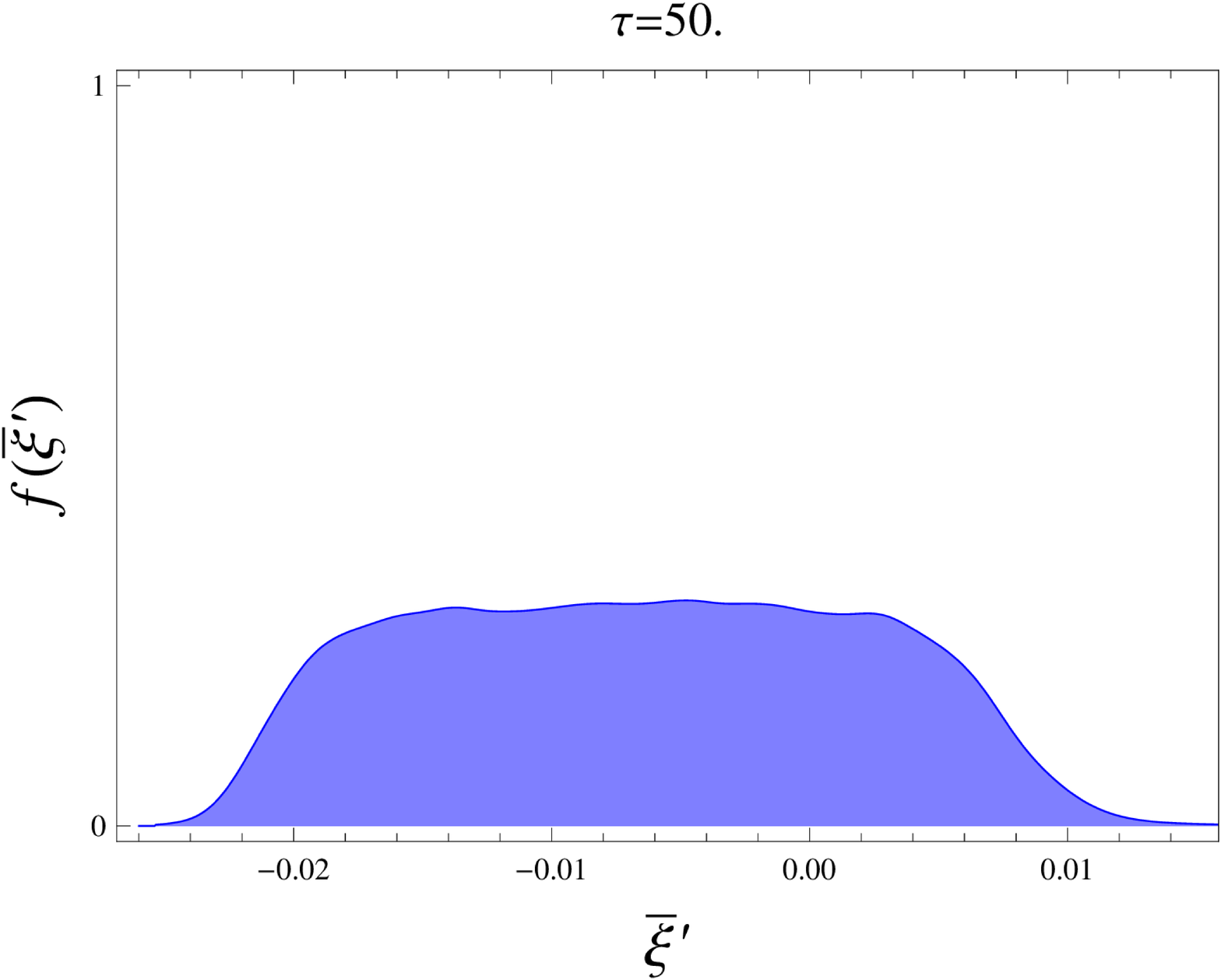}
\caption{\label{fig2.2}Case \emph{(ii)}. Evolution of the velocity distribution function.}
\end{figure}
The presence of well-marked peaks in the intensity spectrum is limited to a maximum $\ell$-value, beyond which energy transfer to new modes is almost suppressed and the saturation level of the excited modes is decreasing. The reason for such behavior is due to a progressive decrease of resonant particle population. The velocity distribution function (see \figref{fig2.2}) behaves similarly to the previous case, with density flattening followed by a uniform plateau formation. However, with respect to case \emph{(i)}, the velocity broadening of the distribution function is more extended as a consequence of stable modes.

\subsubsection{Case \emph{(iii)}: $\ch\simeq15$ and $476\leqslant\ell_{res}\leqslant525$}
\begin{figure}[ht!]
\includegraphics[width=.34\textwidth,clip]{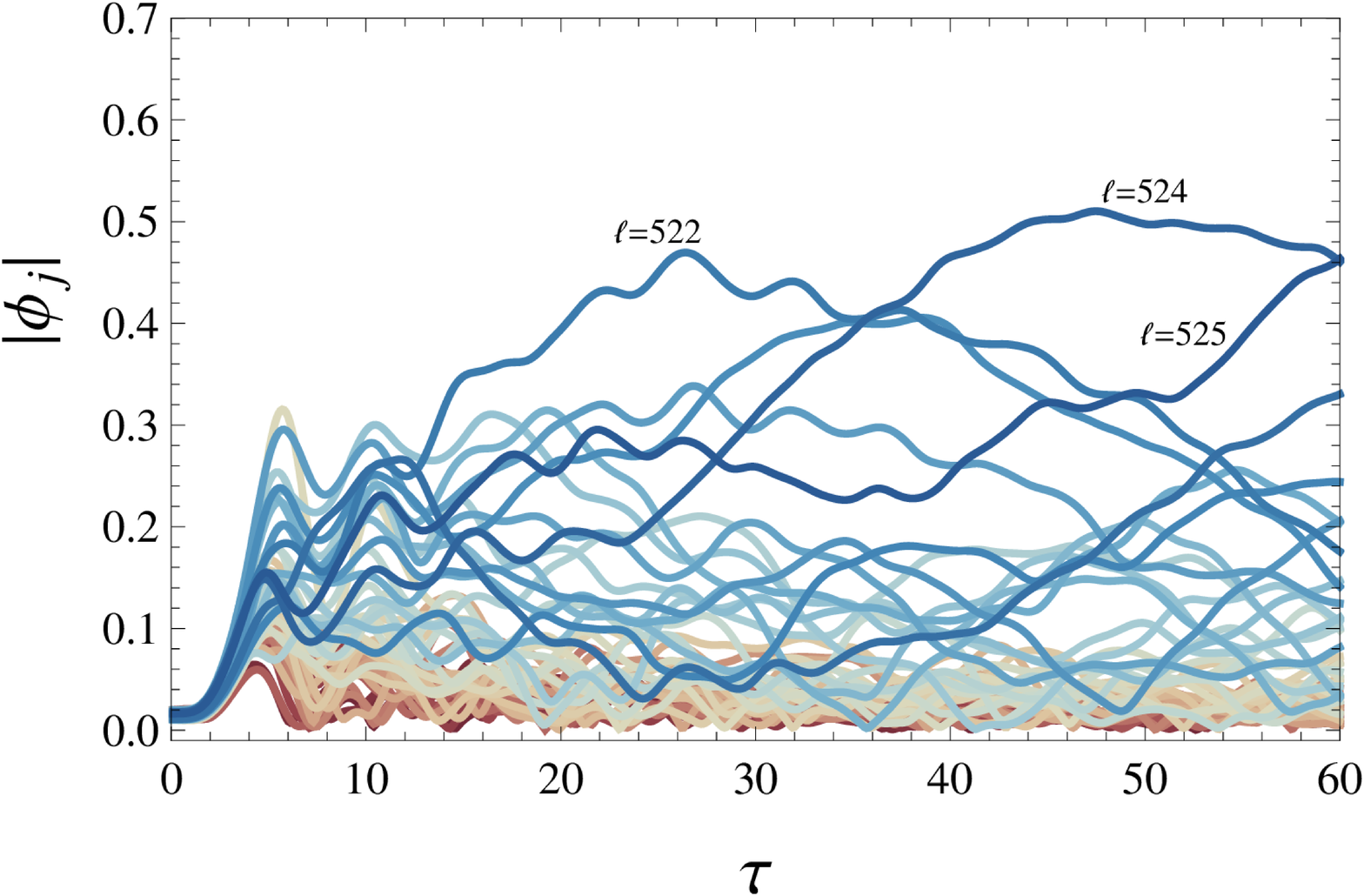}
\includegraphics[width=.3\textwidth,clip]{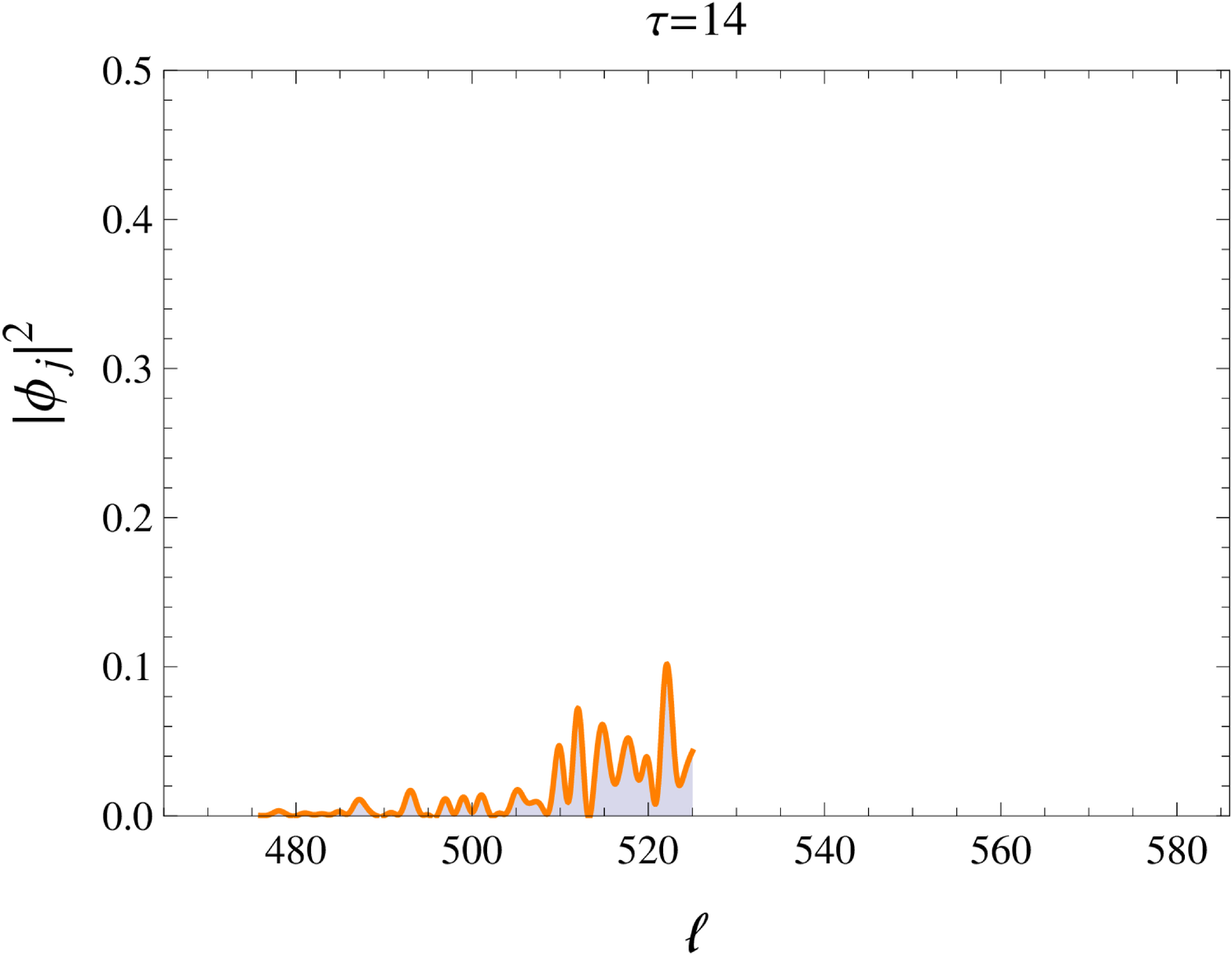}
\includegraphics[width=.3\textwidth,clip]{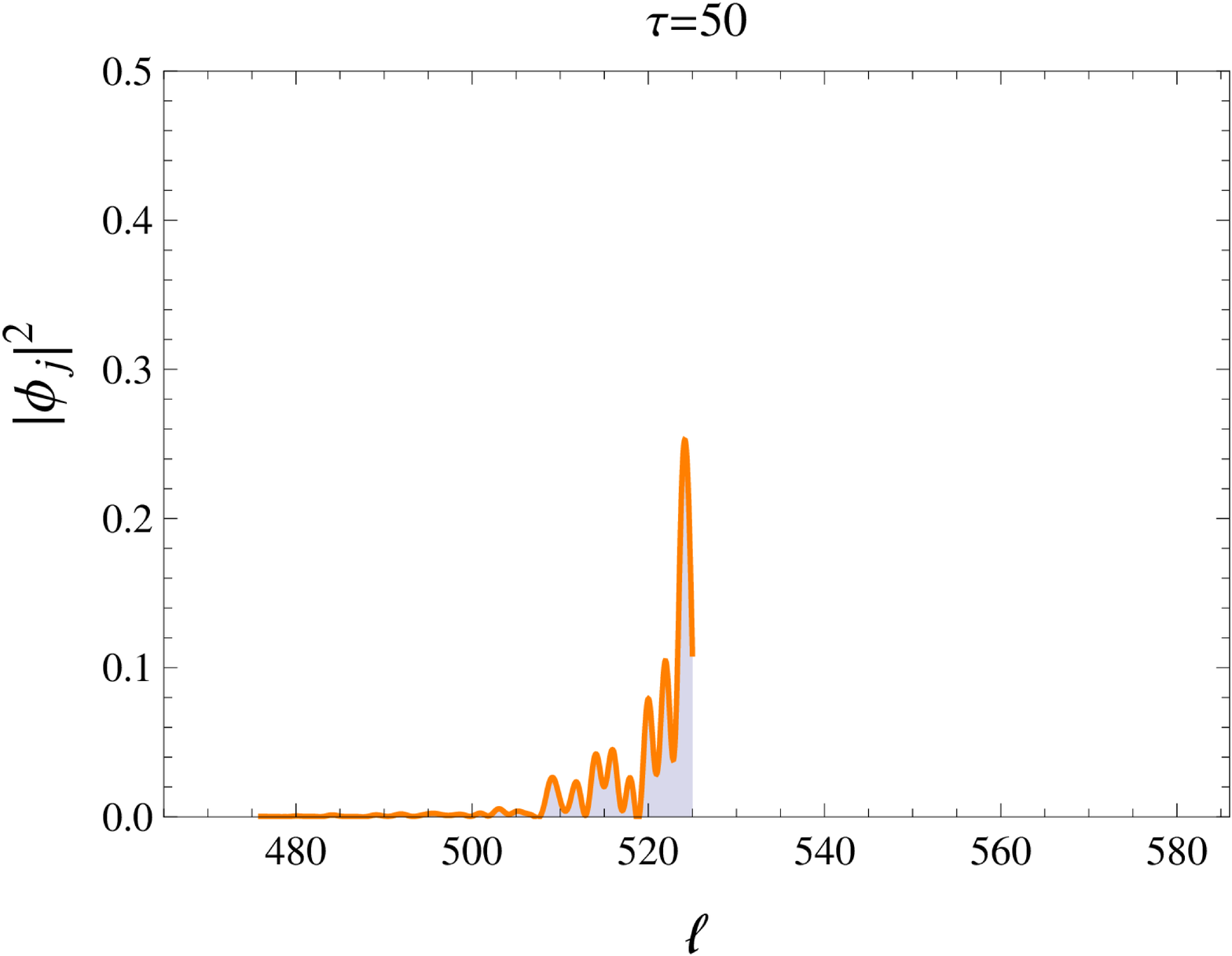}
\caption{\label{fig3.1}Case \emph{(iii)}. Left-hand panel: temporal evolution of $|\phi_j|$. Central and right-hand panels: intensity spectrum at different times.}
\end{figure}

\begin{figure}[ht!]
\includegraphics[width=.23\textwidth,clip]{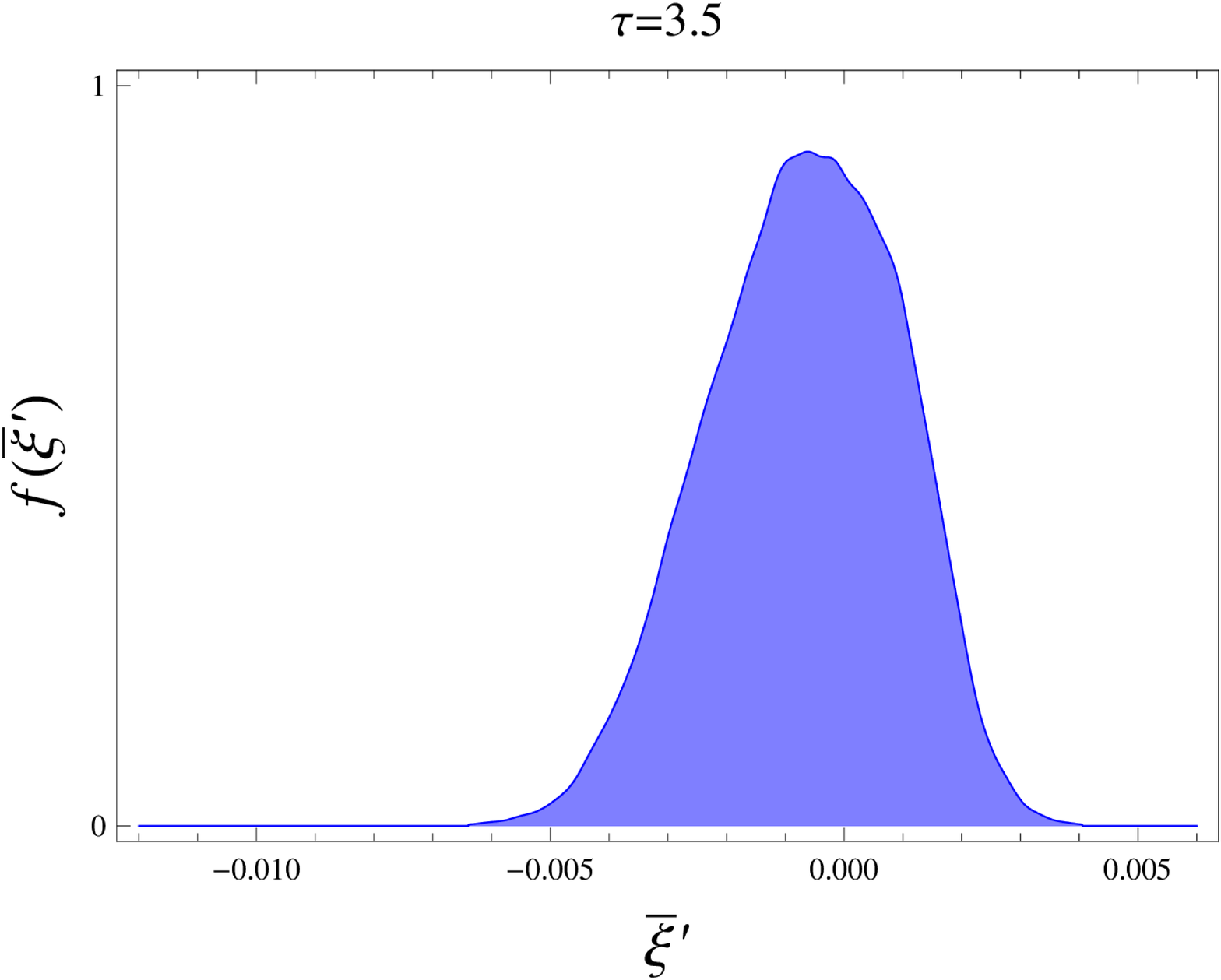}
\includegraphics[width=.23\textwidth,clip]{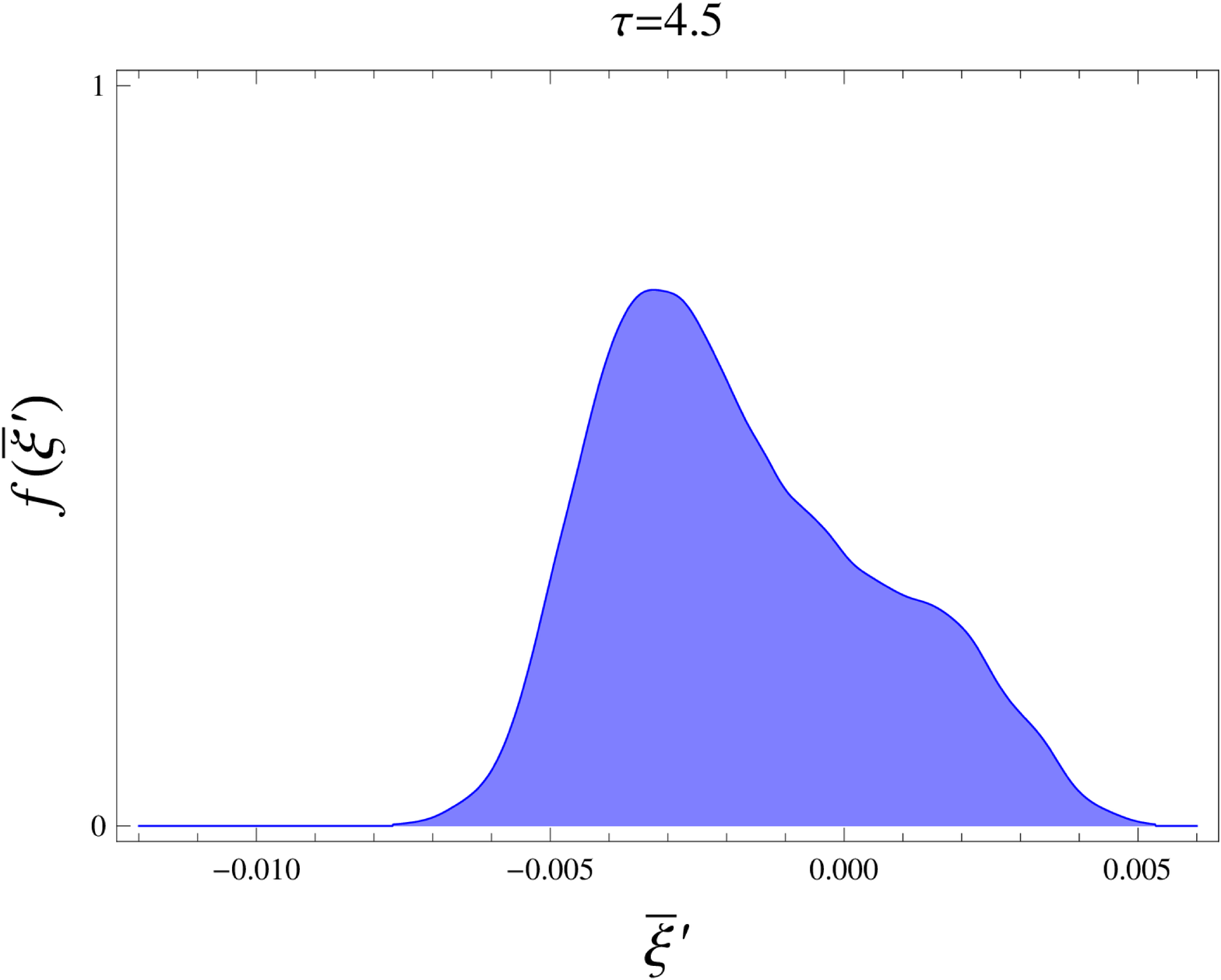}
\includegraphics[width=.23\textwidth,clip]{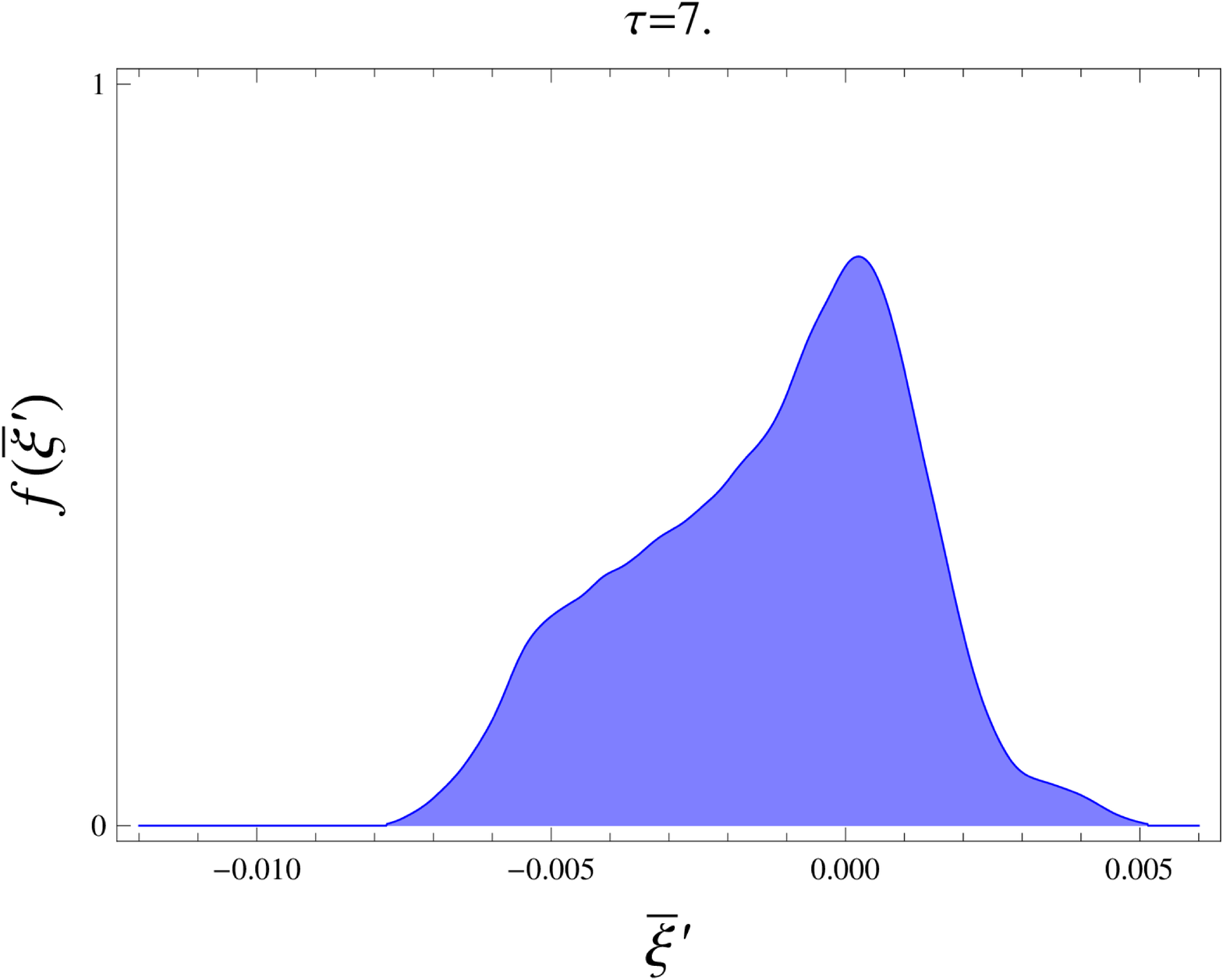}
\includegraphics[width=.23\textwidth,clip]{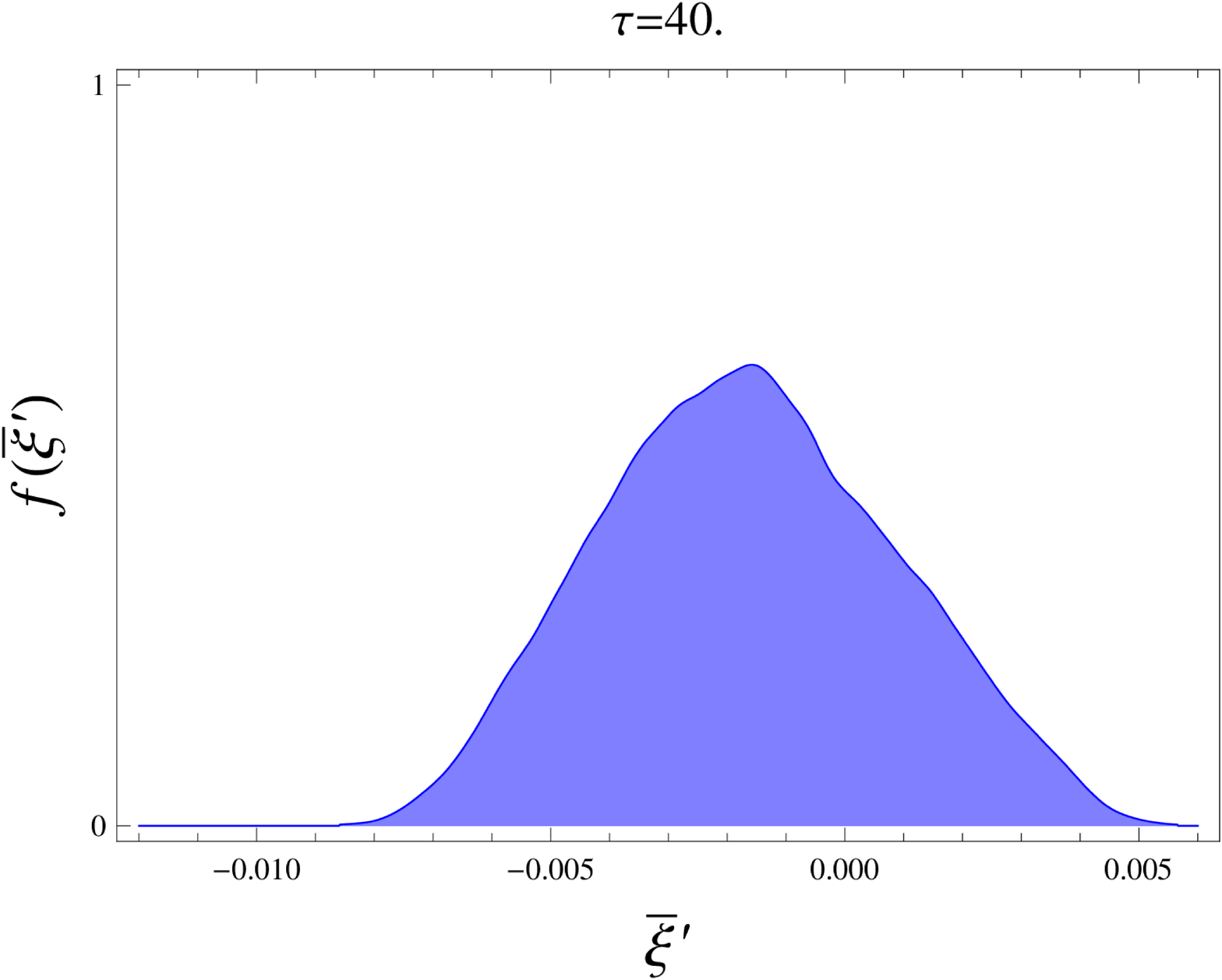}
\caption{\label{fig3.2}Case \emph{(iii)}. Evolution of the velocity distribution function.}
\end{figure}

This case, when the Chirikov parameter is significantly greater then one and without the inclusion of linear stable modes, is characterized by $Q\simeq3.6$ and $K=4.5$ ($\tau_s=4.5/(2\pi) \simeq 0.72$, $\tau_B\simeq16$). The spectrum, thus, is not broad and it is expected that the quasi-linear paradigm
is not appropriate in this case. In fact, in \figref{fig3.1} we can observe a morphology of mode saturation typical of isolated resonances and the intensity spectrum is characterized by a single dominant peak, which is progressively shifted up to the largest $\ell$-value in the simulation domain. Such behavior of the fluctuation spectrum is clearly reflected by the velocity distribution function in \figref{fig3.2}: after initial density flattening, the particle distribution function shows coherent behavior related to structures due to wave-particle trapping. In this scenario, a uniform plateau is never formed and the asymptotic evolution resembles that of an isolated resonance more than the typical morphology of the quasi-linear paradigm.
The explanation for such behavior, again, is due to the fact that we are not dealing, in the present case, with a broad spectrum. In fact, the ratio between the Chirikov parameter and the number of overlapping resonances is of order unity. Thus, a given particle feels the effect of many modes acting as a nearly periodic spectrum. Furthermore, considering the contribution of all particles near a given resonance condition, modes instantaneously transferring their energy to particles are damped, while those receiving energy are amplified. Thus, there exists a mechanism able to create and enforce wave-particle phase synchronization. This is the reason why, in this case, the formation of a plateau in the velocity distribution function is suppressed.

\subsubsection{Case \emph{(iv)}: $\ch\simeq15$ and $466<\ell<476\leqslant\ell_{res}\leqslant525<\ell<585$}
\begin{figure}[ht!]
\includegraphics[width=.34\textwidth,clip]{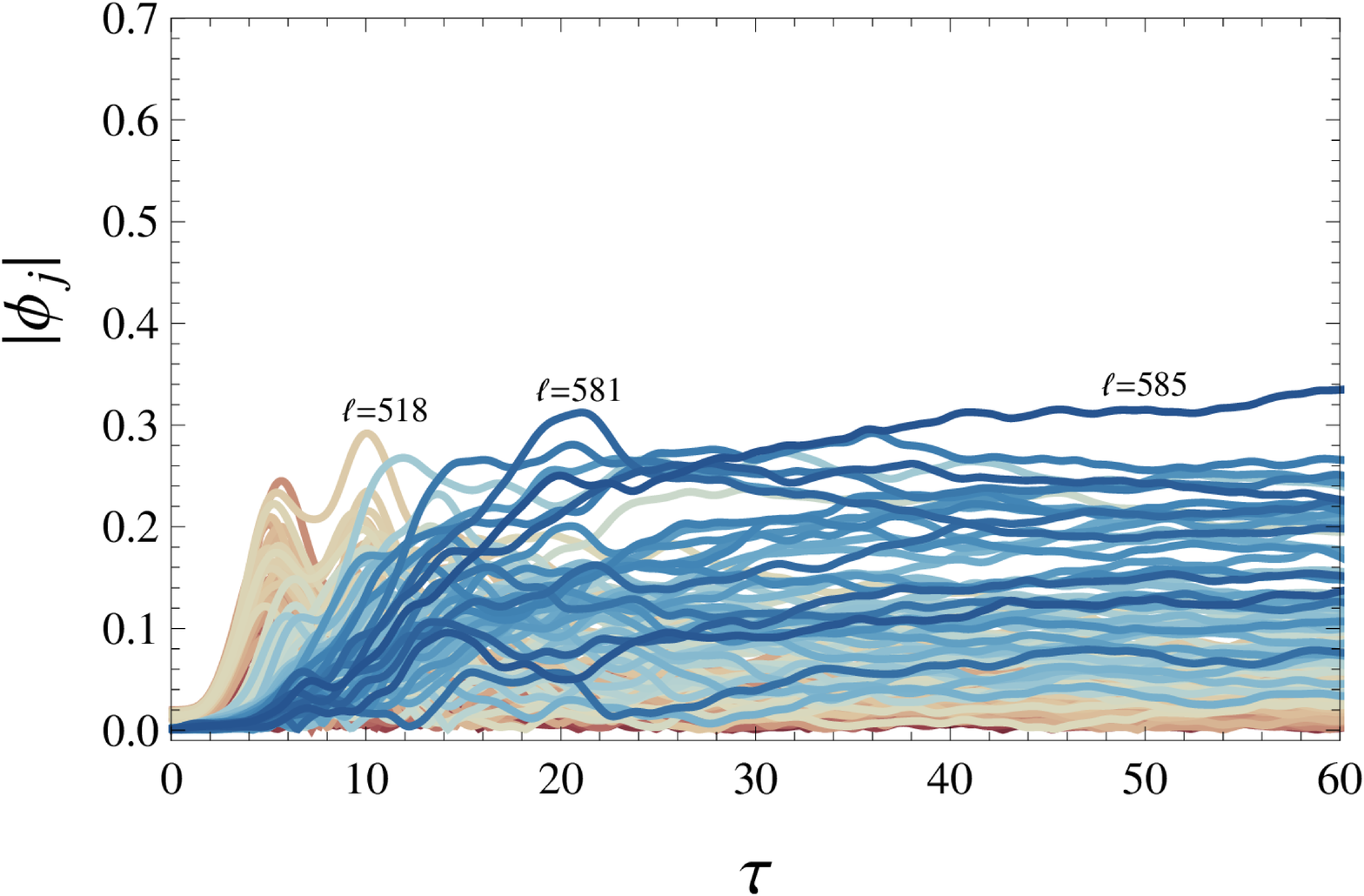}
\includegraphics[width=.3\textwidth,clip]{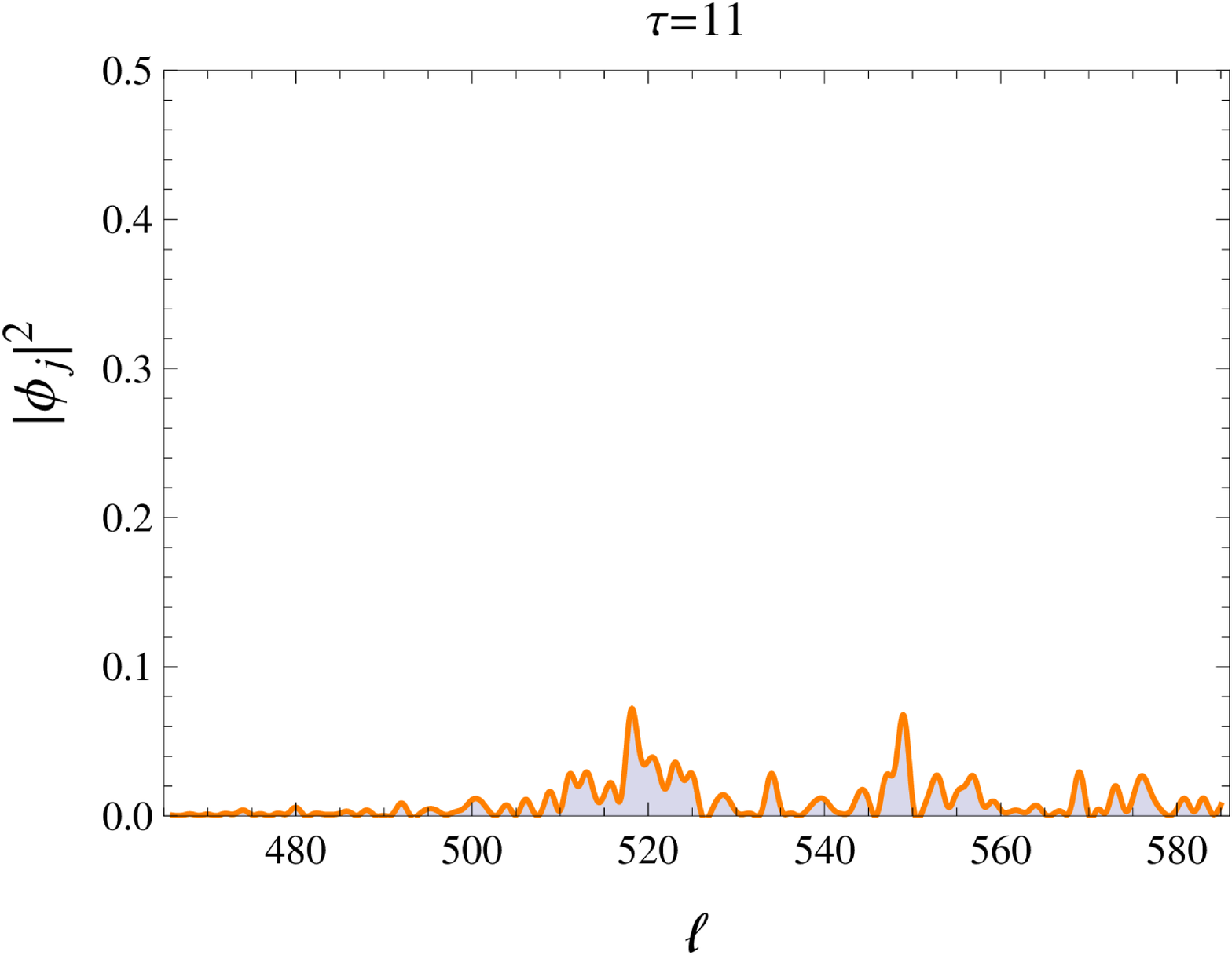}
\includegraphics[width=.3\textwidth,clip]{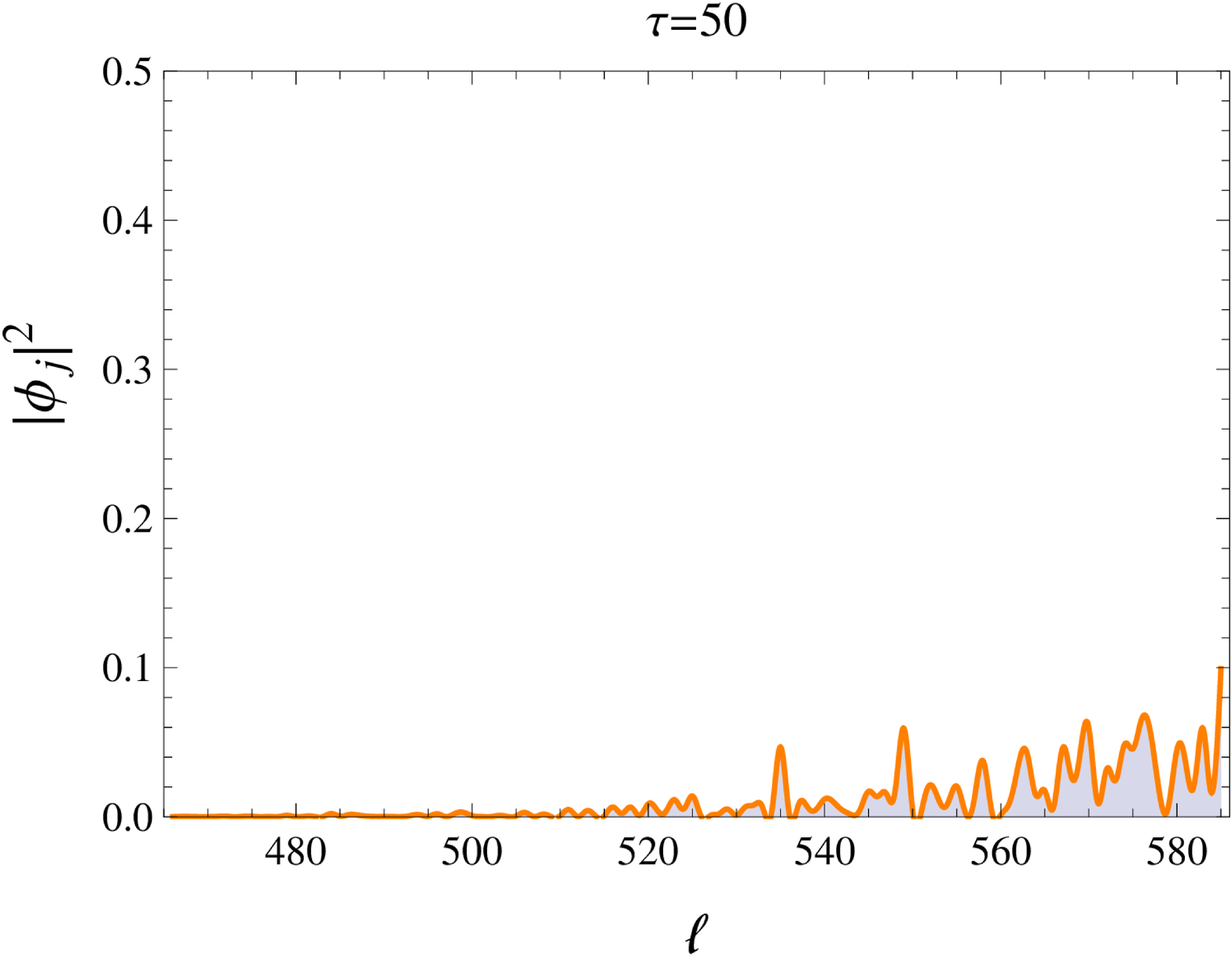}
\caption{\label{fig4.1}Case \emph{(iv)}. Left-hand panel: temporal evolution of $|\phi_j|$. Central and right-hand panels: intensity spectrum at different times.}
\end{figure}

\begin{figure}[ht!]
\includegraphics[width=.23\textwidth,clip]{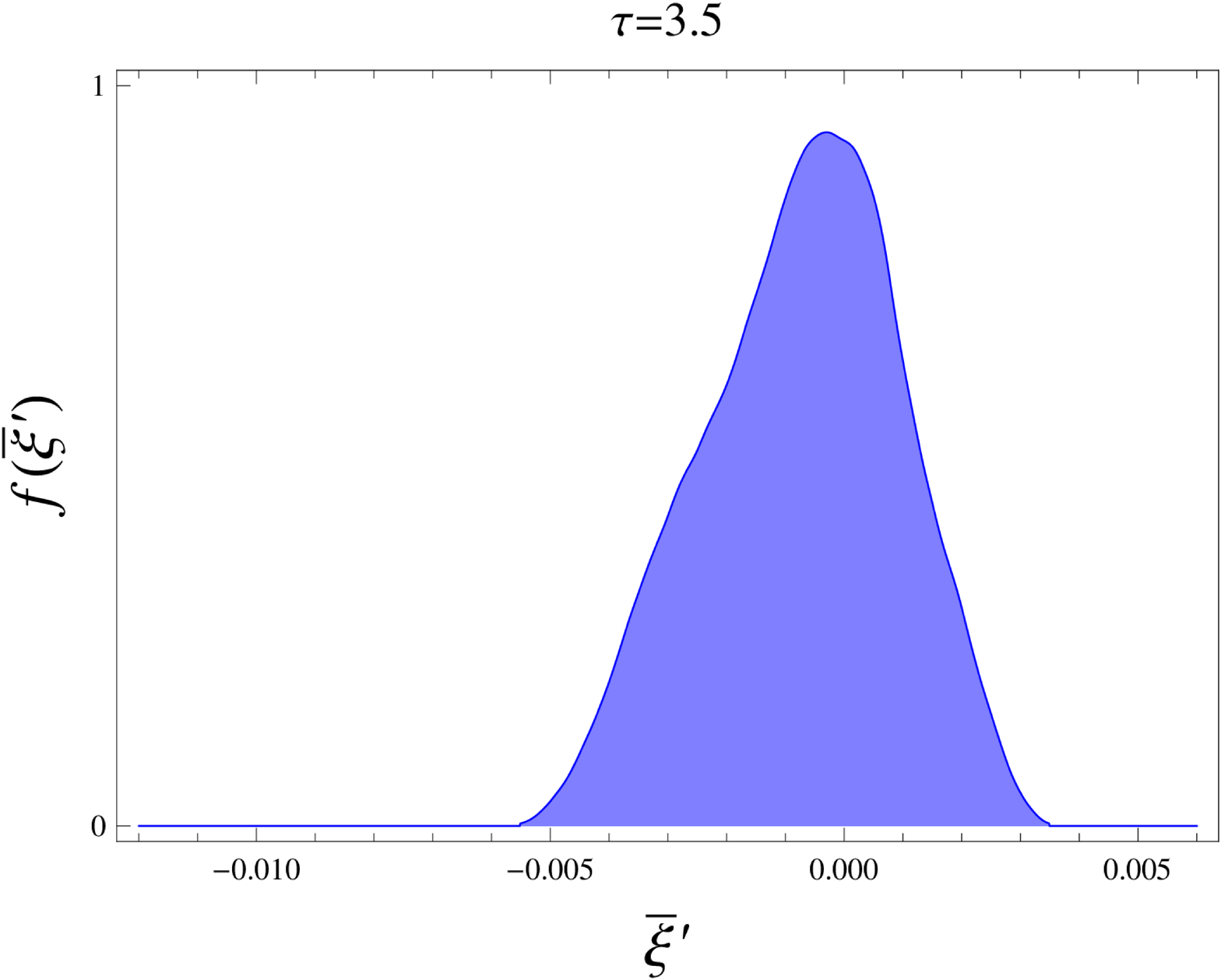}
\includegraphics[width=.23\textwidth,clip]{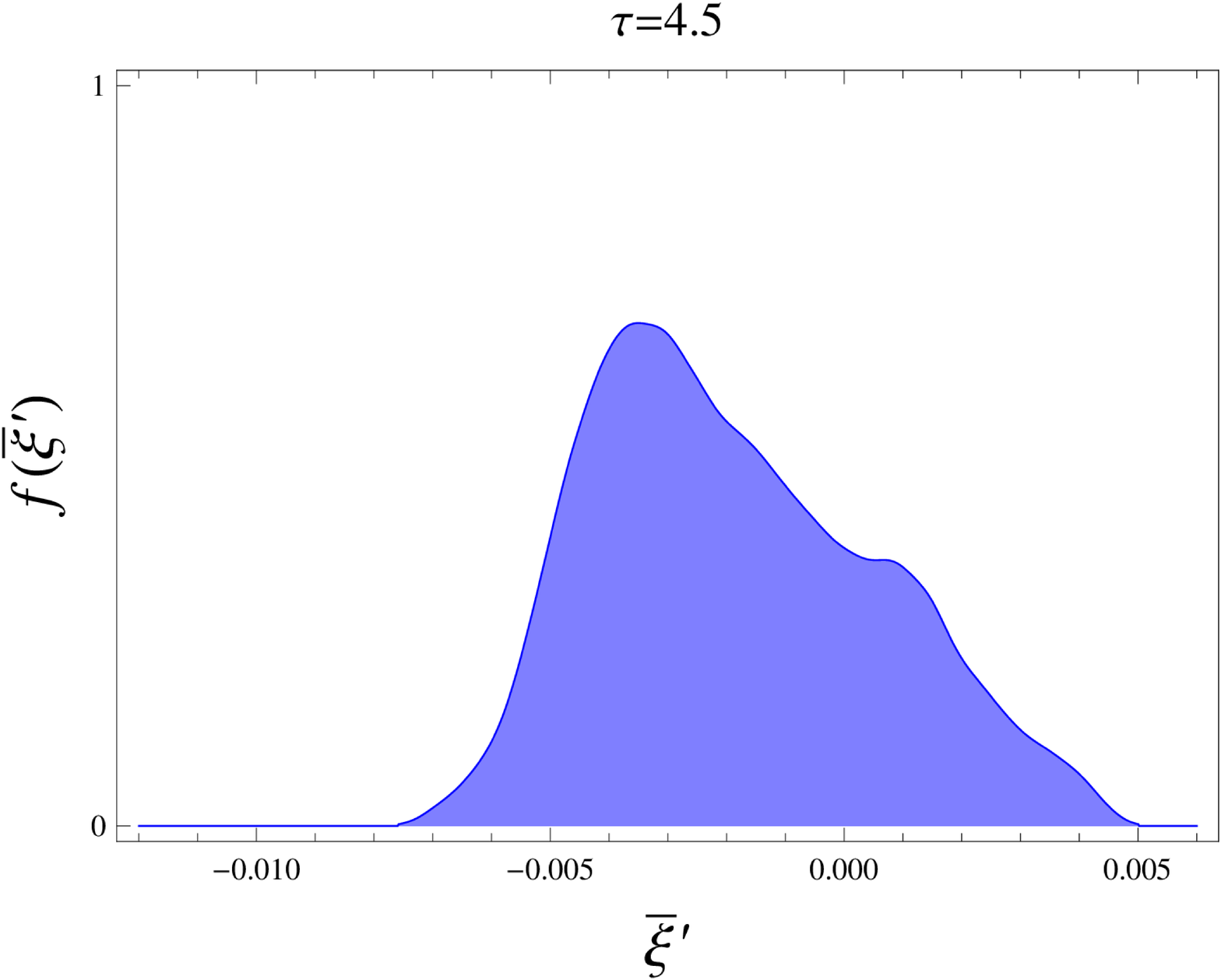}
\includegraphics[width=.23\textwidth,clip]{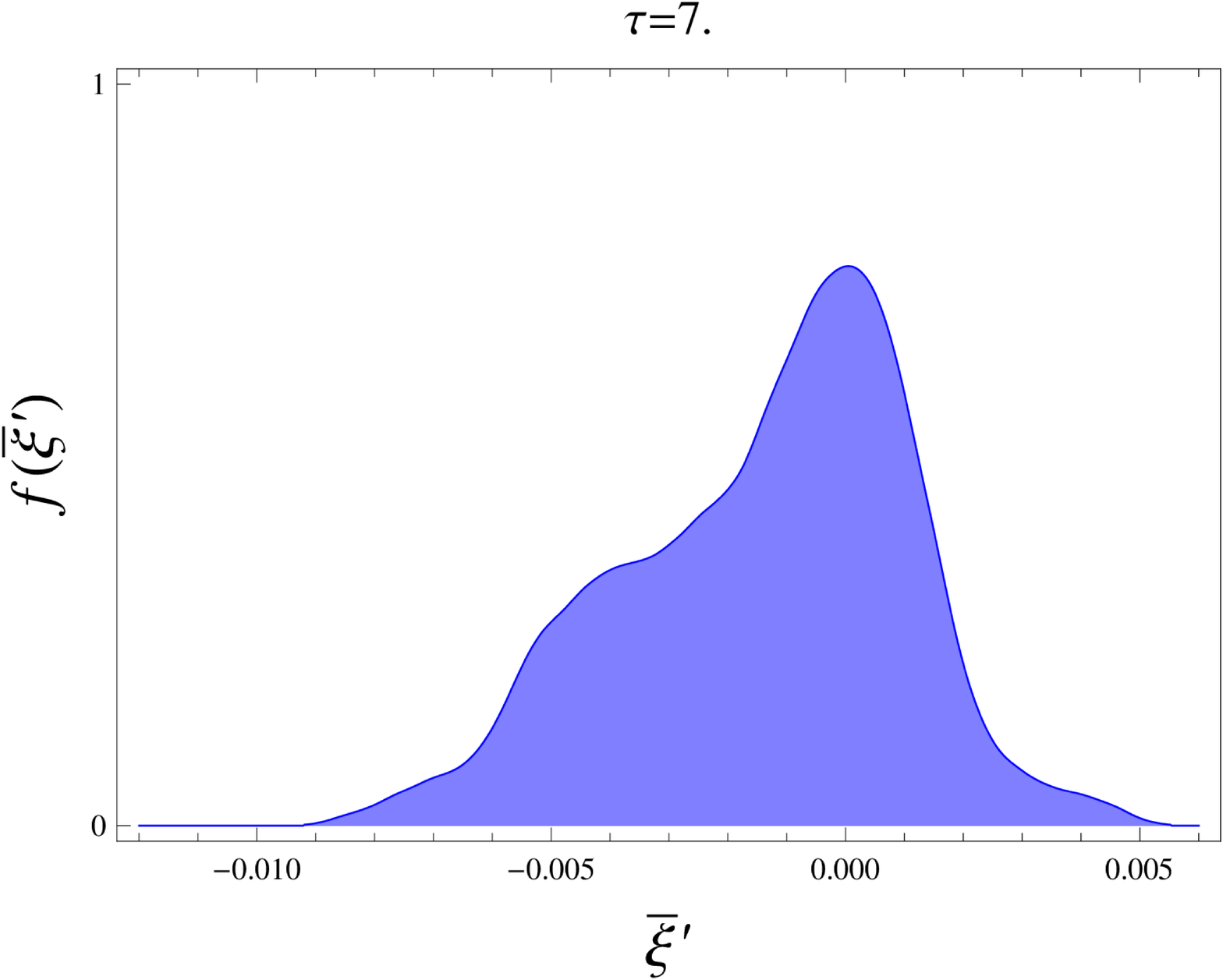}
\includegraphics[width=.23\textwidth,clip]{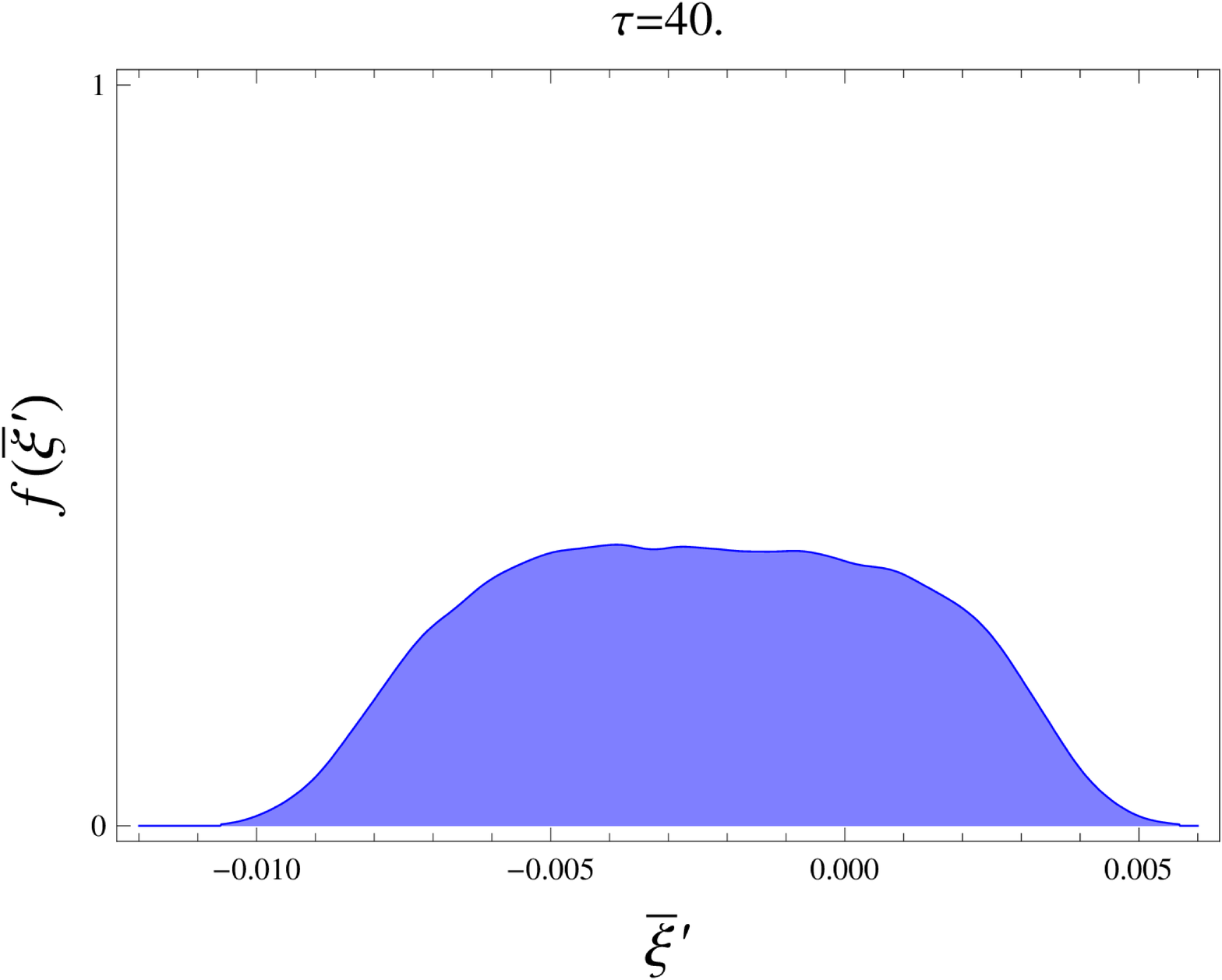}
\caption{\label{fig4.2}Case \emph{(iv)}. Evolution of the velocity distribution function.}
\end{figure}

The conjecture described for case \emph{(iii)} is clearly confirmed in this case, in which the ratio between the Chirikov parameter and the number of resonant beams is still of order unity, but now the presence of linearly stable modes is included in the evolution. As main effect, see \figref{fig4.1}, a broadening of the intensity spectrum is clearly observable and, overall, the plateau in the particle distribution function is restored at sufficiently large times (see \figref{fig4.2}). Indeed, the linearly stable modes, once excited by the nonlinear velocity spread (due to both convective and diffusive transport), reproduce a broad spectrum, significantly enhancing the effective $Q$ value in the system ($Q\simeq7.9$). We note that, in the present case, $K \simeq 1.9$ and, since we  are now dealing with a broad spectrum, this regime is the most appropriate to be compared with the analysis presented in Sec.\ref{dcmodel}. 
This analysis illuminates the relevance that linear stable modes have in determining the evolution of intensity spectrum and particle distribution function.

\subsection{Test particle transport}\label{sims4}

To illustrate the mixed diffusive-convective character of the broad beam relaxation process, we have carried out test particle transport analyses of the simulation results discussed in the previous section. In particular, we analyze the time behavior of  velocity fluctuations. For a given simulation, we extract the self-consistent potential fields and we monitor the evolution of a set of test particles (1000)  evolving under the influence of those fields. The tracers are initialized to represent one single cold beam, among those of which the broad beam initially consists. For a single initial condition (\ie by setting all the test-particle initial velocities equal to $\bar{\xi}'_{\al i}(0)$), we plot the mean square path as function of time:
\begin{align}
\lamal\equiv\langle[\bar{\xi}'(\tau)-\langle\bar{\xi}'(\tau)\rangle]^2\rangle\;,
\end{align}
where the average $\langle...\rangle$ is taken over the tracers. When $\lamal$ shows a linear behavior in $\tau$, we can speak of diffusive transport. This analysis is, in general, non trivial, because $\lam$ is characterized by complicated nonlinear time evolution when tracers are significantly displaced in velocity space (for an in-depth analysis of this problem, see Ref. \cite{VK12}). 

\begin{figure}[ht!]\centering
\includegraphics[width=.3\textwidth,clip]{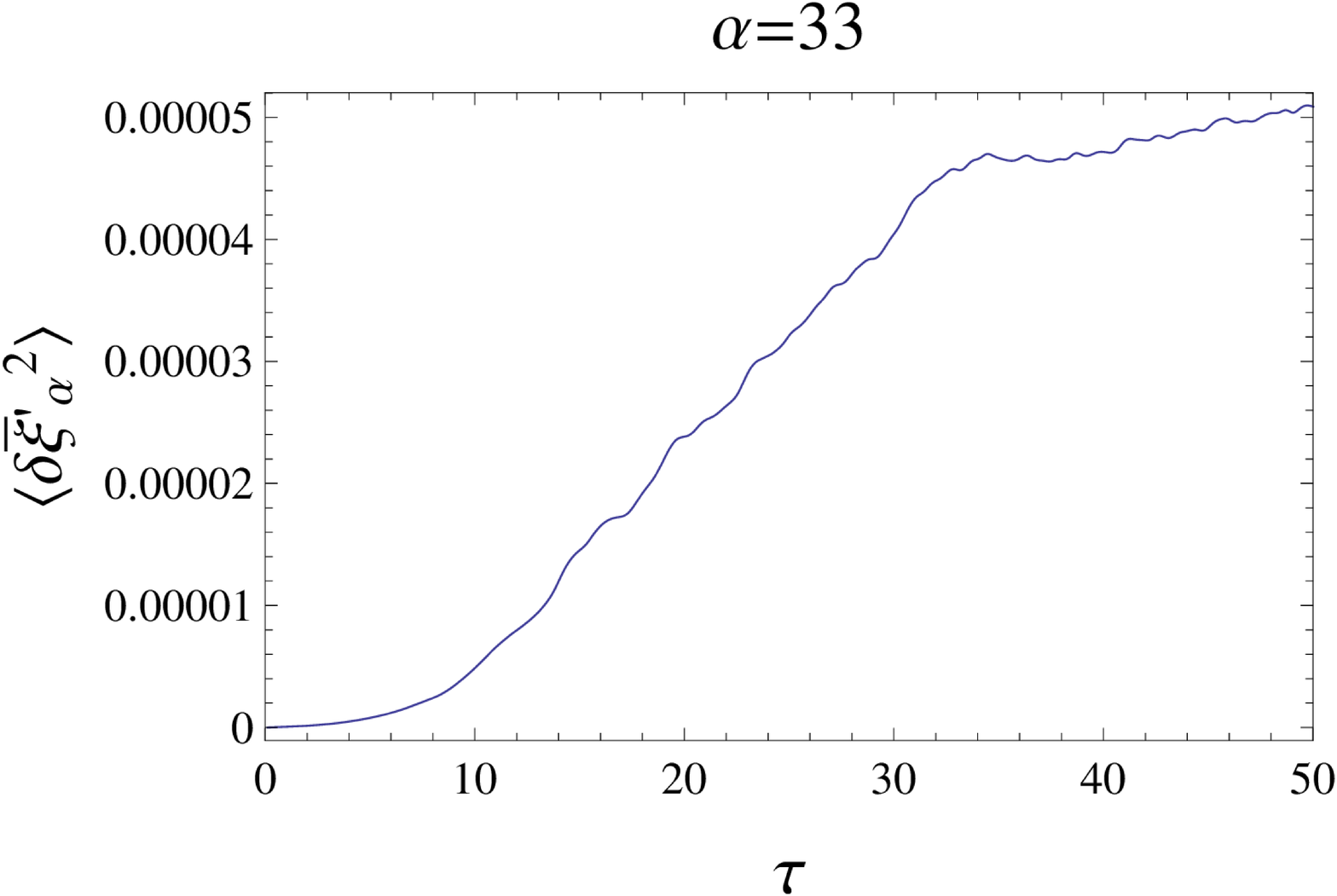}
\includegraphics[width=.3\textwidth,clip]{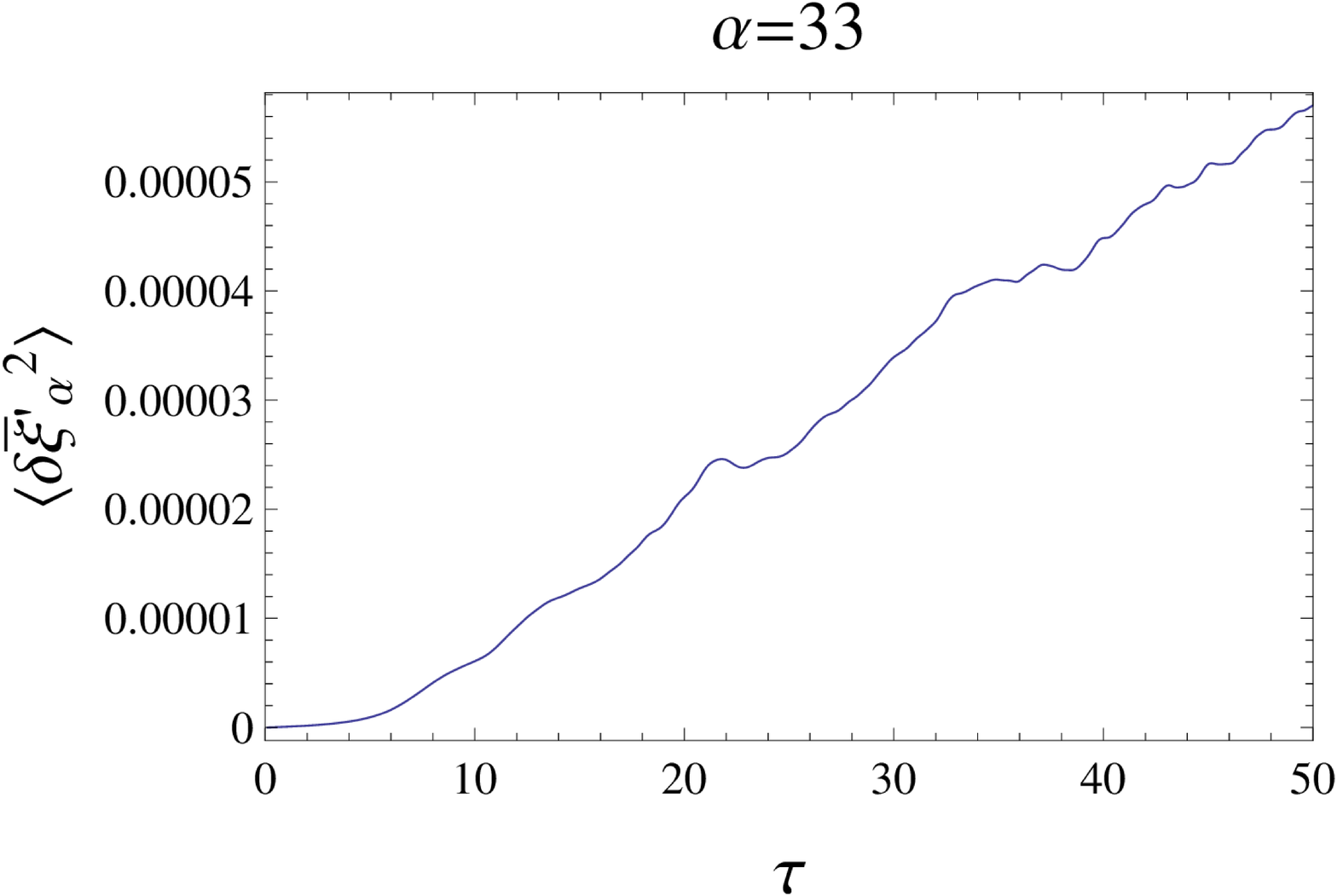}
\caption{\label{LL0102} Temporal evolution of $\langle \delta \bar{\xi}'^2_{33} \rangle(\tau)$ for test particles initialized at $\bar{\xi}'_{33 i}(0)$ (near the central beam): in the left-hand panel for the case \emph{(i)} and in the right-hand panel for case \emph{(ii)}. The linear (diffusive) time behavior is evident.}
\end{figure}

Nonetheless, limiting our attention to specific class of tracers and appropriate time intervals, in cases \emph{(i)} and \emph{(ii)} we can identify a diffusion phase of the evolution, see \figref{LL0102}. There, we show the behavior of $\lamal$ for tracers of the beam $\al=33$ (left panel refers to case \emph{(i)}, while right panel to case \emph{(ii)}). The existence of a clear linear behavior of $\lamal$, \ie of diffusive transport, is clearly recognizable at least for intermediate times. It is worth noting that the presence of linear stable modes (case \emph{(ii)}) does not significantly affect the diffusion process since the corresponding coefficients are very close in the two cases. In this respect, it is interesting to compare the present study with the analysis in Ref.\cite{EE08}, where it is shown that the self-consistency of the model is broken when the plateau is sufficiently broad. Such a study demonstrates that the ratio between the obtained diffusion coefficient to the quasi-linear one can take a wide range of values. However, this ratio tends to unity when the random phase approximation is assumed. This is consistent with our findings in \figref{LL0102}, showing a slight  decreased diffusion coefficient for the broader fluctuation spectrum case, including non-resonant modes.

\begin{figure}[ht!]\centering
\includegraphics[width=.3\textwidth,clip]{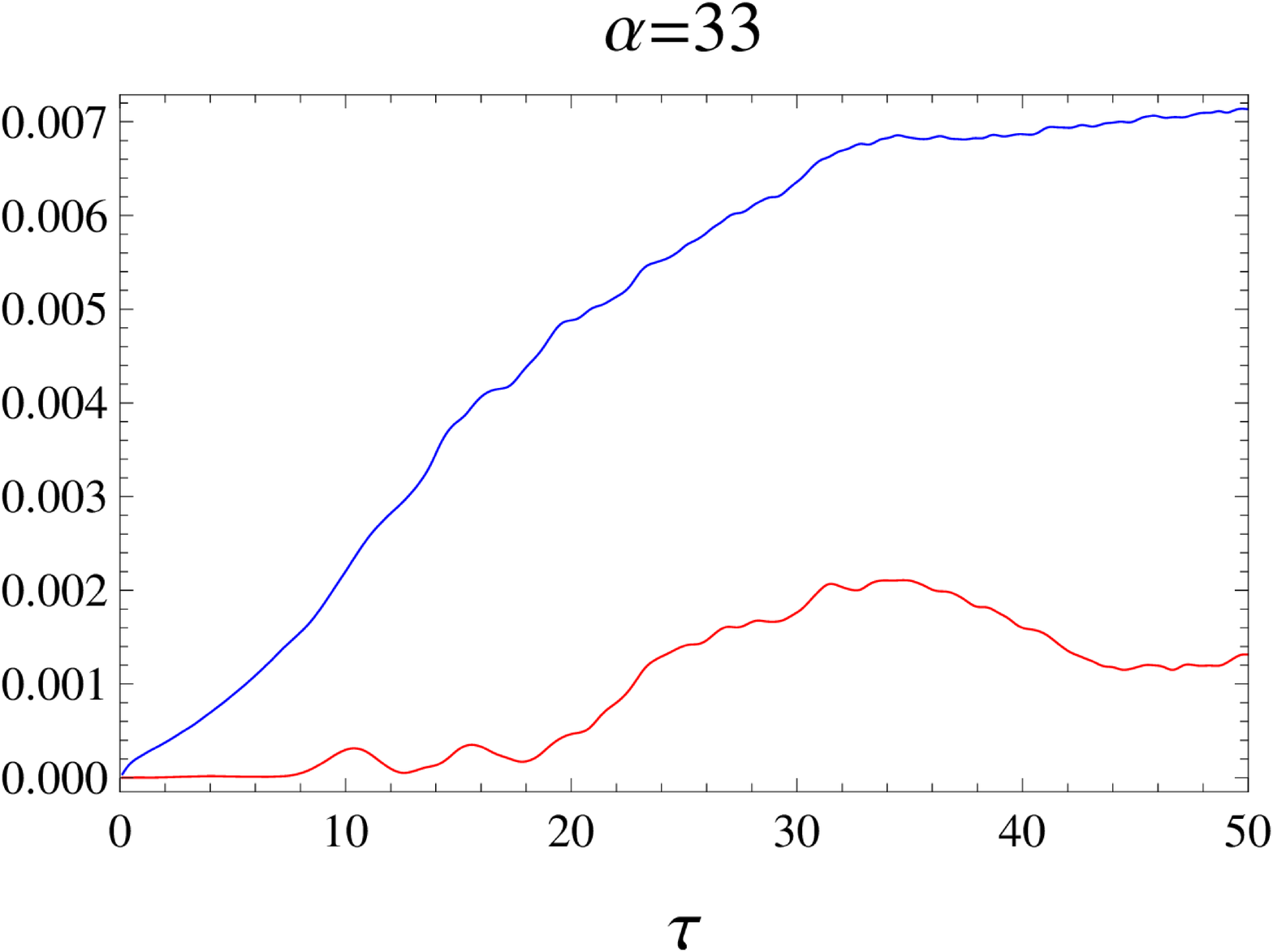}
\includegraphics[width=.3\textwidth,clip]{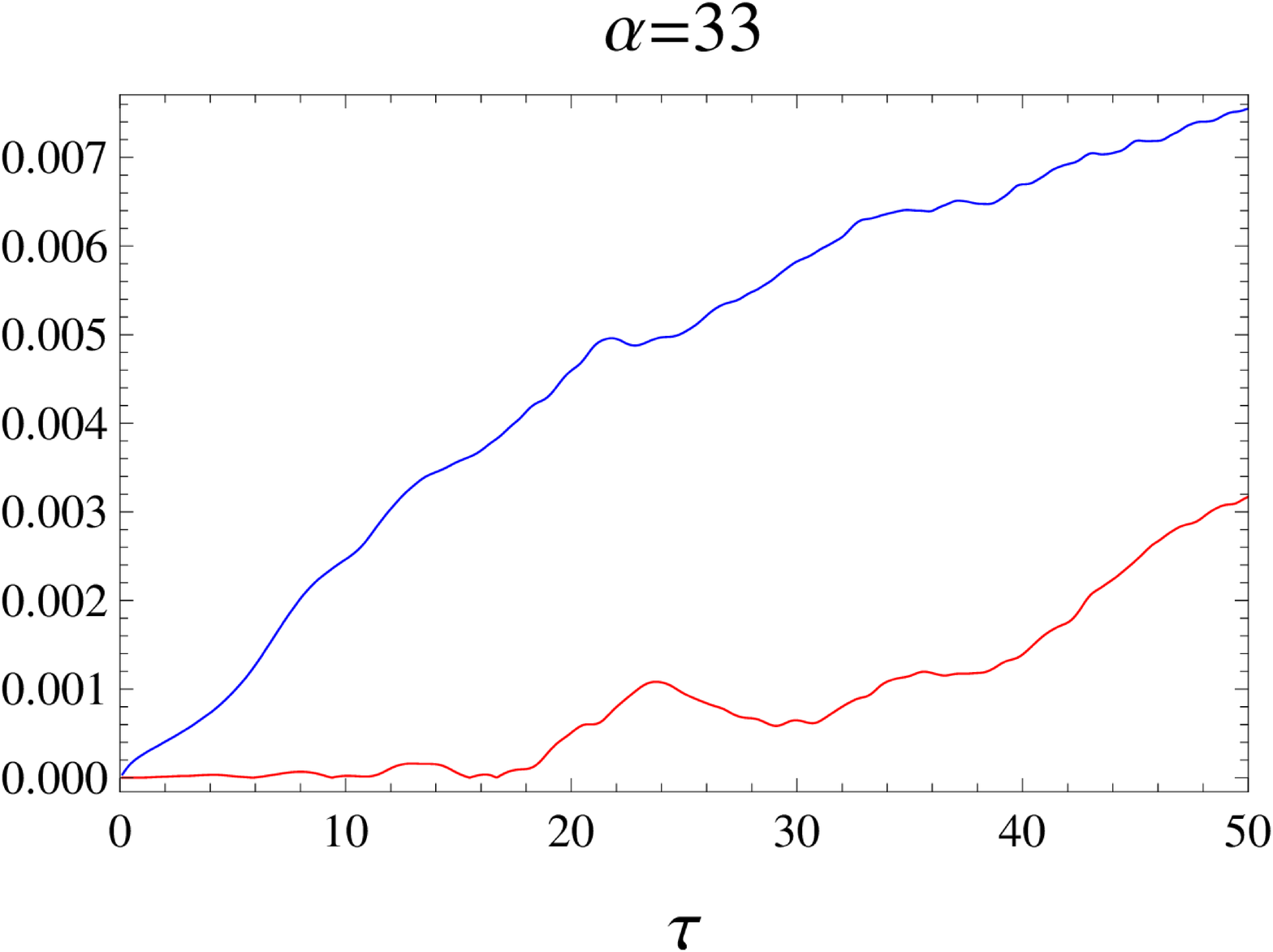}
\caption{\label{LL0102cfr} Overlap of the temporal evolution of $\sqrt{\langle \delta \bar{\xi}'^2_{33} \rangle}$ (blue line) and $\langle\bar{\xi}'_{33} - \bar{\xi}'_{33}(0)\rangle$ (red line), for test particles initialized at $\bar{\xi}'_{33 i}(0)$. Case \emph{(i)} in the left-hand panel and case \emph{(ii)} in the right-hand panel.}
\end{figure}

In order to demonstrate convection in the velocity space in addition to diffusion,  in \figref{LL0102cfr} we plot the quantities $\sqrt{\langle \delta \bar{\xi}'^2_{33} \rangle}$ (blue line) and $\langle\bar{\xi}'_{33} - \bar{\xi}'_{33}(0)\rangle$ (red line) for the same set of tracers considered above (left panel refers to case \emph{(i)}, while right panel to case \emph{(ii)}). The two processes are of the same order of magnitude; and we stress how  convection is enhanced in the presence of linear stable modes. This can be interpreted as a more efficient drag in the velocity space due to the positive power exchange between linear stable modes and particles. In other words, particles can decrease their energy by exciting new modes and, thus, explore (on average) a wider velocity range.

\begin{figure}[ht!]\centering
\includegraphics[width=.3\textwidth,clip]{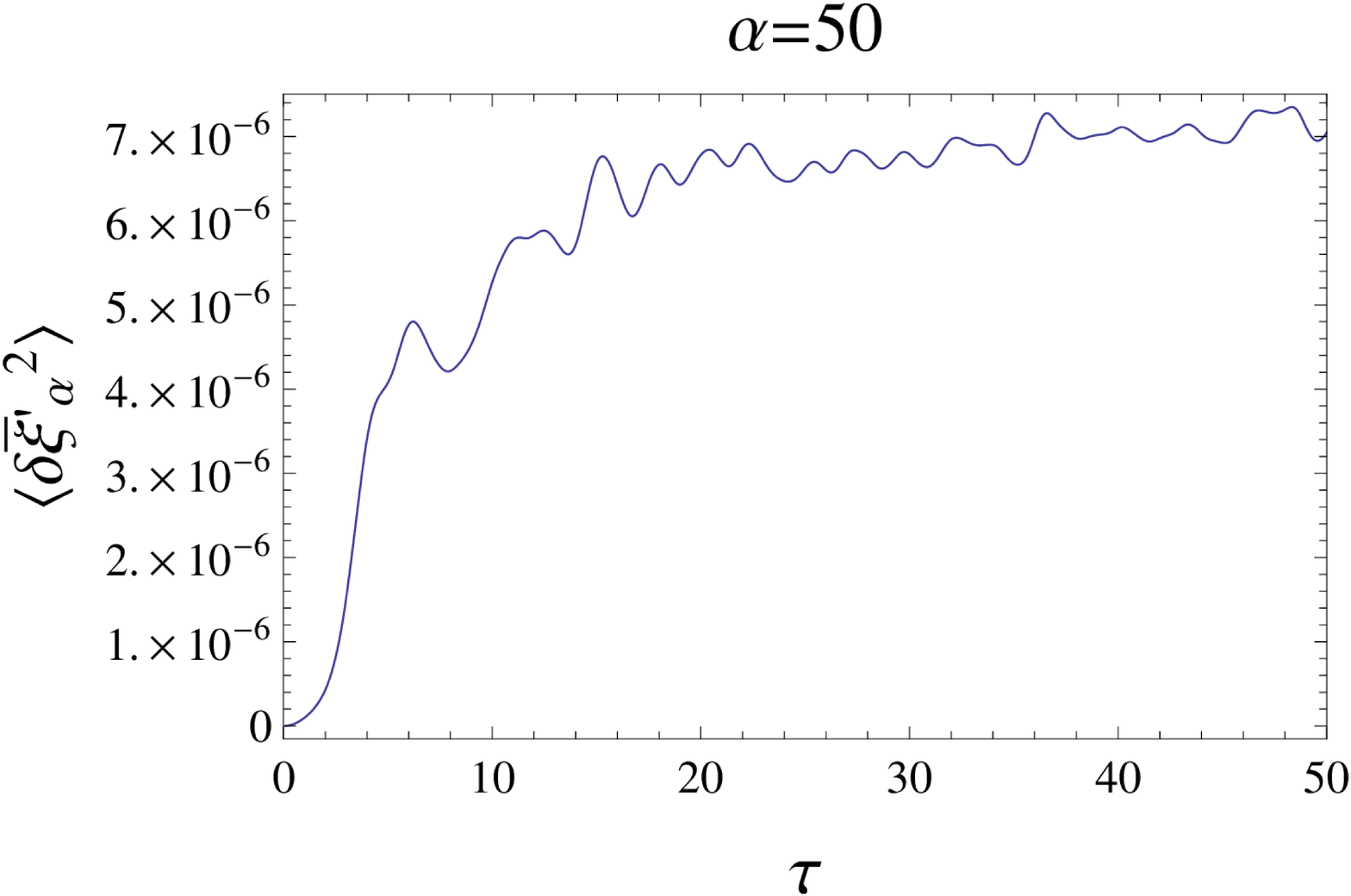}
\includegraphics[width=.3\textwidth,clip]{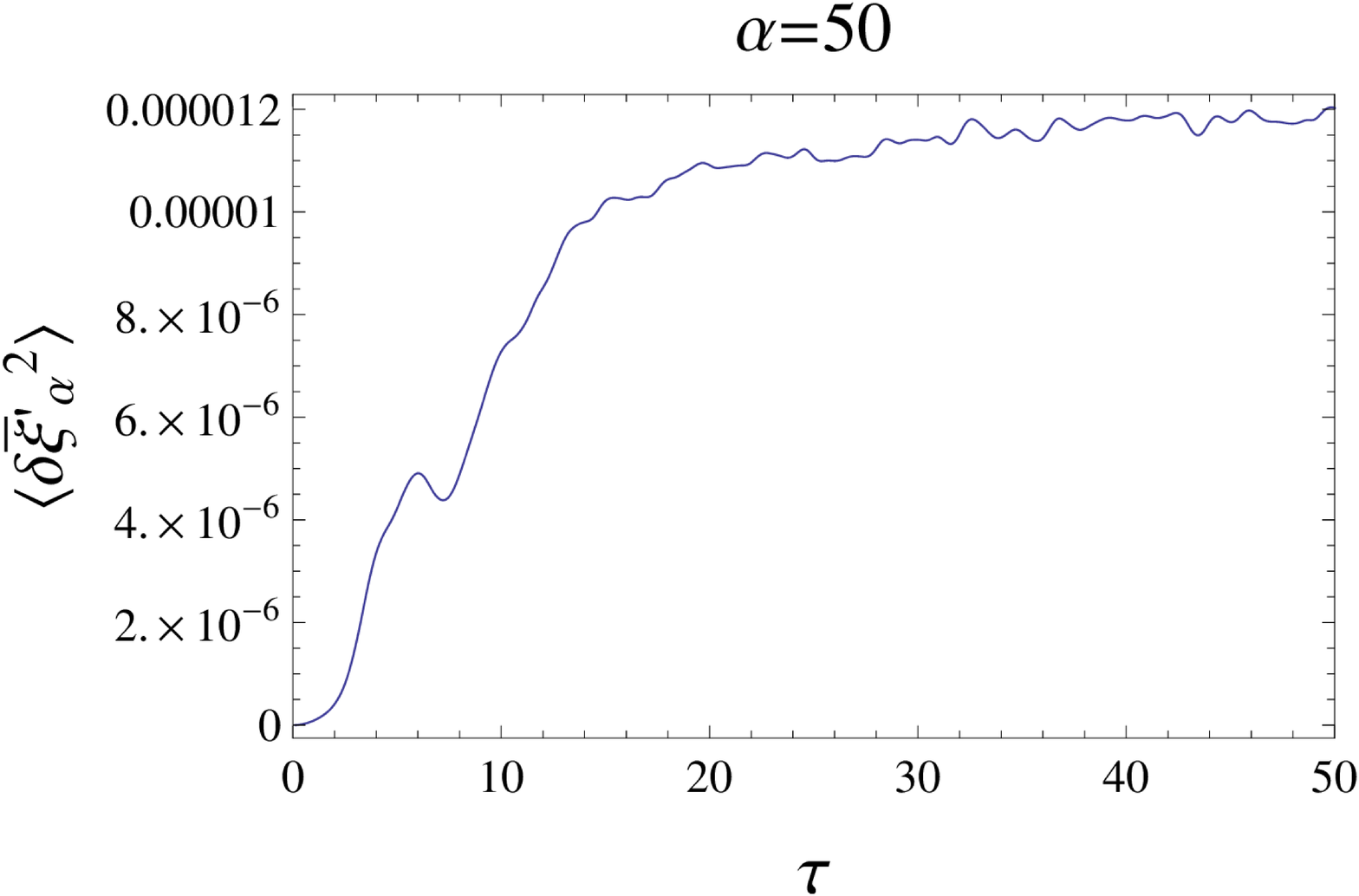}
\caption{\label{LL0304} Time evolution of of $\langle \delta \bar{\xi}'^2_{50} \rangle(\tau)$ for tracers initialized at $\bar{\xi}'_{50 i}(0)$: in the left-hand panel for the case \emph{(iii)} and in the right-hand panel for case \emph{(iv)}.}
\end{figure}

\begin{figure}[ht!]\centering
\includegraphics[width=.3\textwidth,clip]{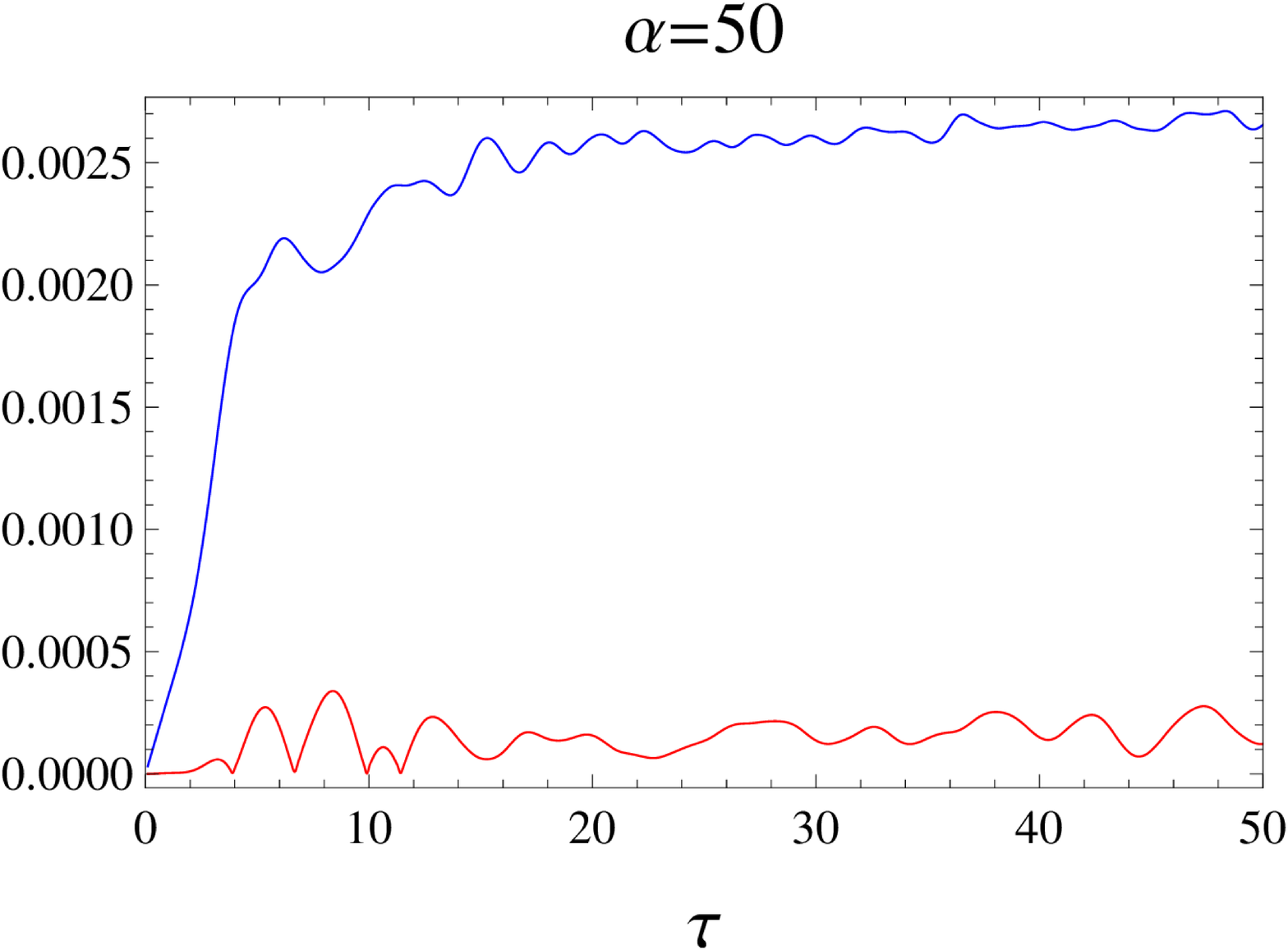}
\includegraphics[width=.3\textwidth,clip]{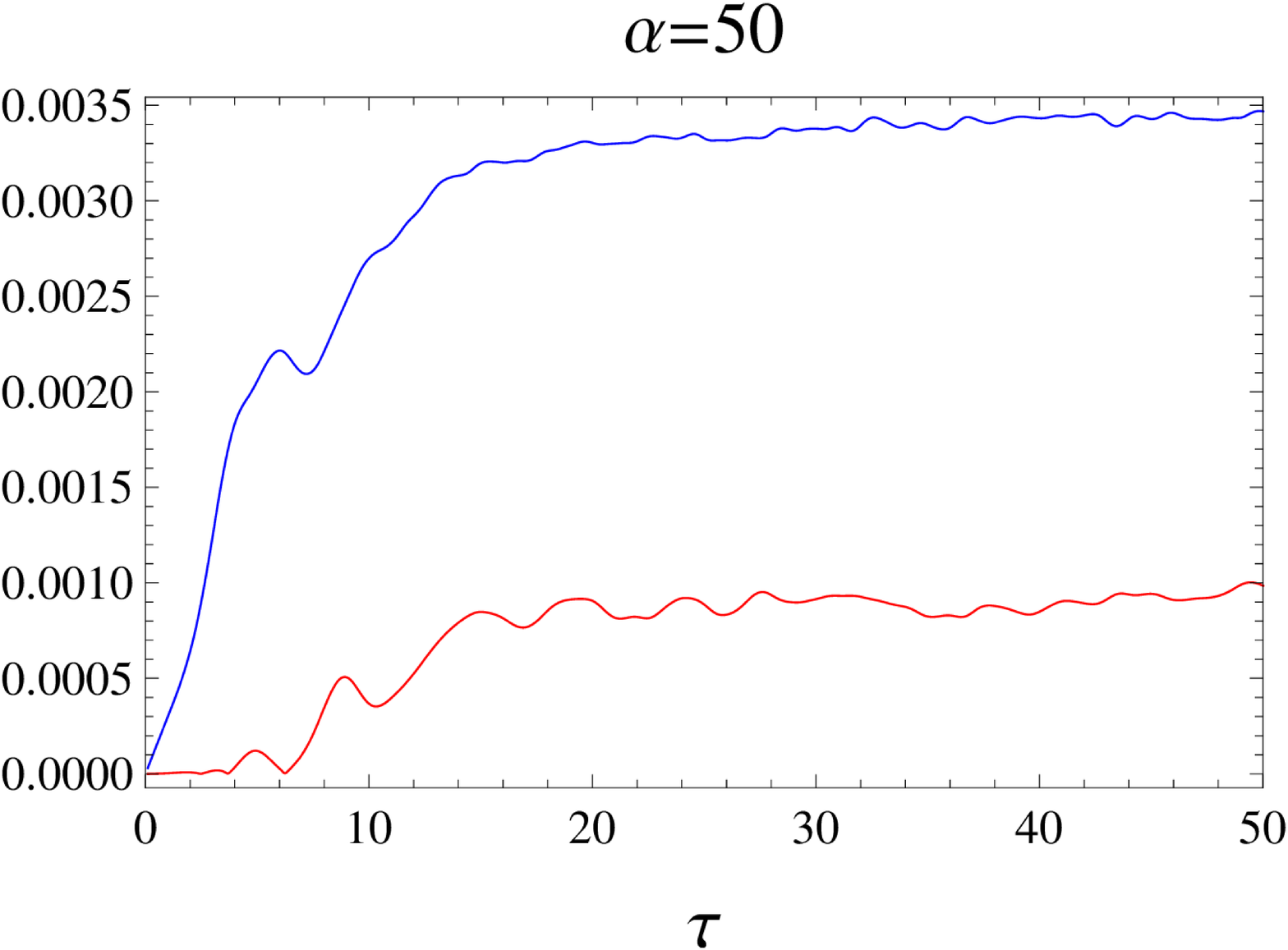}
\caption{\label{LL0304cfr} Overlap of the evolution of $\sqrt{\langle \delta \bar{\xi}'^2_{50} \rangle}$ (blue line) and $\langle\bar{\xi}' - \bar{\xi}'_{50}(0)\rangle$ (red line), for test particles initialized at $\bar{\xi}'_{50 i}(0)$. Case \emph{(iii)} in the left-hand panel and case \emph{(iv)} in the right-hand panel.}
\end{figure}

Let us now repeat this analysis for cases \emph{(iii)} and \emph{(iv)}, corresponding to a ``nearly periodic'' spectrum in the presence or absence of linear stable modes, respectively. For these cases, we consider tracers representing beam $\al=50$, because it is the directly influenced by the stable spectrum. We see in \figref{LL0304} how the case \emph{(iii)} (left-hand panel) does not suggest a diffusive process. This is clearly due to the coherent structures even in the late time evolution of the system, which introduce a strong dependence of transport on velocity and time. As soon as linear stable modes are accounted for (case \emph{(iv)}, right-hand panel), an almost linear behavior of $\langle \delta \bar{\xi}'^2_{50} \rangle$ is recovered in the early evolution of the system ($10\lesssim\tau\lesssim20$). 
The physics underling these two cases is better elucidated when analyzing the behaviors in \figref{LL0304cfr}. In case \emph{(iii)} (left-hand panel), the existence of a significant number of trapped tracers strongly reduces the transport processes in the velocity space. The latter is instead restored to the order of magnitude of $\sqrt{\langle \delta \bar{\xi}'^2_{50} \rangle}$ when stable mode are accounted for. Consistent with the analysis of the previous subsection, all these considerations suggest that the stable part of the spectrum is crucial for diffusion and convection phenomena. Finally, we stress how the restored convection can be regarded to some extent as a prediction of the model discussed in Sec.\ref{dcmodel}, since this case fulfills all the required assumptions (\ie a broad spectrum and $K\gtrsim1$).

\subsection{A toy-model of diffusion-convection relaxation}\label{sims5}

Theoretical analyses of Secs. \ref{botparadigm} and \ref{dcmodel}, and numerical simulation results of Sec.\ref{sims} show that, when ${\cal S} > 1$, the relaxation of a broad beam in cold plasma can be due to both diffusion and convection processes. The value of $Q$, meanwhile, controlled by the width of the fluctuation spectrum, can determine whether coherent periodic behaviors are to be expected, as reflection of wave-particle trapping in a ``narrow'' (nearly periodic) spectrum. 
As the nonlinearity parameter $K \simeq {\cal S}/Q \simeq {\cal S}^2/\Delta \ell$, from Eqs. (\ref{SQ}) and (\ref{KNL}), one may conclude that the convective relaxation discussed in Secs. \ref{botparadigm} and \ref{dcmodel} is relatively unimportant in case \emph{(i)} and/or \emph{(ii)}, where $K<1$. Here, we show that, even in this ``standard'' weak-turbulence limit, significant non-diffusive behavior can be expected, consistent with our numerical simulation results.
In order to illustrate this, we derive a toy-model of diffusion-convection under the assumption that the spectrum be sufficiently broad, \ie that $Q \gg 1$ as well as ${\cal S} > 1$ (the case \emph{(i)} and/or \emph{(ii)} previously treated). 

The goal is to capture the self-consistent evolution of fluctuation spectrum and beam distribution function. Thus, the first model equation should render Eq.(\ref{1}), that is the destabilization of Langmuir waves by the beam particles; while the second one must reflect the nonlinear beam relaxation due to ``weak'' turbulence as described in Eq.(\ref{Diff}). Thus, we can write
\begin{eqnarray}
\label{eq:toymodel}
\partial_{\tau} \mathcal{E} & = & \mathcal{E} \partial_{\bar \xi'}G_{0} \; , \nonumber \\
\partial_{\tau} G_{0} & = & \partial_{\bar \xi'} \left(\mathcal{E}\partial_{\bar \xi'} G_{0}\right) \; ,
\end{eqnarray}
where the dimensionless function $\mathcal{E}$ is connected with the fluctuation intensity and $G_{0}$ with the $k=0$ Fourier component of the beam distribution function, $f_0$. This model can be properly derived from the Vlasov-Poisson system, as represented in the Fourier space under the assumptions above. In particular,
it can be derived from the Dupree's quasilinear equations including trapping \cite{dupree66} in the limit of vanishing broadening with respect to the variation of $f_0$ and 
the fluctuation intensity spectrum, \ie $R \sim 2 \pi \ch (\Delta \omega_k/\gamma_k)^3 \ll 1$ (cf. Sec. 3). By inspection of Eq.(\ref{eq:toymodel}) and comparison with Eqs.(\ref{1}) and (\ref{Diff}), it is possible to show
\begin{eqnarray}
\label{ppaiaoip}
\mathcal{E}(\bar \xi',\tau) & = & \left. 2\pi \etab \frac{m}{\bar\ell \Delta \ell} \left| \phi (k,\tau) \right|^2\right|_{k=(2\pi\bar\ell/L)(1-\etab\bar\ell\bar \xi')} \; , \\
G_0(\bar \xi',\tau) & = & \frac{\etab}{\bar\ell^2} \omp L \left. \frac{f_0(v,\tau)}{n_B}\right|_{v=(\omp L/2\pi\bar\ell)(1+\etab\bar\ell\bar \xi')}\; , \label{eq:rescal}
\end{eqnarray}
where we recall that $m$ is the number of modes in the Langmuir fluctuation spectrum and $\bar\ell$ the reference (central) mode number.
By direct substitution, it is readily verified that Eqs.(\ref{eq:toymodel}) admit the solution
\begin{equation}
\label{eq:G0sol}
G_{0} (\bar \xi',\tau) = \bar{G_{0}}(\bar \xi') + \partial_{\bar \xi'}\mathcal{E} (\bar \xi',\tau) \; ,
\end{equation}
with $\mathcal{E}(\bar \xi',\tau)$ satisfying 
\begin{equation}
\label{eq:toye}
\partial_{\tau} \mathcal{E} = \mathcal{E} \partial_{\bar \xi'} \bar{G_{0}} + \mathcal{E}\partial_{\bar \xi'}^{2}\mathcal{E} \; .
\end{equation}
Here, $\bar{G_{0}}(\bar \xi')$ represents the initial beam distribution function and the source of instability.

To illustrate the ability of Eqs.(\ref{eq:toymodel}) to capture essential features of the broad beam relaxation for $Q\gg1$ as well as ${\cal S}>1$, we solve Eq.(\ref{eq:toye}) numerically and compare
$\mathcal{E}(\bar \xi',\tau)$ and $G_{0}(\bar \xi',\tau)$, obtained from Eq. \reff{eq:G0sol}, with the numerical simulation results for case \emph{(i)} discussed in \figref{fig1.1} and \figref{fig1.2}. 
\begin{figure}[ht!]
\includegraphics[width=.23\textwidth,clip]{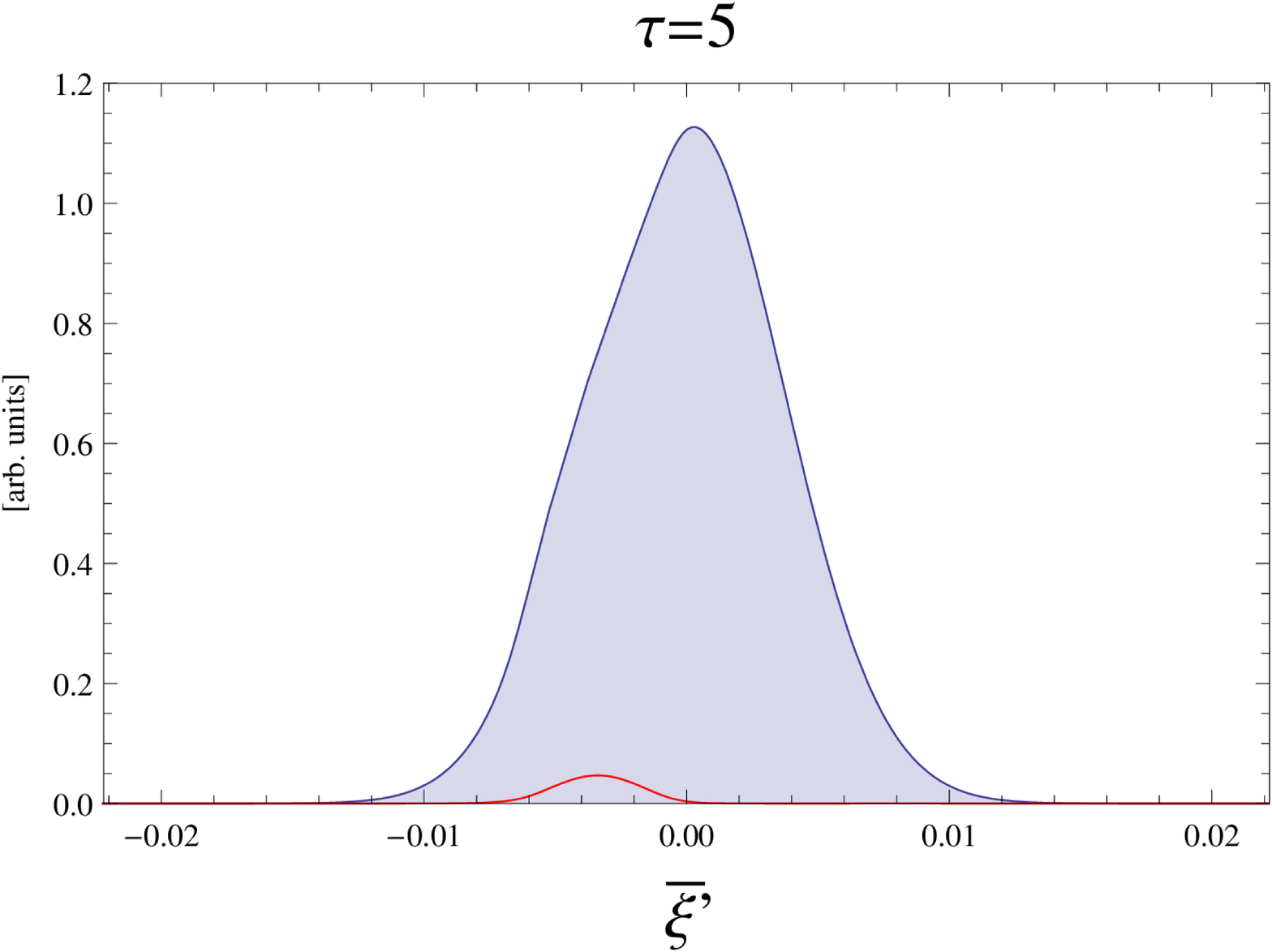}
\includegraphics[width=.23\textwidth,clip]{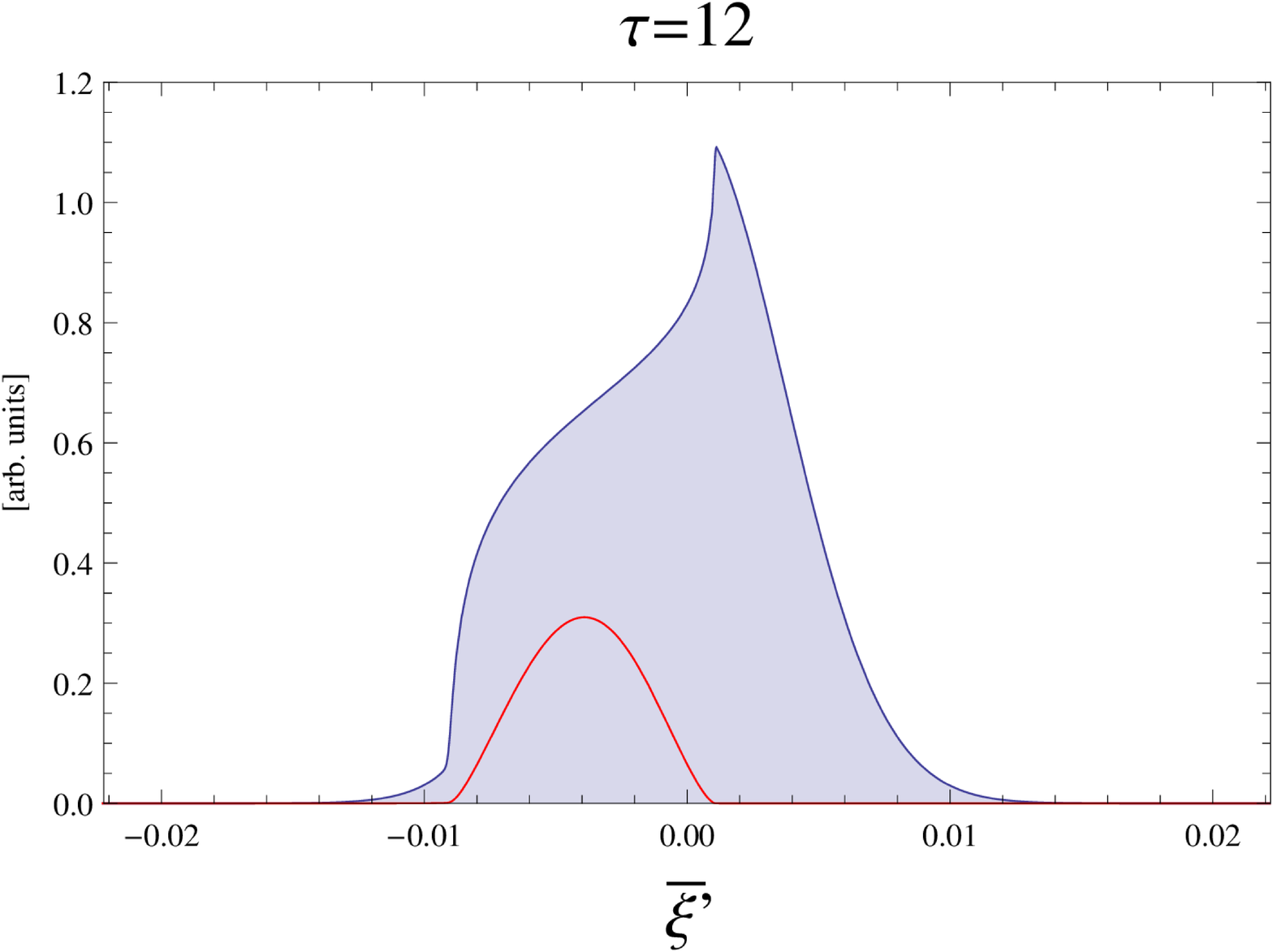}
\includegraphics[width=.23\textwidth,clip]{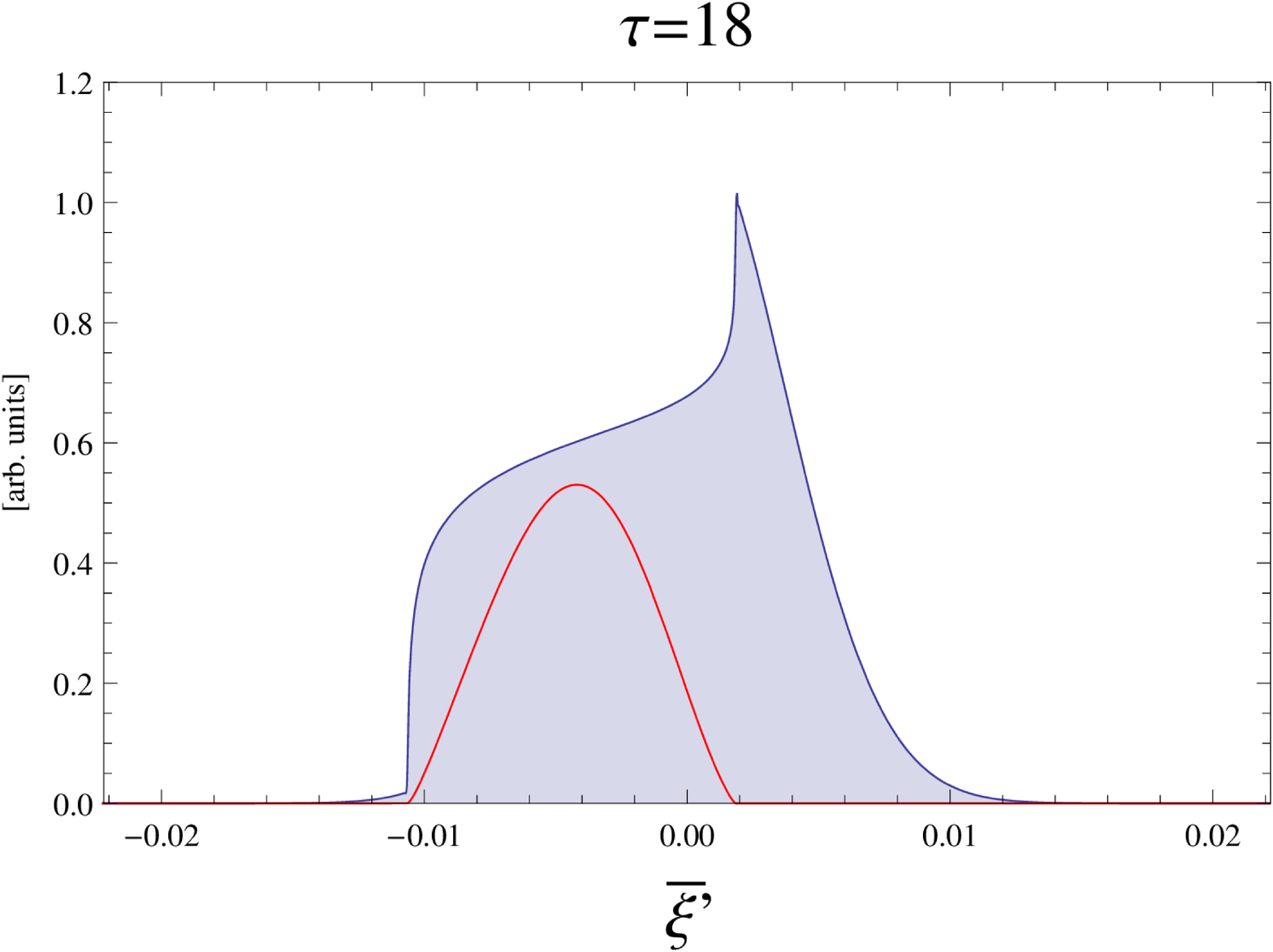}
\includegraphics[width=.23\textwidth,clip]{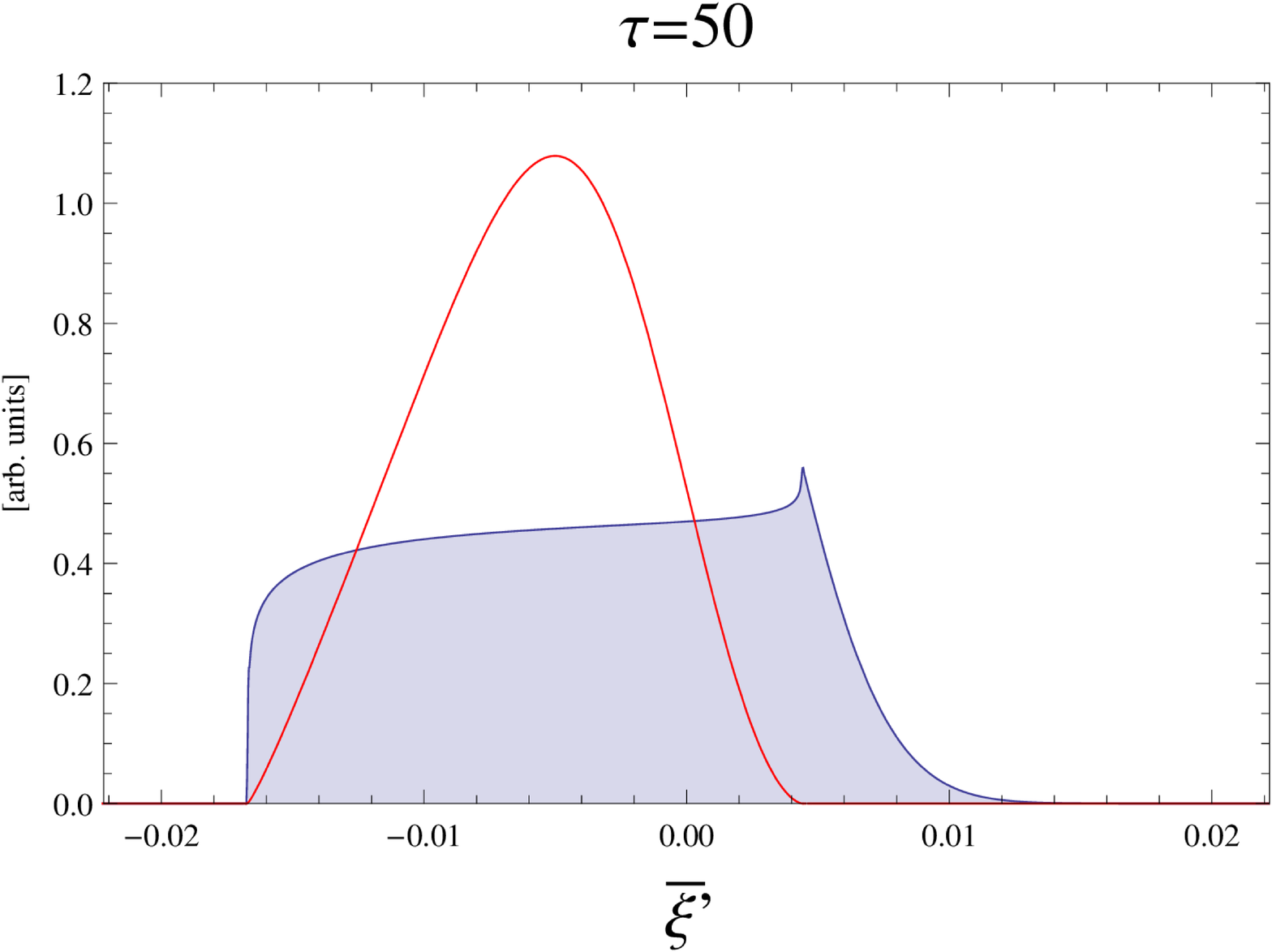}
\caption{\label{figtoymodel} Evolution of the velocity distribution function $G_0$ (blue filled line) and the spectral density $\mathcal{E}$ (red line) as a function of the normalized velocity $\bar{\xi}^{'}$ at different stages of the evolution of \figref{fig1.2}.}
\end{figure}
The initial conditions for $\bar{G_{0}}(\bar \xi')$ are given in order to properly represent the Gaussian distribution of the simulation (see \figref{fig1}) and, for convenience, we also assign a Gaussian $\mathcal{E}(\bar \xi',0)$ profile resembling the initial mode amplitude. The behavior of $\mathcal{E}$ (red line) and $G_{0}$ (blue filled line) for different time steps is shown in \figref{figtoymodel}, well reproducing both the early stages of evolution and the plateau formation at later $\tau$.

Note that $\mathcal{E}(\bar \xi',\tau)$ represents the intensity spectrum evolution discussed in Sec. \ref{sims3} and plays the role of an ``effective diffusion coefficient''. In this respect, we can compare, for fixed times, the behavior of $\mathcal{E}$ with the normalized spectrum evaluated from the right-hand side of \eref{ppaiaoip} 
and the numerical simulation results for $\left| \phi (k,\tau) \right|$.
\begin{figure}[ht!]\centering
\includegraphics[width=.4\textwidth,clip]{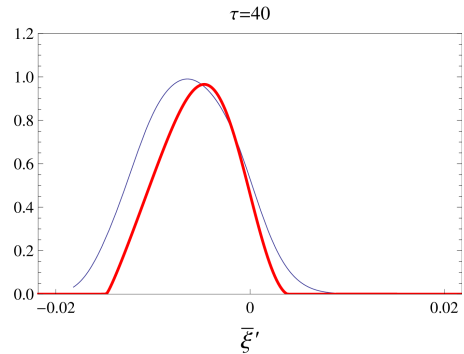}
\caption{\label{diffoeff} Plot of $\mathcal{E}$ numerically estimated from \eref{ppaiaoip} (blue line) and the theoretical values of $\mathcal{E}$ calculated from \eref{eq:toye} (red line), versus the normalized velocity $\bar{\xi}^{'}$ at a fixed time $\tau=40$.}
\end{figure}
This is shown in \figref{diffoeff} for $\tau=50$, comparing the numerical solution of \eref{eq:toye} and the value of $\mathcal{E}$ from \eref{ppaiaoip} evaluated with numerical simulation results for the fluctuation spectrum of case \emph{(i)}. On can easily see that there is a good agreement between numerical simulation results and the toy-model proposed here.

One interesting implication of Eq.(\ref{eq:toymodel}) is to illustrate how non-diffusive behavior may arise in such a system, and how this is a consequence of the self-consistent evolution of fluctuation intensity on the same time scale of particle transport; that is, of the well-known ``conservation relation between the particle
distribution function and the spectral density generated by the instability'' (cf., e.g., p. 152 of A. Hasegawa's book \cite{Hasegawa}). The crucial role of fastest growing modes was already pointed out is Sec. \ref{dcmodel}, since such fluctuations play a key role in the nonlinear evolution. Fastest growing modes are resonant near an inflection point of $G_0$, in the neighborhood of which Eqs.(\ref{eq:toymodel}) can be cast as
\begin{equation}
\partial_\tau G_0 \simeq \partial_{\bar \xi'} {\cal E} \partial_{\bar \xi'} G_0 \simeq \left( G_0 - \bar G_0 \right)  \partial_{\bar \xi'} G_0 \; .
\label{eq:mburgers}\end{equation}
After sufficiently long time that the inflection point of $G_0$ has evolved to a region where the initial beam distribution $\bar{G_{0}}(\bar \xi')$ is sufficiently small, Eq.(\ref{eq:mburgers}) can be cast as the inviscid Burgers equation \cite{Burgers}. This has the formal solution $G_0 = {\cal G}_0\big(\Xi(\bar \xi',\tau)\big)$, with $\Xi(\bar \xi',\tau)$ obtained from
\begin{equation}
\bar \xi' = \Xi - \tau {\cal G}_0(\Xi) \; ; \label{eq:Xi}
\end{equation} and it is known to generate a propagating shock solution, which is visible in the plateau formation at lower speed in the rightmost panel of \figref{figtoymodel}. It is also worthwhile noting that, away from the $\bar{G_{0}}(\bar \xi')$ localization region, Eq. (\ref{eq:toye}) for ${\cal E} \equiv \exp {\cal W}$ reduces to the heat equation with exponential nonlinearity \cite{ovsiannikov59}
\begin{equation}
\partial_{\tau} \mathcal{W} = \partial_{\bar \xi'} \bar{G_{0}} + \partial_{\bar \xi'} \left( e^\mathcal{W} \partial_{\bar \xi'} \mathcal{W} \right) \; .
\end{equation}

\section{Summary and Conclusions}
\label{conclusions}

In this work, we have considered an extension of the familiar quasi-linear diffusion to relaxation processes involving ``broad" (composed of many individual singular) beams of supra-thermal particles in one-dimensional cold plasma, and studied the consequences upon the corresponding generalized kinetic model. For the case of large overlap (large nonlinearity) parameter, we showed a direct generalization employing the convective degree of freedom. The convection process, which we discuss, is induced internally through the intrinsic wave-particle nonlinearities and competes with apparently similar processes due to external force feedback to the velocity distribution. The resulting kinetic model accounting for these generalizations relies on the Klein-Kramers equation for bivariate probability density function and includes the quasi-linear diffusion through the usual entropy-based operators.  

We argued that convective relaxations in the beam-plasma system may occur on intermediate (meso-) time scales and represent kind of ``violent'' relaxation phase. This phase completed, the relaxation process proves to be diffusive quasi-linear and in the long run ($t\rightarrow+\infty$) leads to the formation of the familiar ``plateau'' \cite{Vedenov,Pines,Kadomtzev} in the velocity distribution. Mathematically, the relaxation problem for a broad beam is described by a self-consistent system of coupled nonlinear differential equations, Eqs. (\ref{KKT}) and (\ref{Clos}), which generalize their quasi-linear relatives by directly taking into account the amplification of the resonance domain through the nonlinear interaction and the communication among the beams. Our analysis is complementary to that of Laval and Pesme in Ref. \cite{Laval}, since we consider modifying the convective rather than the diffusion term. Our analysis also applies in a different parameter range; \ie a regime where the fluctuation spectrum is not arbitrarily broad, in order to allow for non-perturbative beam-coupling.

In our numerical studies, we adopt the Hamiltonian formulation of the problem described in Ref. \cite{CFMZJPP}, where the broad supra-thermal particle beam is discretized as superposition of $n \gg 1$ cold beams self-consistently evolving in the presence of $m \geqslant n$ modes nearly degenerate with Langmuir waves. Essential element of this analysis is the crucial role played by wave-particle nonlinearity in determining the non-diffusive feature of supra-thermal particle transport, which self-consistently evolves with the fluctuation intensity spectrum. Thus, parameters other than Chirikov play important roles, such as the ratio of wave-particle trapping time to the autocorrelation time (which is related with the ratio of wave-number spectrum width to the Chirikov parameter), and the nonlinearity parameter, straightforwardly related to the former two.

We performed direct computer simulations of the relaxation dynamics in different regimes, \ie different values of parameters. In this respect, we highlight two cases, namely \emph{(i)} and \emph{(iii)}, characterized by respectively ``broad'' and ``nearly periodic'' (``narrow'') fluctuation spectrum consisting of linearly unstable modes only; and two corresponding cases, these being \emph{(ii)} and \emph{(iv)}, in which the modes of the linear stable spectrum have been accounted for as well. These simulations have elucidated the crucial role played by the linear stable spectrum for transport phenomena through the diffusive and the convective relaxation dynamics. 

Numerical simulation results for the ``narrow'' spectrum exhibit the coherent behavior of spectrum intensity and particle distribution function typical of persistent phase space structures due to wave particle trapping, which suppress convective transport. The plateau in the particle distribution function, in this case, is not formed, unless modes of the linear stable spectrum are accounted for, which not only enhance diffusion but convection as well, because of enhanced de-trapping rate and drag in velocity space.

Numerical simulation results for the ``broad'' spectrum cases, meanwhile, show that a plateau in the distribution function is always formed time asymptotically. Controlling the wave-number spectrum width by including or not the modes of the linear stable spectrum yields behaviors of the diffusion coefficient that are consistent with earlier analyses by Escande et al. \cite{CD92,EE08,EEbook}. Further to this, convective transport on temporal and velocity meso-scales is observed even for small nonlinearity parameter.
This persistent mixed diffusion-convection relaxation is due to the self-consistent evolution of fluctuation intensity on the same time scale of particle transport,
as expected and as demonstrated by an analytical toy-model, whose solution is in good qualitative and quantitative agreement with numerical simulations.
This toy-model provides a valuable tool, allowing the identification of respective roles of convection and diffusion in the velocity space relaxation; in particular,
elucidating how convection processes are not negligible during the relaxation of a bump-on-tail initial profile, especially during the meso-time-scales. 

In conclusion, the present study provides understanding and insights into the mixed diffusion-convection relaxation of a broad beam in a cold one-dimensional plasma in the presence of  weak Langmuir turbulence of varying spectrum width. In particular, crucial roles are played by wave-particle nonlinearity and the self-consistent evolution of particle distribution function with the fluctuation intensity spectrum. These results are of general interest as well as practical importance, in the light of their possible implications as paradigm for Alfv\'enic fluctuation-induced supra-thermal particle transport in fusion plasmas near marginal stability.
\vspace{5mm}\\
{\small\textbf{Acknowledgments:} This work has been carried out within the framework of EUROfusion (European Consortium for the Development of Fusion Energy) as Enabling Research Project ER15-ENEA-03 ``NLED'' (Theory and simulation of energetic particle dynamics and ensuing collective behaviors in fusion plasmas) and has received funding from the Euratom research and training programme 2014-2018 under grant agreement No 633053. The views and opinions expressed herein do not necessarily reflect those of the European Commission.}
\vspace{5mm}\\
{\small\textbf{Author Contributions}: N. Carlevaro, A. V. Milovanov, G. Montani and F. Zonca conceived this work, its motivation, scope and overall structure. A. V. Milovanov conceived the generalization of the quasi-linear theory based on the idea of convective amplification of the resonant domain, using the notion of a propagating instability and the formalism of Klein-Kramers equation.  A. V. Milovanov, G. Montani, M. V. Falessi and F. Zonca developed the theoretical framework for the description and interpretation of the dynamics of the system. N. Carlevaro wrote the code and performed numerical simulations.  N. Carlevaro, D. Terzani and M. V. Falessi analyzed numerical simulation results. D. Terzani, N. Carlevaro and M. V. Falessi analyzed numerically the toy-model. All authors have read and approved the final manuscript.}
\vspace{5mm}\\
{\small\textbf{Conflicts of Interest}: The authors declare no conflict of interest.}

\bibliographystyle{mdpi}
\bibliography{PLASMA}

\begin{thebibliography}{-------}
\providecommand{\natexlab}[1]{#1}

\bibitem[{Vedenov} \em{et~al.}(1961){Vedenov}, {Velikhov}, and
  {Sagdeev}]{Vedenov}
{Vedenov}, A.A.; {Velikhov}, E.P.; {Sagdeev}, R.Z.
\newblock Nonlinear oscillations of rarified plasma.
\newblock {\em Nucl. Fusion} {\bf 1961}, {\em 1},~82--100.

\bibitem[{Drummond} and {Pines}(1962)]{Pines}
{Drummond}, W.E.; {Pines}, D.
\newblock Non-linear Stability of Plasma Oscillations.
\newblock {\em Nucl. Fusion Suppl. Part.} {\bf 1962}, {\em 3},~1049--1057.

\bibitem[{Zaslavsky}(1970)]{Zaslavsky}
{Zaslavsky}, G.
\newblock {\em Statistical Irreversibility in Nonlinear Systems}; Nauka,  1970.

\bibitem[{Sagdeev} and {Zaslavsky}(1988)]{Sagdeev}
{Sagdeev}, R.; {Zaslavsky}, G.
\newblock {\em Introduction to the Nonlinear Physics. From Pendulum to
  Turbulence and Chaos}; Nauka,  1988.

\bibitem[{Kadomtsev}(1988)]{Kadomtzev}
{Kadomtsev}, B.
\newblock {\em Collective phenomena in plasma. 2}; Nauka,  1988.

\bibitem[{Bernstein} \em{et~al.}(1957){Bernstein}, {Greene}, and
  {Kruskal}]{Kruskal}
{Bernstein}, I.; {Greene}, J.; {Kruskal}, M.
\newblock Exact nonlinear plasma oscillations.
\newblock {\em Phys. Rev.} {\bf 1957}, {\em 108},~546.

\bibitem[{O'Neil}(1965)]{oneil65}
{O'Neil}, T.
\newblock Collisionless Damping of Nonlinear Plasma Oscillations.
\newblock {\em Phys. Fluids} {\bf 1965}, {\em 8},~2255.

\bibitem[{Mazitov}(1965)]{mazitov65}
{Mazitov}, R.
\newblock Damping of plasma waves.
\newblock {\em J. App. Mech. Tech. Phys.} {\bf 1965}, {\em 6},~22--25.

\bibitem[{O'Neil} \em{et~al.}(1971){O'Neil}, {Winfrey}, and {Malmberg}]{OWM71}
{O'Neil}, T.; {Winfrey}, J.; {Malmberg}, J.
\newblock Nonlinear Interaction of a Small Cold Beam and a Plasma.
\newblock {\em Phys. Fluids} {\bf 1971}, {\em 14},~1204--1212.

\bibitem[{O'Neil} and {Malmberg}(1968)]{OM68}
{O'Neil}, T.; {Malmberg}, J.
\newblock Transition of the Dispersion Roots from Beam-Type to Landau-Type
  Solutions.
\newblock {\em Phys. Fluids} {\bf 1968}, {\em 11},~1754--1760.

\bibitem[{Shapiro}(1963)]{Sh63}
{Shapiro}, V.
\newblock Nonlinear theory of the interaction of a monoenergetic beam with a
  plasma.
\newblock {\em Sov. Phys. JEPT} {\bf 1963}, {\em 17},~416--423.

\bibitem[{Onishchenko} \em{et~al.}(1970){Onishchenko}, {Linetskii},
  {Matsiborko}, {Shapiro}, and {Shevchenko}]{OL70}
{Onishchenko}, I.; {Linetskii}, A.; {Matsiborko}, N.; {Shapiro}, V.;
  {Shevchenko}, V.
\newblock Contribution to the Nonlinear Theory of Excitation of a Monochromatic
  Plasma Wave by an Electron Beam.
\newblock {\em JETP Letters} {\bf 1970}, {\em 12},~281--285.

\bibitem[{Shapiro} and {Shevchenko}(1971)]{SS71}
{Shapiro}, V.; {Shevchenko}, V.
\newblock Contribution to the Nonlinear Theory of Relaxation of a
  ``Monoenergetic'' Beam in a Plasma.
\newblock {\em Sov. Phys. JEPT} {\bf 1971}, {\em 33},~555--561.

\bibitem[{Thompson}(1971)]{Th71}
{Thompson}, J.
\newblock Nonlinear Evolution of Collisionless Electron Beam-Plasma Systems.
\newblock {\em Phys. Fluids} {\bf 1971}, {\em 14},~1532--1541.

\bibitem[{Levin} \em{et~al.}(1972){Levin}, {Lyubarski{\v i}}, {Onishchenko},
  {Shapiro}, and {Shevchenko}]{L72}
{Levin}, M.; {Lyubarski{\v i}}, M.; {Onishchenko}, I.; {Shapiro}, V.;
  {Shevchenko}, V.
\newblock Contribution to the Nonlinear Theory of Kinetic Instability of an
  Electron Beam in Plasma.
\newblock {\em Sov. Phys. JEPT} {\bf 1972}, {\em 35},~898--901.

\bibitem[{Matsiborko} \em{et~al.}(1972){Matsiborko}, {Onishchenko}, {Shapiro},
  and {Shevchenko}]{Ma72}
{Matsiborko}, N.; {Onishchenko}, I.; {Shapiro}, V.; {Shevchenko}, V.
\newblock On non-linear theory of instability of a mono-energetic electron beam
  in plasma.
\newblock {\em Plasma Phys.} {\bf 1972}, {\em 14},~591--600.

\bibitem[{Dupree}(1966)]{dupree66}
{Dupree}, T.H.
\newblock A Perturbation Theory for Strong Plasma Turbulence.
\newblock {\em Phys. Fluids} {\bf 1966}, {\em 9},~1773--1782.

\bibitem[{Berk} and {Breizman}(1990{\natexlab{a}})]{BB90a}
{Berk}, H.; {Breizman}, B.
\newblock Saturation of a single mode driven by an energetic injected beam. I.
  Plasma wave problem.
\newblock {\em Phys. Fluids B} {\bf 1990}, {\em 2},~2226--2234.

\bibitem[{Berk} and {Breizman}(1990{\natexlab{b}})]{BB90b}
{Berk}, H.; {Breizman}, B.
\newblock Saturation of a single mode driven by an energetic injected beam. II.
  Electrostatic ``universal'' destabilization mechanism.
\newblock {\em Phys. Fluids B} {\bf 1990}, {\em 2},~2235--2245.

\bibitem[{Berk} and {Breizman}(1990{\natexlab{c}})]{BB90c}
{Berk}, H.; {Breizman}, B.
\newblock Saturation of a single mode driven by an energetic injected beam.
  III. Alfv{\'e}n wave problem.
\newblock {\em Phys. Fluids B} {\bf 1990}, {\em 2},~2246--2252.

\bibitem[{Chen} and {Zonca}(2007)]{CZ07}
{Chen}, L.; {Zonca}, F.
\newblock Theory of Alfv{\'e}n waves and energetic particle physics in burning
  plasmas.
\newblock {\em Nucl. Fusion} {\bf 2007}, {\em 47},~S727--S734.

\bibitem[{Chen} and {Zonca}(2013)]{CZ13}
{Chen}, L.; {Zonca}, F.
\newblock On nonlinear physics of shear Alfv{\'e}n waves.
\newblock {\em Phys. Plasmas} {\bf 2013}, {\em 20},~055402.

\bibitem[{Breizman} and {Sharapov}(2011)]{BS11}
{Breizman}, B.; {Sharapov}, S.
\newblock Major minority: energetic particles in fusion plasmas.
\newblock {\em Plasma Phys. Contr. Fusion} {\bf 2011}, {\em 53},~054001.

\bibitem[{Chen} and {Zonca}(2016)]{ZCrmp}
{Chen}, L.; {Zonca}, F.
\newblock Physics of Alfv\'en waves and energetic particles in burning plasmas.
\newblock {\em Rev. Mod. Phys.} {\bf 2016}, {\em 88},~015008.

\bibitem[Fasoli \em{et~al.}(2007)Fasoli, Gormenzano, Berk, Breizman, Briguglio,
  Darrow, Gorelenkov, Heidbrink, Jaun, Konovalov, Nazikian, Noterdaeme,
  Sharapov, Shinohara, Testa, Tobita, Todo, Vlad, and Zonca]{Fa07}
Fasoli, A.; Gormenzano, C.; Berk, H.; Breizman, B.; Briguglio, S.; Darrow, D.;
  Gorelenkov, N.; Heidbrink, W.; Jaun, A.; Konovalov, S.; Nazikian, R.;
  Noterdaeme, J.M.; Sharapov, S.; Shinohara, K.; Testa, D.; Tobita, K.; Todo,
  Y.; Vlad, G.; Zonca, F.
\newblock Chapter 5: Physics of energetic ions.
\newblock {\em Nucl. Fusion} {\bf 2007}, {\em 47},~S264--S284.

\bibitem[{Heidbrink}(2008)]{H08}
{Heidbrink}, W.
\newblock Basic physics of Alfv{\'e}n instabilities driven by energetic
  particles in toroidally confined plasmas.
\newblock {\em Phys. Plasmas} {\bf 2008}, {\em 15},~055501.

\bibitem[{Lauber}(2013)]{La13}
{Lauber}, P.
\newblock Super-thermal particles in hot plasmas - Kinetic models, numerical
  solution strategies, and comparison to tokamak experiments.
\newblock {\em Phys. Rept.} {\bf 2013}, {\em 533},~33--68.

\bibitem[{Vlad} \em{et~al.}(2013){Vlad}, {Briguglio}, {Fogaccia}, {Zonca},
  {Fusco}, and {Wang}]{VB13}
{Vlad}, G.; {Briguglio}, S.; {Fogaccia}, G.; {Zonca}, F.; {Fusco}, V.; {Wang},
  X.
\newblock Electron fishbone simulations in tokamak equilibria using XHMGC.
\newblock {\em Nucl. Fusion} {\bf 2013}, {\em 53},~083008.

\bibitem[{Wang} \em{et~al.}(2011){Wang}, {Briguglio}, {Chen}, {Di Troia},
  {Fogaccia}, {Vlad}, and {Zonca}]{WB11}
{Wang}, X.; {Briguglio}, S.; {Chen}, L.; {Di Troia}, C.; {Fogaccia}, G.;
  {Vlad}, G.; {Zonca}, F.
\newblock An extended hybrid magnetohydrodynamics gyrokinetic model for
  numerical simulation of shear Alfv{\'e}n waves in burning plasmas.
\newblock {\em Phys. Plasmas} {\bf 2011}, {\em 18},~052504.

\bibitem[{Schneller} \em{et~al.}(2013){Schneller}, {Lauber}, {Bilato},
  {Garc{\'{\i}}a-Mu{\~n}oz}, {Br{\"u}dgam}, {G{\"u}nter}, and {the ASDEX
  Upgrade Team}]{ML13}
{Schneller}, M.; {Lauber}, P.; {Bilato}, R.; {Garc{\'{\i}}a-Mu{\~n}oz}, M.;
  {Br{\"u}dgam}, M.; {G{\"u}nter}, S.; {the ASDEX Upgrade Team}.
\newblock Multi-mode Alfv{\'e}nic fast particle transport and losses: numerical
  versus experimental observation.
\newblock {\em Nucl. Fusion} {\bf 2013}, {\em 53},~123003.

\bibitem[{Schneller} \em{et~al.}(2012){Schneller}, {Lauber}, {Br{\"u}dgam},
  {Pinches}, and {G{\"u}nter}]{ML12}
{Schneller}, M.; {Lauber}, P.; {Br{\"u}dgam}, M.; {Pinches}, S.; {G{\"u}nter},
  S.
\newblock Double-resonant fast particle-wave interaction.
\newblock {\em Nucl. Fusion} {\bf 2012}, {\em 52},~103019.

\bibitem[{Gorelenkov} \em{et~al.}(2014){Gorelenkov}, {Pinches}, and
  {Toi}]{GPT14}
{Gorelenkov}, N.; {Pinches}, S.; {Toi}, K.
\newblock Energetic particle physics in fusion research in preparation for
  burning plasma experiments.
\newblock {\em Nucl. Fusion} {\bf 2014}, {\em 54},~125001.

\bibitem[{Zonca} \em{et~al.}(2015{\natexlab{a}}){Zonca}, {Chen}, {Briguglio},
  {Fogaccia}, {Milovanov}, {Qiu}, {Vlad}, and {Wang}]{ZC15ppcf}
{Zonca}, F.; {Chen}, L.; {Briguglio}, S.; {Fogaccia}, G.; {Milovanov}, A.;
  {Qiu}, Z.; {Vlad}, G.; {Wang}, X.
\newblock Energetic particles and multi-scale dynamics in fusion plasmas.
\newblock {\em Plasma Phys. Contr. Fusion} {\bf 2015}, {\em 57},~014024.

\bibitem[{Zonca} \em{et~al.}(2015{\natexlab{b}}){Zonca}, {Chen}, {Briguglio},
  {Fogaccia}, {Vlad}, and {Wang}]{ZC15njp}
{Zonca}, F.; {Chen}, L.; {Briguglio}, S.; {Fogaccia}, G.; {Vlad}, G.; {Wang},
  X.
\newblock Nonlinear dynamics of phase space zonal structures and energetic
  particle physics in fusion plasmas.
\newblock {\em New J. Phys.} {\bf 2015}, {\em 17},~013052.

\bibitem[{Pinches} \em{et~al.}(2015){Pinches}, {Chapman}, {Lauber}, {Oliver},
  {Sharapov}, {Shinohara}, and {Tani}]{Pi15}
{Pinches}, S.; {Chapman}, I.; {Lauber}, P.; {Oliver}, H.; {Sharapov}, S.;
  {Shinohara}, K.; {Tani}, K.
\newblock Energetic ions in ITER plasmas.
\newblock {\em Phys. Plasmas} {\bf 2015}, {\em 22},~021807.

\bibitem[{Berk} \em{et~al.}(1995){Berk}, {Breizman}, {Fitzpatrick}, and
  {Wong}]{BB95b}
{Berk}, H.; {Breizman}, B.; {Fitzpatrick}, J.; {Wong}, H.
\newblock Line broadened quasi-linear burst model [fusion plasma].
\newblock {\em Nucl. Fusion} {\bf 1995}, {\em 35},~1661--1668.

\bibitem[{Berk} \em{et~al.}(1996){Berk}, {Breizman}, {Fitzpatrick}, {Pekker},
  {Wong}, and {Wong}]{BB96b}
{Berk}, H.; {Breizman}, B.; {Fitzpatrick}, J.; {Pekker}, M.; {Wong}, H.;
  {Wong}, K.
\newblock Nonlinear response of driven systems in weak turbulence theory.
\newblock {\em Phys. Plasmas} {\bf 1996}, {\em 3},~1827--1838.

\bibitem[{Ghantous} \em{et~al.}(2012){Ghantous}, {Gorelenkov}, {Berk},
  {Heidbrink}, and {Van Zeeland}]{GG12}
{Ghantous}, K.; {Gorelenkov}, N.; {Berk}, H.; {Heidbrink}, W.; {Van Zeeland},
  M.
\newblock 1.5D quasilinear model and its application on beams interacting with
  Alfv{\'e}n eigenmodes in DIII-D.
\newblock {\em Phys. Plasmas} {\bf 2012}, {\em 19},~092511.

\bibitem[{Ghantous} \em{et~al.}(2014){Ghantous}, {Berk}, and
  {Gorelenkov}]{GBG14}
{Ghantous}, K.; {Berk}, H.; {Gorelenkov}, N.
\newblock Comparing the line broadened quasilinear model to Vlasov code.
\newblock {\em Phys. Plasmas} {\bf 2014}, {\em 21},~032119.

\bibitem[Lauber(2015)]{lauber15}
Lauber, {\relax Ph}.
\newblock {Local and global kinetic stability analysis of Alfv\'en eigenmodes
  in the 15MA ITER scenario}.
\newblock {\em Plasma Phys. Control. Fusion} {\bf 2015}, {\em 57},~054011.

\bibitem[Schneller \em{et~al.}(2016)Schneller, Lauber, and
  Briguglio]{schneller16}
Schneller, M.; Lauber, P.; Briguglio, S.
\newblock {Nonlinear energetic particle transport in the presence of multiple
  Alfv\'enic waves in ITER}.
\newblock {\em Plasma Phys. Control. Fusion} {\bf 2016}, {\em 58},~014019.

\bibitem[{Fisch}(2014)]{Fi14}
{Fisch}, N.
\newblock Some Unsolved Challenges In Radio-Frequency Heating and Current
  Drive.
\newblock {\em Fus. Sci. Technol.} {\bf 2014}, {\em 65},~79--87.

\bibitem[{Hay} \em{et~al.}(2015){Hay}, {Schiff}, and {Fisch}]{HSF15}
{Hay}, M.; {Schiff}, J.; {Fisch}, N.
\newblock Maximal energy extraction under discrete diffusive exchange.
\newblock {\em Phys. Plasmas} {\bf 2015}, {\em 22},~102108.

\bibitem[{Levy} \em{et~al.}(2014){Levy}, {Wilks}, {Tabak}, {Libby}, and
  {Baring}]{LW14}
{Levy}, M.; {Wilks}, S.; {Tabak}, M.; {Libby}, S.; {Baring}, M.
\newblock Petawatt laser absorption bounded.
\newblock {\em Nature Comm.} {\bf 2014}, {\em 5},~4149.

\bibitem[{Levy} \em{et~al.}(2013){Levy}, {Wilks}, {Tabak}, and {Baring}]{LW13}
{Levy}, M.; {Wilks}, S.; {Tabak}, M.; {Baring}, M.
\newblock Conservation laws and conversion efficiency in ultraintense
  laser-overdense plasma interactions.
\newblock {\em Phys. Plasmas} {\bf 2013}, {\em 20},~103101.

\bibitem[{Laval} and {Pesme}(1999)]{Laval}
{Laval}, G.; {Pesme}, D.
\newblock Controversies about quasi-linear theory.
\newblock {\em Plasma Phys. Control. Fusion} {\bf 1999}, {\em 41},~A239--A246.

\bibitem[{Laval} and {Pesme}(1984)]{laval84}
{Laval}, G.; {Pesme}, D.
\newblock Self-Consistency Effects in Quasilinear Theory: A Model for Turbulent
  Trapping.
\newblock {\em Phys. Rev: Lett.} {\bf 1984}, {\em 53},~270--273.

\bibitem[{Cary} \em{et~al.}(1992){Cary}, {Doxas}, {Escande}, and {Verga}]{CD92}
{Cary}, J.; {Doxas}, I.; {Escande}, D.; {Verga}, A.
\newblock Enhancement of the velocity diffusion in longitudinal plasma
  turbulence.
\newblock {\em Phys. Fluids B} {\bf 1992}, {\em 4},~2062.

\bibitem[{Elskens} and {Escande}(2003)]{EEbook}
{Elskens}, Y.; {Escande}, D.
\newblock {\em Microscopic Dynamics of Plasmas Chaos}; Taylor Francis Ltd,
  2003.

\bibitem[{Escande} and {Elskens}(2008)]{EE08}
{Escande}, D.; {Elskens}, Y.
\newblock Self-consistency vanishes in the plateau regime of the bump-on-tail
  instability.
\newblock {\em arXiv:0807.1839} {\bf 2008}.

\bibitem[{Carlevaro} \em{et~al.}(2015){Carlevaro}, {Falessi}, {Montani}, and
  {Zonca}]{CFMZJPP}
{Carlevaro}, N.; {Falessi}, M.; {Montani}, G.; {Zonca}, F.
\newblock Nonlinear physics and energetic particle transport features of the
  beam-plasma instability.
\newblock {\em J. Plasma Phys.} {\bf 2015}, {\em 81},~495810515.

\bibitem[{Zaslavsky} and {Chirikov}(1972)]{Chirikov_USP}
{Zaslavsky}, G.; {Chirikov}, B.
\newblock Stochastic instability of non-linear oscillations.
\newblock {\em Phys. Uspekhi} {\bf 1972}, {\em 14},~549--568.

\bibitem[{Zaslavsky}(2002)]{Report}
{Zaslavsky}, G.
\newblock Chaos, fractional kinetics, and anomalous transport.
\newblock {\em Phys. Repts.} {\bf 2002}, {\em 371},~461--580.

\bibitem[{Chirikov}(1979)]{Ch79}
{Chirikov}, B.
\newblock A universal instability of many-dimensional oscillator systems.
\newblock {\em Phys. Rep.} {\bf 1979}, {\em 52},~263--379.

\bibitem[{Shapiro}(1963)]{shapiro63}
{Shapiro}, V.
\newblock Nonlinear Theory of the Interaction of a Monoenergetic Beam With a
  Plasma.
\newblock {\em Sov. Phys. JETP} {\bf 1963}, {\em 17},~416--423.

\bibitem[{Milovanov} and {Iomin}(2012)]{Mio12}
{Milovanov}, A.; {Iomin}, A.
\newblock Localization-delocalization transition on a separatrix system of
  nonlinear Schr{\"o}dinger equation with disorder.
\newblock {\em Eur. Phys. Lett.} {\bf 2012}, {\em 100},~10006.

\bibitem[{Milovanov} and {Iomin}(2014)]{Mio14}
{Milovanov}, A.; {Iomin}, A.
\newblock Topological approximation of the nonlinear Anderson model.
\newblock {\em Phys. Rev. E} {\bf 2014}, {\em 89},~062921.

\bibitem[{Metzler} and {Klafter}(2004)]{Metzler2004}
{Metzler}, R.; {Klafter}, J.
\newblock The restaurant at the end of the random walk: Recent developments in
  the description of anomalous transport by fractional dynamics.
\newblock {\em J. Phys. A: Math. Gen.} {\bf 2004}, {\em 37},~R161--R208.

\bibitem[{Wax}(1954)]{Wax}
{Wax}, N.
\newblock {\em Selected papers on noise and stochastic processes}; Dover, New
  York,  1954.

\bibitem[{van Kampen}(1981)]{Kampen}
{van Kampen}, N.G.
\newblock {\em Statistic Processes in Physics and Chemistry}; North-Holland,
  Amsterdam,  1981.

\bibitem[{Risken}(1989)]{Risken}
{Risken}, H.
\newblock {\em The Fokker-Planck equation}; Springer-Verlag, Berlin,  1989.

\bibitem[{Milovanov} and {Rasmussen}(2014)]{PLA}
{Milovanov}, A.; {Rasmussen}, J.
\newblock A mixed SOC-turbulence model for nonlocal transport and
  L{\'e}vy-fractional Fokker--Planck equation.
\newblock {\em Phys. Lett. A} {\bf 2014}, {\em 378},~1492--1500.

\bibitem[{Milovanov} and {Rasmussen}(2015)]{JPP}
{Milovanov}, A.; {Rasmussen}, J.
\newblock Self-organized criticality revisited: non-local transport by
  turbulent amplification.
\newblock {\em J. Plasma Phys.} {\bf 2015}, {\em 81},~495810606.

\bibitem[{Volokitin} and {Krafft}(2012)]{VK12}
{Volokitin}, A.; {Krafft}, C.
\newblock Velocity diffusion in plasma waves excited by electron beams.
\newblock {\em Plasma Phys. Control. Fusion} {\bf 2012}, {\em 54},~085002.

\bibitem[{Hasegawa}(1975)]{Hasegawa}
{Hasegawa}, A.
\newblock {\em Plasma Instabilities and Nonlinear Effects}; Springer-Verlag,
  1975.

\bibitem[{Burgers}(1948)]{Burgers}
{Burgers}, J.
\newblock A mathematical model illustrating the theory of turbulence.
\newblock {\em Adv. Appl. Mech.} {\bf 1948}, {\em 1},~171--199.

\bibitem[{Ovsiannikov}(1959)]{ovsiannikov59}
{Ovsiannikov}, L.
\newblock Group properties of nonlinear heat equations [in Russian].
\newblock {\em Doklady AN SSSR} {\bf 1959}, {\em 125},~492.

\end{thebibliography}

\end{document}